\pdfoutput=1
\documentclass[11pt,twoside,a4paper,cmspaper,final,collab]{cms-tdr}
\def\svnVersion{796d111}\def\svnDate{2026/11/24}\def\cmsCernNoTag{CERN-EP-2026-100}\def\cmsCernDate{\today}\def\cmsMessage{Published in the Journal of High Energy Physics as \href{http://dx.doi.org/10.1007/JHEP11(2026)076}{\doi{10.1007/JHEP11(2026)076}.}}
\begin{document}\cmsNoteHeader{HIG-24-006}

\newlength\cmsFigWidth
\setlength\cmsFigWidth{0.4\textwidth}
\newlength\cmsTabSkip\setlength\cmsTabSkip{1.6ex}
\newlength\cmsVecKern
\providecommand{\PGgst}{\HepParticle{\PGg}{}{\ast}}
\newcommand{\qq}{\ensuremath{\Pq\Pq}\xspace}
\newcommand{\mgg}{\ensuremath{m_{\PGg\PGg}}\xspace}
\newcommand{\mH}{\ensuremath{m_\PH}\xspace}
\newcommand{\mjj}{\ensuremath{m_{\mathrm{jj}}}\xspace}
\newcommand{\mll}{\ensuremath{m_{\Pell\Pell}}\xspace}
\newcommand{\V}{{\HepParticle{V}{}{}}\Xspace}
\newcommand{\f}{{\HepParticle{f}{}{}}\Xspace}
\newcommand{\fG}{\ensuremath{f_{a3}^{\Pg\Pg\PH}}\xspace}
\newcommand{\fai}{\ensuremath{f_{ai}}\xspace}
\newcommand{\fC}{\ensuremath{f_{a3}}\xspace}
\newcommand{\fB}{\ensuremath{f_{a2}}\xspace}
\newcommand{\fL}{\ensuremath{f_{\Lambda 1}}\xspace}
\newcommand{\fLZg}{\ensuremath{f_{\Lambda 1}^{\PZ\PGg}}\xspace}
\newcommand{\kappaf}{\ensuremath{\kappa_\f}\xspace}
\newcommand{\kappaftilde}{\ensuremath{\widetilde{\kappa}_\f}\xspace}

\newcommand{\DzVBF}{\ensuremath{\mathcal{D}^\mathrm{VBF}_\mathrm{0-}}\xspace}
\newcommand{\DnnVBFbkg}{\ensuremath{\mathcal{D}^\mathrm{VBF}_\mathrm{NNbkg}}\xspace}
\newcommand{\DnnVBFbsm}{\ensuremath{\mathcal{D}^\mathrm{VBF}_\mathrm{NNBSM}}\xspace}
\newcommand{\DVHhbkg}{\ensuremath{\mathcal{D}^\mathrm{\VH had}_\mathrm{NNbkg}}\xspace}
\newcommand{\DVHhbsm}{\ensuremath{\mathcal{D}^\mathrm{\VH had}_\mathrm{NNBSM}}\xspace}

\newcommand{\DVHbkg}{\ensuremath{\mathcal{D}_\mathrm{STXS}}\xspace}
\newcommand{\DVHbsm}{\ensuremath{\mathcal{D}_\mathrm{BSM}}\xspace}

\newcommand{\DVHlbkg}{\ensuremath{\mathcal{D}^\mathrm{\PW\PH lep}_\mathrm{STXS}}\xspace}
\newcommand{\DVHlbsm}{\ensuremath{\mathcal{D}^\mathrm{\PW\PH lep}_\mathrm{BSM}}\xspace}
\newcommand{\DVHllbkg}{\ensuremath{\mathcal{D}^\mathrm{\PZ\PH lep}_\mathrm{STXS}}\xspace}
\newcommand{\DVHllbsm}{\ensuremath{\mathcal{D}^\mathrm{\PZ\PH lep}_\mathrm{BSM}}\xspace}
\newcommand{\DVHmbkg}{\ensuremath{\mathcal{D}^\mathrm{\VH MET}_\mathrm{STXS}}\xspace}
\newcommand{\DVHmbsm}{\ensuremath{\mathcal{D}^\mathrm{\VH MET}_\mathrm{BSM}}\xspace}
\newcommand{\DzggH}{\ensuremath{\mathcal{D}_{0-}^{\Pg\Pg\PH}}\xspace}
\newcommand{\DCPggH}{\ensuremath{\mathcal{D}_{CP}^{\ggH}}}
\newcommand{\DggHbkg}{\ensuremath{\mathcal{D}^\mathrm{\ggH}_\mathrm{STXS}}\xspace}
\newcommand{\Dggjetsbkg}{\ensuremath{\mathcal{D}^\mathrm{\ggH+2jets}_\mathrm{bkg}}\xspace}
\newcommand{\Dggjetsbsm}{\ensuremath{\mathcal{D}^\mathrm{\ggH+2jets}_\mathrm{BSM}}\xspace}

\newcommand{\ggH}{\ensuremath{\Pg\Pg\PH}\xspace}
\newcommand{\qqH}{\ensuremath{\PQq\PQq\PH}\xspace}
\newcommand{\VhadH}{\ensuremath{\V(\qq)\PH}\xspace}
\newcommand{\VlepH}{\ensuremath{\V(\mathrm{lep})\PH}\xspace}
\newcommand{\VH}{\ensuremath{\PV\PH}\xspace}
\newcommand{\WH}{\ensuremath{\PW\PH}\xspace}
\newcommand{\ZH}{\ensuremath{\PZ\PH}\xspace}
\newcommand{\VBF}{\ensuremath{\mathrm{VBF}}\xspace}
\newcommand{\ttH}{\ensuremath{\PQt\PAQt\PH}\xspace}
\newcommand{\tH}{\ensuremath{\PQt\PH}\xspace}
\newcommand{\Hff}{\ensuremath{\PH\f\f}\xspace}
\newcommand{\Hgg}{\ensuremath{\PH\Pg\Pg}\xspace}
\newcommand{\HVV}{\ensuremath{\PH\PV\PV}\xspace}
\newcommand{\Htt}{\ensuremath{\PH\PQt\PQt}\xspace}
\newcommand{\muf}{\ensuremath{\mu_{\Pf}}\xspace}
\newcommand{\muV}{\ensuremath{\mu_{\PV}}\xspace}

\newcommand{\seff}{\ensuremath{\sigma_{\text{eff}}}\xspace}
\newcommand{\dNLL}{\ensuremath{-2\Delta\ln{\text{L}}}\xspace}

\newcommand{\Hgamgam} {\ensuremath{\PH\to\PGg\PGg}\xspace}
\newcommand{\Hllll} {\ensuremath{\PH\to4\Pell}\xspace}
\newcommand{\HTT} {\ensuremath{\PH\to\PGt\PGt}\xspace}
\newcommand{\Zee} {\ensuremath{\PZ/\PGgst\to\Pe\Pe}\xspace}

\newcommand{\gamplusjet}{\ensuremath{\PGg\!+\!\text{jet}}\xspace}
\newcommand{\jetplusjet}{\ensuremath{\text{jet}\!+\!\text{jet}}\xspace}

\newcommand{\MELA}{\textsc{mela}\xspace}
\providecommand{\JHUGen}{\textsc{JHUGen}\xspace}
\providecommand{\cmsTable}[1]{\resizebox{\textwidth}{!}{#1}}
\newcommand{\MINLO} {{\textsc{powheg+MiNLO}}\xspace}

\cmsNoteHeader{HIG-24-006}
\title{Constraints on anomalous Higgs boson couplings to vector bosons and fermions using the \texorpdfstring{$\PGg\PGg$}{gamma gamma} final state in proton-proton collisions at \texorpdfstring{$\sqrt{s} = 13\TeV$}{sqrt(s) = 13 TeV}}

\date{\today}

\abstract{Possible anomalous couplings of the Higgs boson to vector bosons and fermions are studied using Higgs boson candidates decaying to a pair of photons. The study is based on proton-proton collision data at $\sqrt{s} = 13\TeV$ collected by the CMS experiment, corresponding to an integrated luminosity of 138\fbinv. Events with Higgs boson candidates produced via gluon fusion, electroweak vector boson fusion and in association with a vector boson, are categorized using matrix element techniques and multivariate discriminants. The $CP$ properties of the Higgs boson couplings to gluons through loops of heavy particles, as well as the tensor structure of its interactions with two electroweak bosons, are investigated. The results are interpreted in terms of the fractional contributions of anomalous Higgs boson couplings to the total production cross section of each process and are found to be consistent with the standard model expectations.}
\hypersetup{
  pdfauthor={CMS Collaboration},
  pdftitle={Constraints on anomalous Higgs boson couplings to vector bosons and fermions using the gamma gamma final state in proton-proton collisions at sqrt(s) = 13 TeV},
  pdfsubject={CMS},
  pdfkeywords={CMS, Higgs boson, anomalous couplings}}

\maketitle

\section{Introduction}
\label{sec:intro}

The properties of the Higgs boson (\PH), with a mass of approximately 125\GeV,
discovered by the ATLAS and CMS experiments at the LHC~\cite{Aad:2012tfa,Chatrchyan:2012xdj,Chatrchyan:2013lba}, after extensive studies,
have been found to be consistent with the standard model (SM)
predictions ~\cite{StandardModel67_1, Englert:1964et,Higgs:1964ia,Higgs:1964pj,Guralnik:1964eu,StandardModel67_2,StandardModel67_3}.
In particular, nonzero spin assignments of the \PH have been
excluded~\cite{Khachatryan:2014kca,Aad:2015mxa}, and its spin-parity
quantum numbers are consistent with $J^{PC} = 0^{++}$~\cite{
  Chatrchyan:2012jja,Chatrchyan:2013mxa,Khachatryan:2014kca,Aad:2015mxa,Khachatryan:2015mma,Khachatryan:2016tnr,Sirunyan:2017tqd,Sirunyan:2019twz,
  Sirunyan:2019htt,Chatrchyan:2020htt,CMS-HIG-19-009, CMS-HIG-20-006,
  CMS-HIG-21-013,Aad:2013xqa,Aad:2015mxa,Aad:2016nal,Aaboud:2017oem,Aaboud:2017vzb,Aaboud:2018xdt,Aad:2020mnm,HttAtlas,ATLAS:2021pkb}.

However, the limited precision of current studies allows for small anomalous
couplings of the \PH with two electroweak (EW) gauge bosons (\HVV, $\PV=\PW,\PZ$) or gluons
(\Hgg). In the case where the Hgg interaction is a purely loop-induced process
dominated by the top quark contribution, the access to the interaction between
the Higgs boson and fermions (\Hff) is also provided~\cite{PhysRevLett.76.4468}.

In this paper, we report on a search for anomalous effects, including
possible signs of $CP$ violation, using
the \PH decay into a pair of photons, \Hgamgam. The data are produced during proton-proton
($\Pp\Pp$) collisions at the LHC at \texorpdfstring{$\sqrt{s} = 13\TeV$}{sqrt(s) = 13 TeV} and are collected
with the CMS detector~\cite{CMSdetector,CMS:2023gfb}.

\begin{figure*}[!tbp]
  \centering
  \subfloat[]{\includegraphics[width=0.35\textwidth]{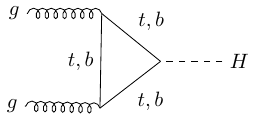}}
  \subfloat[]{\includegraphics[width=0.35\textwidth]{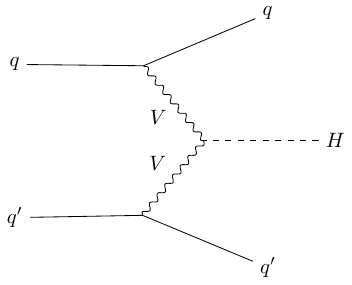}}\\
  \subfloat[]{\includegraphics[width=0.35\textwidth]{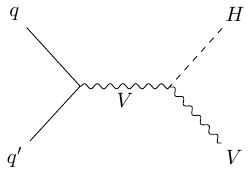}}
  \subfloat[]{\includegraphics[width=0.35\textwidth]{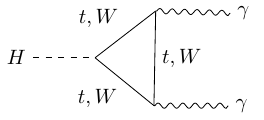}}
  \caption
  {
    LO SM Feynman diagrams for the (a) \ggH, (b) \VBF, and (c) \VH production processes, as well as for the (d) $\Hgamgam$ decay mode.
    \label{fig:Feyn}
  }
\end{figure*}

We use
the same parameterization of the interactions as in previous
CMS studies of anomalous couplings~\cite{Khachatryan:2014kca,Chatrchyan:2013mxa,
  Khachatryan:2015mma,Khachatryan:2016tnr,Sirunyan:2017tqd,Sirunyan:2019twz,Sirunyan:2019htt,
  Chatrchyan:2020htt, CMS-HIG-19-009,CMS:2024bua}. The dominant production
processes employed in this study are EW vector boson fusion
(\VBF) and associated production with a weak vector boson ($\PV\PH$), which probe anomalous \HVV interactions,
and gluon fusion (\ggH), which gives access to \Hff interactions in the
top quark dominated loop. The Feynman diagrams for these processes and
for the \Hgamgam decay are shown in Fig.~\ref{fig:Feyn}.

Previous studies were performed using the \Hllll
channel~\cite{CMS-HIG-19-009}, where $\Pell$ denotes an electron or muon.
In this channel, both production and decay information are employed to
probe anomalous \HVV couplings, with the dominant sensitivity arising
from the production mechanism. Complementary searches for anomalous \Hgg and \HVV couplings
in different final states, with alternative decay modes, can be employed
without significantly reducing the sensitivity to the underlying production dynamics,
such as the
\HTT  ~\cite{Sirunyan:2019htt,HIG-20-007} and
$\PH\to\PW\PW$ channels~\cite{CMS:2024bua}.
The \Hgamgam
channel can substantially improve the sensitivity to such anomalous couplings,
because of the clean, fully
reconstructible final state with two photons.

The structure of the paper is as follows. Section~\ref{sec:pheno} introduces the
phenomenological framework used to parametrize anomalous \HVV and
\Hgg interactions, followed by a description of the analysis
strategy adopted for the extraction of the effective cross section fractions in Section~\ref{sec:analysis_strategy}.
A brief overview of the CMS detector is presented in Section~\ref{sec:detectorAndReco}, while
Section~\ref{sec:samples} summarizes the data and simulated samples employed in the study.
Section~\ref{sec:reconstruction} then outlines the reconstruction of photons, jets, and all other
relevant physics objects.
The production and decay kinematics, together with the discriminants used
in the analysis, are discussed in Section~\ref{sec:kinematics}. Section~\ref{sec:event_selection} describes in detail
the event selection and categorization strategies developed for both the \HVV
and \Hgg analyses. The statistical methodology and the treatment of systematic
uncertainties are presented in Sections~\ref{sec:sig_bkg} and~\ref{sec:Systematics}, respectively.
Finally, Section~\ref{sec:results} reports the results, with tabulated versions provided in HEPData~\cite{hepdata},
and Section~\ref{sec:Summary} concludes the
paper with a summary of the main findings.

\section{Phenomenology of anomalous couplings and cross sections}
\label{sec:pheno}
In this paper, the formalism used in the measurement of \PH couplings
in earlier CMS
analyses is adopted~\cite{Khachatryan:2014kca,Chatrchyan:2013mxa, Khachatryan:2014kca, Khachatryan:2015mma,Khachatryan:2016tnr,Sirunyan:2017tqd,Sirunyan:2019twz,Sirunyan:2019htt, Chatrchyan:2020htt, CMS-HIG-19-009}.
The theoretical approach is described
in Refs.~\cite{Plehn:2001nj,Hankele:2006ma,Accomando:2006ga,Hagiwara:2009wt,Gao:2010qx,DeRujula:2010ys,
  Bolognesi:2012mm,Ellis:2012xd,Artoisenet:2013puc,Anderson:2013afp,Dolan:2014upa,Greljo:2015sla,Gritsan:2016hjl}.

The anomalous interactions of a spin-0 \PH with two spin-1 gauge bosons $\PV_1\PV_2$,
such as $\PW\PW$, $\PZ\PZ$, $\PZ\PGg$, $\PGg\PGg$, and $\Pg\Pg$,
can be written in terms of a
scattering amplitude that includes three tensor structures with expansion of coefficients up
to $(q^2/\Lambda_{1}^2)$:
\begin{equation}
  \mathcal{A}(\PH\PV_1\PV_2) \sim
  \left[ a_{1}^{\V\V}
    + \frac{\kappa_1^{\V\V}q_{V1}^2 + \kappa_2^{\V\V} q_{\V2}^{2}}{\left(\Lambda_{1}^{\V\V} \right)^{2}}
    \right]
  m_{\V1}^2 \epsilon_{\V1}^* \epsilon_{\V2}^*\\
  + a_{2}^{\V\V}  f_{\mu \nu}^{*(1)}f^{*(2)\mu\nu}
  + a_{3}^{\V\V}   f^{*(1)}_{\mu \nu} {\tilde f}^{*(2)\mu\nu},
  \label{eq:formfact-fullampl-spin0}
\end{equation}

where $q_{i}$, $\epsilon_{{\V}i}$, and $m_{{\V}i}$ are the four-momentum,
polarization vector, and pole mass of the gauge boson, indexed by
$i=1,2$. The field strength tensor of the gauge bosons, and its dual are
$f^{(i){\mu \nu}} = \epsilon_{{\V}i}^{\mu}q_{i}^{\nu} - \epsilon_{{\V}i}^\nu
  q_{i}^{\mu}$ and ${\tilde f}^{(i)}_{\mu \nu}
  = \frac{1}{2} \epsilon_{\mu\nu\rho\sigma} f^{(i)\rho\sigma}$, respectively.  The
coupling coefficients $a_{i}^{\V\V}$ and $\kappa_i^{\V\V}/(\Lambda_{1}^{\V\V})^2$, which
multiply the next term in the $q^2$ expansion for the first tensor
structure, where $\Lambda_{1}$ denotes the
scale of beyond the SM (BSM) physics, are to be determined from data.  The convention
$\epsilon_{0123}=+1$ defines the relative sign of the $CP$-odd and
$CP$-even couplings. The sign in front of the gauge fields in the
covariant derivative defines the sign of the photon field and sets the
sign convention of the $\PZ\PGg$ couplings. The conventions adopted
in this analysis are discussed in Section~\ref{sec:samples}.

In the scattering amplitude $\mathcal{A}(\PH\PV_1\PV_2)$ the couplings $a_{i}^{\V\V}$
can be considered as  momentum dependent form factors that,
in general, can have both real and imaginary parts.
The momentum dependence is introduced in the Higgs boson
couplings to the vector bosons by the term $\kappa_i^{\V\V}/(\Lambda_{1}^{\V\V})^2$.

In Eq.~(\ref{eq:formfact-fullampl-spin0}), the only non-zero SM
contributions at tree level are $a_{1}^{\PW\PW}$ and $a_{1}^{\PZ\PZ}$,
which are assumed to be equal under custodial symmetry.  All other
${\PW\PW}$ and ${\PZ\PZ}$ couplings are considered anomalous
contributions, which can be attributed either to BSM physics or small
contributions arising in the SM from loop effects that cannot be
detected with the current precision~\cite{Davis:2021tiv}. Among the
anomalous contributions, considerations of symmetry and gauge
invariance require
$a_{1}^{\PZ\PGg}=a_{1}^{\PGg\PGg}=a_{1}^{\Pg\Pg}=0$,
$\kappa_1^{\PZ\PZ}=\kappa_2^{\PZ\PZ}$,
$\kappa_1^{\PGg\PGg}=\kappa_2^{\PGg\PGg}=0$,
$\kappa_1^{\Pg\Pg}=\kappa_2^{\Pg\Pg}=0$, and
$\kappa_1^{\PZ\PGg}=0$~\cite{Gritsan:2020pib}. For the $\Pg\Pg$
couplings, the only couplings potentially containing BSM contributions
are $a_{2}^{\Pg\Pg}$, which has an SM contribution via loops, and
$a_{3}^{\Pg\Pg}$, which is zero in the SM.  Therefore, there are a
total of 13 independent
parameters that describe the \HVV coupling and two that describe
the \Hgg coupling.  The $a_{3}^{\V\V}$ couplings are $CP$-odd, and
their presence together with any other $CP$-even couplings would
result in $CP$ violation in a given process.

Since the kinematics of the EW \PH production in the ${\PW\PW}$ and in
$\PZ\PZ$ fusion is very similar, it is experimentally impossible to
distinguish between $a_i^{\PW\PW}$ and $a_i^{\PZ\PZ}$ in the \VBF
process. Therefore, a convention for the relative size of the
$\PH\PW\PW$ and $\PH\PZ\PZ$ couplings has to be chosen, and then the
results can be reinterpreted for any chosen relationship between the
$a_i^{\PW\PW}$ and $a_i^{\PZ\PZ}$ couplings~\cite{Sirunyan:2019twz}.
In the following, we assume custodial symmetry $a_i^{\PW\PW}= a_i^{\PZ\PZ}$.

As in previous measurements in the other \PH decay channels, the
effective cross section ratios \fai are measured, rather than the
direct anomalous couplings $a_i$.  They are defined as
follows~\cite{CMS-HIG-19-009}:
\begin{equation}\begin{aligned}
    \fai & = \frac{\abs{a_i}^2 \sigma_{i}}
             {\abs{a_1}^2 \sigma_{1}
               + \abs{a_2}^2 \sigma_{2} + \abs{a_3}^2 \sigma_{3} +  {\tilde\sigma}_{\Lambda_1}/({\Lambda_1})^2
               +  {\tilde\sigma}_{\Lambda_1}^{\PZ\PGg}/({\Lambda_1}^{\PZ\PGg})^2
             }
         & \sgn\left(\frac{a_{i}}{a_{1}}\right),                                                              \\
    \label{eq:fa_definitions}
  \end{aligned}
\end{equation}
where $\sigma_i$ is the cross section for the process corresponding
to $a_i=1$ with all other anomalous couplings set to zero and $\sgn()$ is the sign function.
This approach is convenient as
most uncertainties, notably theoretical ones and those associated with
the integrated luminosity measurement, cancel in the
ratio.  Moreover, the effective fractions are quantities bounded
between $-1$ and 1, independent of the coupling convention.
The quantities  $\tilde{\sigma}_{\Lambda_1}$ and $\tilde{\sigma}_{\Lambda_1}^{\PZ\PGg}$
represent the effective cross sections associated with the presence of anomalous
interactions parametrized via a dimension-6 operator proportional to  ${\Lambda_1}^2$ in effective field theories, given in units of
fb\,$\TeVns^4$.
These cross sections are computed by setting ${\Lambda_1}=1\TeV$ and switching off all other (SM and anomalous) contributions.
The choice
of the sign for the $a_1$ and $a_2^{\PZ\PGg}$ terms follows the
convention introduced in prior
results~\cite{Khachatryan:2014kca,Sirunyan:2017tqd,Sirunyan:2019twz,CMS-HIG-19-009}.
The other sign conventions follow
the \JHUGen~7.0.2~\cite{Gao:2010qx,Bolognesi:2012mm,Anderson:2013afp,Gritsan:2016hjl}
event generator, as discussed in Section~\ref{sec:samples} and
Ref.~\cite{Davis:2021tiv}.
Since the production cross sections depend on the parton distribution functions (PDFs),
the use of cross section ratios defined at the decay level provides a more model-independent treatment.
For this reason, the cross section ratios for each production process are defined for the $\PH\to\PV\PV\to2\Pe2\mu$
as done in
the \Hllll channel~\cite{Khachatryan:2014kca,CMS-HIG-19-009}.  The numerical values can
be found in Ref.~\cite{CMS-HIG-19-009}, and they are computed using
the \JHUGen event generator.  It is assumed that the couplings in
Eq.~(\ref{eq:formfact-fullampl-spin0}) are constant and real, and
therefore this formulation is equivalent to an effective Lagrangian
formalism.

Unlike the \VBF and \VH production, the \ggH process is
loop-induced, and dominated by the top quark contribution in the SM, with a smaller
contribution from the bottom quark~\cite{Hamilton:2015nsa}. In the SM,
this interaction is $CP$-even. Nevertheless, a $CP$-odd contribution to
the \PH coupling to fermions is still allowed and searches for these
effects have been performed in the \ttH production and
\HTT and \Hllll decay channels.
Thus, a study of the \Hgg
coupling provides complementary information on the nature of the \PH
and serves as an indirect search for new phenomena. Both the CMS and
ATLAS Collaborations have previously searched for $CP$-violation in
the \Hgg coupling, but the constraints are of limited sensitivity~\cite{Chatrchyan:2020htt,CMS-HIG-20-006,CMS-HIG-19-009,ATLAS:2021pkb,HIG-20-007,CMS:2024bua}.

The effective cross section fraction for \Hgg couplings can be defined as
\begin{equation}
  \fG = \frac{\abs{a_3^{\Pg\Pg}}^2 } {\abs{a_2^{\Pg\Pg}}^2  + \abs{a_3^{\Pg\Pg}}^2 }
  \, \sgn\left(\frac{a_{3}^{\Pg\Pg}}{a_{2}^{\Pg\Pg}} \right) ,
  \label{eq:fggH_definitions}
\end{equation}
Assuming that other BSM
interactions do not contribute to the gluon fusion loop, measuring a value of \fG
different from zero in the \ggH process is equivalent to measuring $CP$ violation
in Yukawa interactions, which can be parametrized with the amplitude
\begin{equation}
  \mathcal{A}(\Hff) = - \frac{m_f}{v}
  \bar{\psi}_{\f} \left ( \kappaf  + i \, \kappaftilde  \PGg_5 \right ) {\psi}_{\f},
  \label{eq:ampl-spin0-qq}
\end{equation}
where $\bar{\psi}_\f$ and ${\psi}_\f$ are the Dirac spinors, $m_\f$ is the fermion mass, $v$ is the SM Higgs field vacuum expectation value and
$\kappaf$ and $\kappaftilde$ are the $CP$-even and $CP$-odd Yukawa couplings.
In the SM, the couplings have the values $\kappa_\f=1$ and $\tilde\kappa_\f=0$.
Following Ref.~\cite{Gritsan:2020pib} and as stated in Ref.~\cite{Sirunyan:2017tqd}, the $f_{a3}^{\ggH}$ measurement can be expressed in terms of $f_{CP}^{\Hff}$, \ie, the effective cross section fraction for \Hff couplings, assuming that only the top and bottom quarks contribute to the loop with $\kappa_\PQt=\kappa_\PQb$ and $\tilde\kappa_\PQt= \tilde\kappa_\PQb$:
\begin{eqnarray}
  \abs{f_{CP}^{\Htt}} = \left(1 +2.38\left[  \frac{1}{\abs{f_\mathrm{a3}^{\ggH}}} -1  \right]  \right)^{-1}.
  \label{eq:fai-relationship-hgg-tth}
\end{eqnarray}
where the signs of $f_\mathrm{a3}^{\ggH}$ and $f_{CP}^{\Htt}$ are equal.

In this paper, we present a search for anomalous couplings in which one
coupling parameter is measured at a time, while all other anomalous
coupling parameters are fixed to their SM values.
This approach also maintains the structure of the loop-induced amplitude of the \ggH process,
in particular the dominance of the top quark contribution, only
changing the value of the complex coupling relevant in this amplitude.
The procedure is
consistent with the approach adopted in previous analyses, enabling a meaningful comparison with earlier results.

\section{Analysis strategy}
\label{sec:analysis_strategy}

\begin{figure*}[!tbhp]
  \centering
  \includegraphics[width=0.4\textwidth]{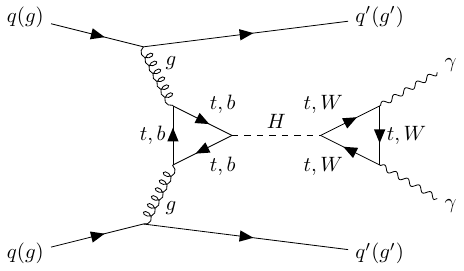}
  \includegraphics[width=0.4\textwidth]{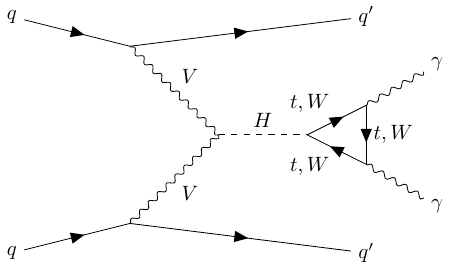}
  \caption
  { Leading-order SM Feynman diagrams for the process in which a Higgs boson decaying into a pair of
    photons is produced via gluon fusion in association with two jets (left) and via vector boson
    fusion (right).
    \label{fig:Feyn_2}
  }
\end{figure*}

The extraction of the \HVV and \Hgg anomalous couplings is done separately,
in two different analyses. The first analysis focuses on the \Hgamgam events
produced through the \VBF and
\VH processes (with \PV decaying freely)  and targets
$\vec{f}_j=(\fB, \fC, \fL, \fLZg)$ parameters. The second one aims at
the production of H in association with two jets through gluon fusion and measures the $\fG$ parameter.
The sensitivity to the \Hgg anomalous coupling is maximal for events with kinematics
similar to those of \VBF production (referred to as '\VBF-like' below),
characterized by the presence of two jets with high energy and large dijet invariant mass (\mjj),
as shown in Fig.~\ref{fig:Feyn_2}.
From the kinematics of these two
jets, and their correlation with the Higgs boson,
it is possible to access the information related to the production vertex, from which \fG can be extracted.
Since most of the sensitivity to BSM effects in \ggH production arises from
\ggH + 2 jets events that significantly overlap with the VBF topology,
two separate analyses are devised, each with its own categorization.
Consequently, an overlap of events between the \Hgg and \HVV analyses is expected,
preventing their combination.

In both cases, events are first required to pass a diphoton preselection,
based on the photon kinematics, shower shape, and isolation variables, described
in Section~\ref{sec:reconstruction}.
Subsequently, dedicated phase space regions are defined using the
properties of the reconstructed diphoton system and any additional
final-state particles to improve the sensitivity of the two analyses and
maximally separate events from various Higgs boson production processes.
Kinematic variables sensitive to variation of the couplings of the \PH either
with vector bosons or gluons (and thus, indirectly, to fermions) are
employed to optimize the sensitivity to anomalous signals.
Discriminants are defined using deep neural networks (DNNs), boosted decision trees (BDTs),
and the matrix element likelihood approach (\MELA)~\cite{Chatrchyan:2012xdj,Chatrchyan:2012jja}, as described in Section~\ref{sec:kinematics}, and are employed for event categorizations.

The goal of the measurement is to extract the four $f_i$ parameters
describing the anomalous \HVV couplings in the \HVV analysis and the $\fG$
parameter in the \Hgg analysis. This is achieved by performing a simultaneous
fit to the resulting diphoton invariant mass \mgg distributions in all analysis categories.

When targeting \HVV couplings through the \VBF and \VH production processes, a similar categorization to the one used in the previous Run~2
CMS simplified template cross section (STXS) measurement in this channel~\cite{HIG-19-015}
is applied.
In the STXS framework, kinematic regions based on the Higgs boson production process and
particle-level quantities are defined.
The definition of these kinematic regions is chosen to minimize the theory dependence,
making the measurements both easier to reinterpret and less affected by potential updates to theoretical predictions.
The categorization is designed to target the individual STXS regions and
is performed using the detector-level equivalents of the particle-level quantities.
As demonstrated in that study, increasing the total number of analysis categories to
target individual STXS bins does not degrade the analysis sensitivity
to the individual production process or the total Higgs boson cross sections.
Events originating from gluon fusion, \tH, and \ttH production,
which are treated as resonant backgrounds in this analysis, are
categorized according to the same STXS binning as defined in the STXS framework.
Categories targeting events produced via \VBF and \VH processes, which might
contain anomalous coupling contributions targeted by this analysis, are instead reoptimized to maximize the
sensitivity to a non-SM coupling of the Higgs boson to vector bosons.

The analysis targeting \Hgg couplings does not use the STXS binning.
The \Hgg analysis selects events with at least two jets, requiring
transverse momenta (\pt) greater than 40 and 30\GeV for the leading and
subleading jets, respectively.
The sensitivity
to anomalous interactions is then optimized by using several kinematic
discriminants to define categories.

\section{The CMS detector}
\label{sec:detectorAndReco}

The CMS detector comprises a silicon pixel and strip tracker, a lead
tungstate crystal electromagnetic calorimeter (ECAL), and a
brass/scintillator hadron calorimeter (HCAL), each composed of a barrel and
two endcap sections, all within a superconducting solenoid of
6\unit{m} internal diameter, providing a magnetic field of
3.8\unit{T}.  Extensive forward calorimetry complements the pseudorapidity ($\eta$)
  coverage
provided by the barrel and endcap detectors.  Outside the solenoid are
the gas-ionization detectors for muon measurements, which are embedded
in the steel flux-return yoke.

Events of interest are selected using a two-tiered trigger system. The
first level, composed of custom hardware processors, uses information
from the calorimeters and muon detectors to select events at a rate of
around 100\unit{kHz} within a fixed latency of about
4\mus~\cite{CMS:2020cmk}.
The second level, known as the
high-level trigger, consists of a farm of processors running a version
of the full event reconstruction software optimized for fast
processing, and reduces the event rate to a few \unit{kHz} before
data storage~\cite{Khachatryan:2016bia,CMS:2024aqx}.

More detailed descriptions of the CMS detector, together with
a definition of the coordinate system used and the relevant kinematic variables,
can be found in Refs.~\cite{CMSdetector,CMS:2023gfb}.

\section{Data samples and simulated events}
\label{sec:samples}

The data samples used in this analysis correspond to integrated
luminosities of 36.3, 41.5, and 59.8\fbinv collected in 2016, 2017, and 2018, respectively,
for a total of 138\fbinv recorded
by the CMS experiment during LHC Run~2 $\Pp\Pp$ collisions at
$\sqrt{s} = 13\TeV$~\cite{CMS-LUM-17-003,CMSlumi2017,CMSlumi2018}.
The 2016 data are split into pre-VFP and post-VFP periods to account for
a change in the pixel detector readout configuration. In this
section, the data sets and simulated event samples for all four periods
are described.

Events are selected using a diphoton high-level trigger with
asymmetric photon \pt thresholds of 30 (30) and 18 (22)\GeV in 2016
(2017 and 2018) data.  A calorimetric selection is applied at trigger
level, based on the shape of the electromagnetic shower, the isolation
of the photon candidate, and the ratio of the hadronic and
electromagnetic energy deposits of the shower. The $\RNINE$ variable
is defined as the energy sum of the $3{\times}3$ crystals centered on
the most energetic crystal in the candidate electromagnetic cluster
divided by the energy of the candidate.
The value of $\RNINE$ is used
to identify photons undergoing a conversion in the material upstream
of the ECAL.  Unconverted photons typically have narrower transverse
shower profiles, resulting in higher values of the $\RNINE$ variable,
compared to converted photons.  The trigger efficiency is measured
with \Zee events using the ``tag-and-probe"
technique~\cite{CMS:2011aa}.  The efficiency, measured in bins
of $\pt$, $\RNINE$, and $\eta$, is used to weight the simulated events
to replicate the trigger efficiency observed in data.  In the \pt range of interest,
the trigger efficiency ranges from 90\% for low-$\RNINE$ photons in the endcaps to 99--100\% for high-$\RNINE$ photons.

Monte Carlo (MC) simulation is used to model signal and background processes
and their reconstruction in the CMS detector.
All parton-level samples are interfaced with {\PYTHIA}8 version 8.226
(8.230)~\cite{Sjostrand:2014zea} for parton showering and
hadronization, with the CUETP8M1~\cite{Khachatryan:2015pea}
(CP5~\cite{Sirunyan:2019dfx}) tune used for the simulation of
2016~(2017 and 2018)~data. The PDFs are
taken from the NNPDF~3.0~\cite{NNPDF3} (3.1~\cite{NNPDF31}) set, when
simulating 2016~(2017 and 2018)~data. The production cross sections
and branching fractions recommended by the LHC Higgs Working
Group are used, computed at next-to-next-to-leading order
in quantum chromodynamics (QCD) and including next-to-leading order  electroweak
(EW) corrections, as reported in Ref.~\cite{deFlorian:2016spz}.
In the analysis targeting the measurement of
the \HVV anomalous couplings,
the relative fraction of each STXS bin for
each inclusive production process at the particle level is taken from
simulation and used to compute the SM production
cross section.
The response of the CMS detector is simulated using the \GEANTfour
package~\cite{Agostinelli:2002hh}. This includes the simulation of
the multiple $\Pp\Pp$ interactions (pileup) taking place in each bunch
crossing.  These can occur at the nominal bunch crossing (in-time
pileup) or at the crossing of previous and subsequent bunches
(out-of-time pileup).
Simulated
out-of-time pileup is limited to a window of $[-12, +3]$ bunch
crossings around the nominal, in which the effects on the observables
reconstructed in the detector are most relevant.  Simulated events are
weighted to reproduce the distribution of the number of interactions.
The average number of interactions per bunch
crossing in the 2016 (2017 and 2018) data sets is 23 (32).

In the analysis targeting the measurement of
the \HVV anomalous couplings, SM \ggH, \VBF and \ttH production processes
are generated using \MGvATNLO (version 2.4.2) at
next-to-leading order (NLO) accuracy~\cite{Alwall:2014hca} in perturbative QCD. Events produced via the associated production of the
Higgs boson with a vector boson (\WH and \ZH) are generated with \POWHEG
2.0~\cite{Nason:2004rx,Frixione:2007vw,Alioli:2008tz,Nason:2009ai,Alioli:2010xd,Hartanto:2015uka}
at NLO  accuracy in perturbative QCD.
For each production process, events are
generated with three different \PH masses, $\mH=120$, 125, and 130\GeV, to account for the dependence of the cross section and signal line shape on the Higgs boson mass.  Events produced via the
gluon fusion and \ttH mechanism are considered as a resonant background in this analysis.
Following the formalism discussed in Section~\ref{sec:pheno}, samples
with the SM and anomalous \PH couplings in \VBF and \VH production are
generated with the~\JHUGen program at leading order (LO).
After including parton shower effects, the \VBF and \VH \JHUGen SM simulations are compared with the
NLO QCD SM samples produced
by \MGvATNLO (\VBF) or \POWHEG (\VH). To account for the small differences observed,
the~\JHUGen SM simulation is reweighted as a function of selected discriminant variables, described in Section~\ref{sec:kinematics},
in order to match the NLO prediction.
In particular, a reweighting is applied as a function of the diphoton transverse momentum,
as well as of the DNN output discriminants:
\DnnVBFbkg for the \VBF process, and  \DVHhbkg,  \DVHhbsm for the VH production.
This procedure reweights the \JHUGen samples to reproduce the NLO predictions
in the DNN input variables used for the event categorization.
The reweighting reaches values of the order of ~10\% at high transverse momentum,
where the differences between NLO and LO predictions are largest, as reported in Ref.~\cite{deFlorian:2016spz}.
The reweighted samples are treated with an associated NLO systematic uncertainty to account for this effect.
The same
weights are applied to describe the kinematics of the \VBF and \VH processes
with anomalous coupling effects.
Moreover, the expected yields are
scaled to match the SM theoretical predictions for the inclusive cross
sections and \Hgamgam branching fraction from
Ref.~\cite{deFlorian:2016spz}.

In the \Hgg analysis, the samples with the SM and anomalous Higgs boson couplings for
the \ggH production process are generated for a Higgs boson produced with 2 jets at NLO
in QCD using \MINLO~\cite{Nason_2020}. This generator is the state-of-the-art
simulation for \ggH+jets processes, which are matched to the parton shower.
Three models are considered: pure $CP$-even ($\fG = 0$), pure $CP$-odd
($\fG = 1$), and an equal mixture of the two ($\fG = 0.5$).
Signal processes (\ggH + 2 jets) are also studied with~\JHUGen samples
for the same three $CP$ models and with \MGvATNLO and \POWHEG for SM contributions.
However, it was observed that \MINLO better describes the processes because of its
treatment of jets and reduced occurrence of negative weights.
The production of the \PH through \VBF, \VH, and \ttH are
considered resonant background processes
in the \Hgg analysis. Concerning SM processes, \MGvATNLO samples are used for the \VBF and
\ttH, \POWHEG samples for \ZH, and \JHUGen samples for \WH.
The production of \VBF, \ZH and \WH is also simulated using \JHUGen at LO QCD for both SM and BSM
processes. The \JHUGen and \MGvATNLO (for \VBF) or \POWHEG (for \ZH) simulations are explicitly compared after parton showering in the SM case, and no significant
differences are found in the relevant kinematic observables. Therefore, the \JHUGen simulation is adopted to describe the kinematics in the \VBF and \VH production processes with anomalous couplings,
with expected yields taken from the reference SM simulation. In the \Hgg analysis only the samples generated with  $\mH=125\GeV$ are used for each process, being the only mass value simulated in the \MINLO signal samples.

In addition, Higgs boson samples produced via \VBF and generated with \POWHEG are used
to train the multivariate discriminants described in Section~\ref{sec:kinematics}.
This choice ensures that an independent dataset is used for the training of the \VBF
production process with respect to the statistical inference. In contrast, for the \ggH analysis
the training is performed using statistically independent subsets of the same samples.

The dominant source of background events in this analysis is SM
diphoton production.  A smaller component comes from \gamplusjet
or \jetplusjet events, in which jets are misidentified as photons.
The diphoton background is generated
with the \textsc{sherpa} (version 2.2.4)
generator~\cite{Gleisberg:2008ta}.  It includes the Born processes
with up to 3 additional jets, as well as the box processes at LO accuracy.
The \gamplusjet and \jetplusjet backgrounds are
simulated at LO with {\PYTHIA}8, after applying a filter at the
generator level to enrich the production of jets with a high
electromagnetic activity.
A sample of Drell--Yan (DY) events is generated with \MGvATNLO, and is used
both to derive corrections for simulation and validation purposes.

In the final fits of the analysis, the background is estimated directly
from the \mgg distribution in data.

A summary of the MC simulation used in the analysis is shown in Table~\ref{tab:MC}.

\begin{table}[ht!]
  \centering
  \caption{
    Summary of the MC simulation used in the analysis
  }

  \begin{tabular}{ll}

    \hline

    \multicolumn{2}{c}{HVV analysis}             \\
    {Process}  & {Event generator} \\
    \hline

    SM \ggH \VBF \ttH & \MGvATNLO                \\
    SM \ZH \WH        & \POWHEG                  \\
    BSM \VBF \ZH \WH  & \JHUGen                  \\

    \hline
    \multicolumn{2}{c}{ggH analysis}             \\
    {Process}  & {Event generator} \\
    \hline

    SM \ggH + 2 Jets  & \MINLO                   \\
    SM \VBF \ttH      & \MGvATNLO                \\
    SM \ZH            & \POWHEG                  \\
    SM \WH            & \JHUGen                  \\
    BSM \ggH + 2 Jets & \JHUGen                  \\
  \end{tabular}
  \label{tab:MC}
\end{table}

\section{Event reconstruction}
\label{sec:reconstruction}
The particle-flow (PF) algorithm~\cite{ParticleFlow} aims to
reconstruct and identify each individual particle (PF candidate) in an
event, with an optimized combination of information from the various
elements of the CMS detector.  The energy of photons is obtained from
the ECAL measurement. The energy of electrons is determined from a
combination of the electron momentum at the primary interaction vertex
as measured by the tracker, the energy of the corresponding ECAL
cluster, and the energy sum of all bremsstrahlung photons spatially
compatible with originating from the electron track~\cite{ElectronPhotonRECO}.  The energy of
muons is obtained from the curvature of the corresponding track.  The
energy of charged hadrons is determined from a combination of their
momentum measured in the tracker and the matching ECAL and HCAL energy
deposits, corrected for zero-suppression effects and for the response
function of the calorimeters to hadronic showers.  Finally, the energy
of neutral hadrons is obtained from the corresponding corrected ECAL
and HCAL energies.

For each event, hadronic jets are clustered from these reconstructed
particles using the infrared and collinear-safe anti-\kt
algorithm~\cite{AntiKt, Cacciari:2011ma} with a distance parameter of
0.4.  Jet momentum is determined as the vectorial sum of all particle
momenta in the jet, and is found from simulation to be, on average,
within 5 to 10\% of the true momentum over the whole transverse
momentum spectrum and detector acceptance.  Pileup interaction
can contribute with additional tracks and calorimetric energy
deposits to the jet momentum.  To mitigate this effect, charged
particles identified to be originating from pileup vertices are
discarded and an offset correction is applied to account for remaining
contributions.  Jet energy corrections are derived from simulation to
bring the measured response of jets to that of particle level jets on
average.  In situ measurements of the momentum balance in dijet,
$\text{photon}$+$\text{jet}$, $\PZ$+$\text{jet}$, and multijet events are
used to account for any residual differences in the jet energy scale
between data and simulation~\cite{JECperformance}.  The jet energy
resolution amounts typically to 15--20\% at 30\GeV, 10\% at 100\GeV,
and 5\% at 1\TeV~\cite{JECperformance}.  Additional selection criteria
are applied to each jet to remove jets potentially dominated by
anomalous contributions from various subdetector components or
reconstruction failures.

The missing transverse momentum vector \ptvecmiss is computed as the
negative vector \pt sum of all the PF candidates in an event, its magnitude is referred to
as the missing transverse energy (MET),
denoted as \ptmiss~\cite{METperformance}.  The \ptvecmiss
is modified to account for corrections to the energy scale of the
reconstructed jets in the event.

Higgs boson candidates are built from pairs of photon candidates, which
are reconstructed from energy clusters in the ECAL not linked to
charged-particle tracks (with the exception of converted photons).
The photon energies are corrected for the containment of
electromagnetic showers in the clustered crystals and the energy
losses of converted photons with a multivariate regression technique
based on simulation~\cite{HggMass}.  The ECAL energy scale in data is
corrected using $\PZ\to\EE$ simulated events smeared to reproduce the
energy resolution measured in data.  The offline diphoton selection
criteria are similar to, but more stringent than, those used in the
trigger. These offline criteria are called ``preselection'' and are described in the following.

The \pt-leading and \pt-subleading photons must have \pt greater than 35 and 25\GeV, respectively, and lie within the ECAL fiducial region, excluding the barrel-endcap transition region ($1.44 <\abs{\eta} < 1.57$).
Additional identification requirements are imposed on the shower shape variables, such as the lateral spread of the electromagnetic shower ($\sigma_{i\eta i\eta}$) and the $R_9$ variable.
Events with energy deposits inconsistent with a single photon are rejected. The ratio of hadronic to electromagnetic energy ($H/E$) is also required to be small, ensuring minimal contamination from hadronic showers.
An electron veto is applied to reject photon candidates matched to a track consistent with originating from an electron, with exceptions made for tracks compatible with reconstructed photon conversions.

Isolation requirements are based on PF quantities that include: photon isolation ($\text{Iso}_\text{ph}$), defined as the scalar sum of the transverse momenta
of PF photon candidates within a cone of radius $\Delta R = \sqrt{(\Delta \eta)^2 + (\Delta \phi)^2} = 0.3$, where $\phi$ is the azimuthal angle in radians; and track isolation ($\text{Iso}_\text{track}$), similarly defined but excluding tracks within an inner cone of $\Delta R = 0.04$ to suppress contributions from photon conversions.
Both photons are required to satisfy at least one of the following conditions: $R_9 > 0.8$, $\text{Iso}_\text{ch}/\pt^{\PGg} < 0.3$, or $\text{Iso}_\text{ch} < 20\GeV$, where $\text{Iso}_\text{ch}$ denotes the charged-hadron isolation, computed as the scalar \pt sum of charged hadrons within a cone of radius $\Delta R = 0.3$. The preselection criteria are summarized in Table~\ref{tab:preselcuts}.

\begin{table}[htbp]
  \caption{List of the \Hgamgam preselection requirements. The EB is the ECAL barrel region, with $\abs{\eta} < 1.44$,
    while EE is the ECAL endcap region, with $1.57 < \abs{\eta} < 2.5$.}
  \centering
  \begin{tabular}{llllll}

                                                      & $R_9$        & $H/E$   & $\sigma_{i\eta i\eta}$ & $\text{Iso}_\text{ph}$ & $\text{Iso}_\text{track}$ \\
    \hline
    \multirow{2}{*}{EB}                               & $[0.5,0.85]$ & $<$0.08 & $<$0.015            & $<4.0\GeV$             & $<$6.0\GeV                \\

                                                      & $>$0.85      & $<$0.08 & \NA                 & \NA                    & \NA                       \\
    \hline
    \multirow{2}{*}{EE}                               & $[0.8,0.90]$ & $<$0.08 & $<$0.035            & $<$4.0\GeV             & $<$6.0\GeV                \\

                                                      & $>$0.90      & $<$0.08 & \NA                 & \NA                    & \NA                       \\

    \multicolumn{6}{c}{Other preselection requirements}                                                                                                   \\
    \hline
    \multicolumn{6}{c}{$R_9>0.8$ or $\text{Iso}_\text{ch}<20\GeV$ or $\text{Iso}_\text{ch}/\pt<0.3$}                                                      \\

    \multicolumn{3}{l}{Leading photon $\pt > 35\GeV$} &
    \multicolumn{3}{l}{Sub-leading photon $\pt > 25\GeV$}                                                                                                 \\

    \multicolumn{6}{c}{$\mgg > 100\GeV$}                                                                                                                  \\
  \end{tabular}
  \label{tab:preselcuts}
\end{table}

Once the preselection has been applied, events must satisfy $\mgg < 180\GeV$, with $\pt/\mgg > 1/3$
for the \pt-leading photon and $\pt/\mgg> 1/4$ for the \pt-subleading photon.

Photons are further required to satisfy a loose identification criterion based on a BDT
classifier, trained to separate photons from jets~\cite{HIG-19-015}.
Inputs to photon identification,
such as shower shape and isolation variables in simulation are
corrected with a chained quantile regression method
based on studies of $\PZ\to\EE$ events. Each variable is corrected
with a separately trained BDT, taking the photon kinematic properties,
per-event energy density, and the previously corrected features as
inputs, to ensure that correlations between the inputs are preserved
and closer to those in data.

\section{Event kinematics and discriminants}
\label{sec:kinematics}

\begin{figure*}[!tbp]
  \centering
  \includegraphics[width=0.49\textwidth]{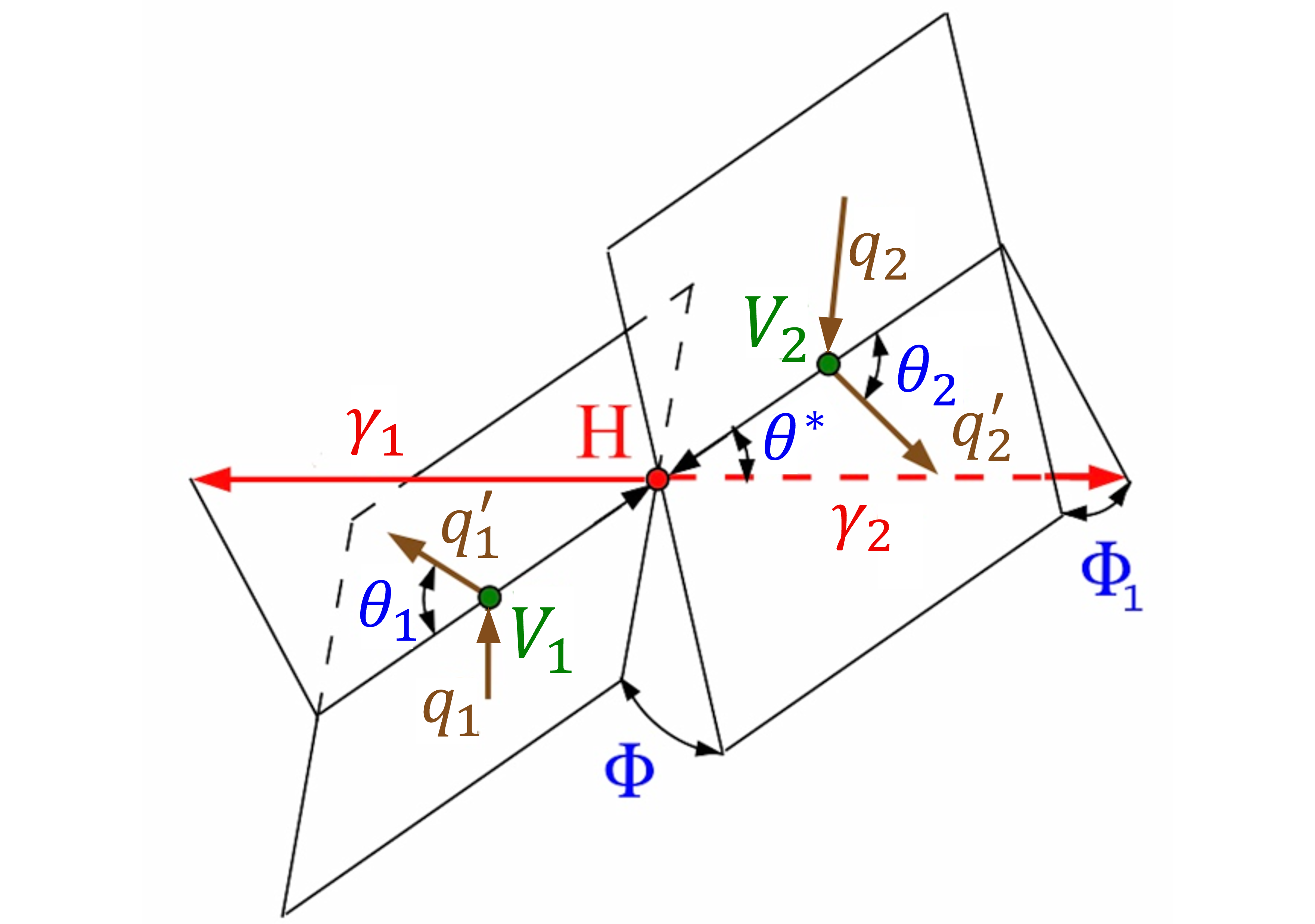}
  \includegraphics[width=0.49\textwidth]{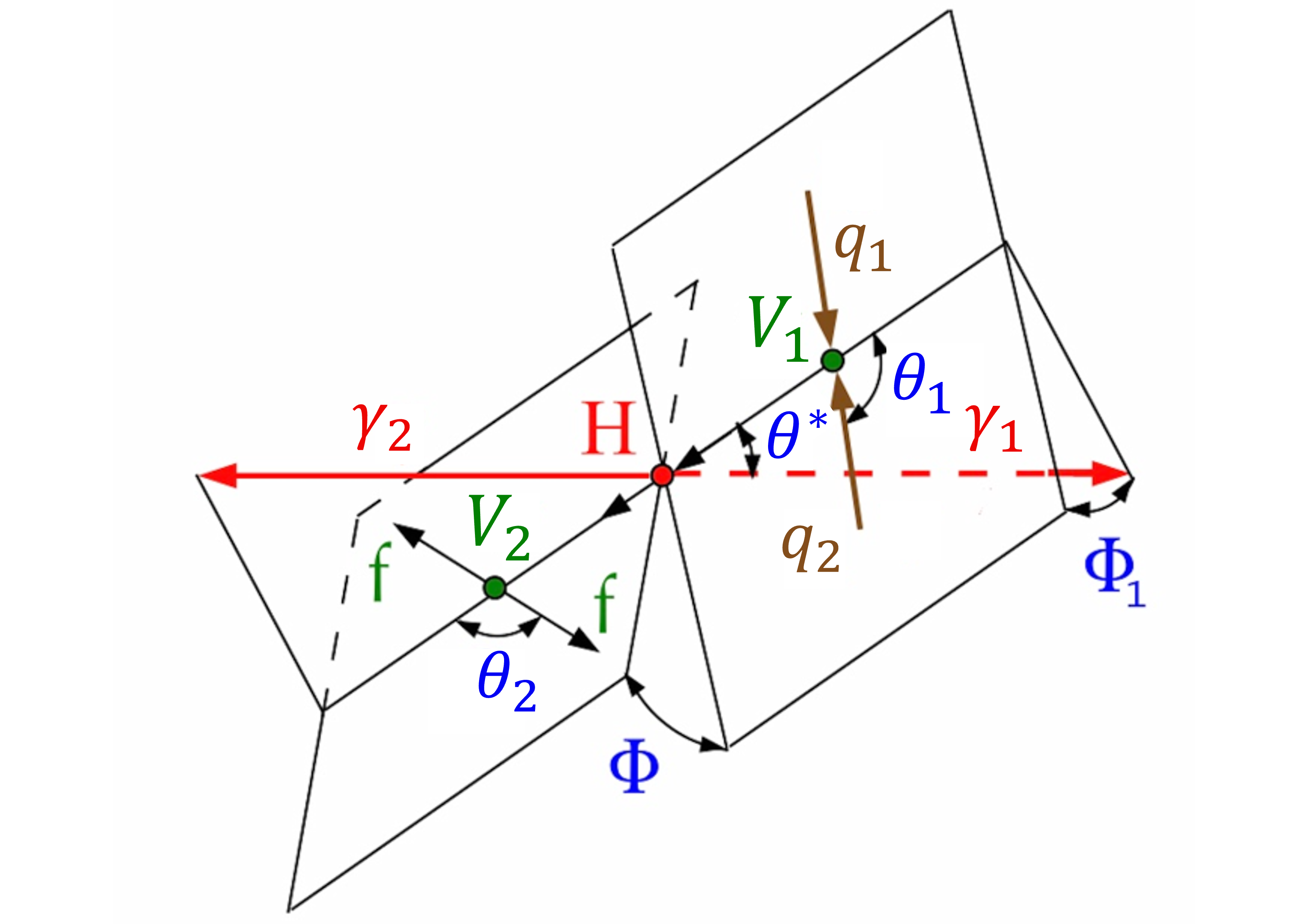}
  \caption{
    Topologies of the \PH production and decay, useful for the
    measurement of \HVV couplings: EW vector boson fusion
    $\Pq_1 \Pq_2\to \PV_1\PV_2 + \Pq_1^{'}\Pq_2^{'} \to \PH + \Pq_1^{'}\Pq_2^{'} \to \PGg_1\PGg_2 + \Pq_1^{'}\Pq_2^{'}$
    (left) and associated
    production $\Pq_1 \Pq_2\to \PV_1 \to \PV_2\PH \to \PGg_1\PGg_2 (\mathrm{ff})$
    (right). The figure on the left is
    valid also to describe gluon fusion events in association with two jets, useful for the
    measurement of \Hgg couplings, when $\PV = \Pg$.
    The incoming partons are shown in brown and the intermediate
    or final-state particles are shown in green and red.  The angles
    characterizing kinematic distributions are shown in blue and are
    defined in the respective rest
    frames~\cite{Gao:2010qx,Anderson:2013afp,Gritsan:2016hjl}.  }
  \label{fig:kinematics}
\end{figure*}

The kinematic distributions of the particles generated in the decay of the
Higgs boson or produced in association with
it are sensitive to the anomalous couplings of the Higgs boson. The squared momentum transfer
of the initial vector bosons or gluons, $p_1^2$ and $p_2^2$, and the five angles indicated in
Fig.~\ref{fig:kinematics} provide the complete
kinematic information of the production and decay of the \PH.
For each production process,
a set of observables can be defined, such as $\mathbf{\Omega}^\text{prod}=\{\theta_1^\text{prod}, \theta_2^\text{prod},
  \theta^{*\text{prod}}, \Phi^\text{prod}, \Phi_1^\text{prod}, p_1^{2,\text{prod}}, p_2^{2,\text{prod}} \}$
(where ``prod'' denotes \VBF, \VH, or \ggH), as shown in Fig.~\ref{fig:kinematics} and
discussed in Ref.~\cite{Anderson:2013afp}. This set of observables
fully characterizes the kinematic distributions of the decay products
(two photons) and the associated particles.

The analysis utilizes both \MELA
and machine-learning (ML) algorithms to optimize the measurement.
The discriminant variables described in this section are used for event selection and categorization.

The \MELA method has been used in earlier analyses
~\cite{Chatrchyan:2012xdj,Chatrchyan:2012jja,Gao:2010qx,Bolognesi:2012mm,Anderson:2013afp,Gritsan:2016hjl}
and is designed to reduce the number of observables to the minimum,
while retaining all essential information.
This approach can be applied to define two types of discriminant variables. One type of discriminant is designed to
separate a signal model from an alternative model, such as
\begin{equation}
  \mathcal{D}_\text{alt}\left(\boldsymbol{\Omega}\right) =  \frac{\mathcal{P}_\text{sig}\left(\boldsymbol{\Omega}\right) }
  {\mathcal{P}_\text{sig}\left(\boldsymbol{\Omega}\right) +\mathcal{P}_\text{alt}\left(\boldsymbol{\Omega}\right) },
  \label{eq:melaD}
\end{equation}
where the probability density $\mathcal{P}$ of a certain process is
calculated using the full kinematic description characterized by
$\boldsymbol{\Omega}$ for both the signal process (denoted ``sig'') and the alternative model (denoted ``alt''). The alternative model could be an
alternative \PH production mechanisms, nonresonant background, or a different \PH
coupling hypothesis (to separate SM and BSM coupling hypotheses).
The \HVV analysis employs the discriminant \DzVBF,
designed to provide the optimal separation between
an SM \PH produced via \VBF and a pseudoscalar \PH.
This discriminant is also found to provide good separation
against all the other \VBF anomalous couplings.
The \Hgg analysis uses \DzggH, designed to
distinguish between
an SM \PH produced via gluon fusion and a pseudoscalar \PH.

The second type of discriminant isolates the interference contribution:
\begin{equation}
  \mathcal{D}_\text{int}\left(\boldsymbol{\Omega}\right) =
  \frac{\mathcal{P}^\text{int}_\text{sig-alt}\left(\boldsymbol{\Omega}\right) }
  {\mathcal{P}_\text{sig}\left(\boldsymbol{\Omega}\right) + \mathcal{P}_\text{alt}\left(\boldsymbol{\Omega}\right) }
  ,
  \label{eq:melaDcp}
\end{equation}
where $\mathcal{P}^\text{int}_\text{sig-alt}$ is the probability distribution of the interference component for a process with a mixture of the SM
and the alternative anomalous contributions.
This type of discriminant, commonly used in this type of analysis,
defined as $\mathcal{D}_{CP}^{\ggH}$, is only used in \Hgg analysis and is $CP$-odd.
An asymmetry of the distribution around zero would indicate $CP$ violation.

In both the \HVV and \Hgg analyses, the \MELA discriminants are paired with additional ML algorithms, as described in Section~\ref{sec:event_selection}.
This allows for a more efficient separation of signal events with modified Higgs boson couplings from
SM-like Higgs boson events via all production processes and from the nonresonant SM background
($\PGg\PGg$, $\PGg$+jet and QCD multijet contributions).

In the \HVV analysis, two different Higgs boson production processes are considered
as potential sources of anomalous couplings to vector bosons, \VBF and \VH, and categories are defined accordingly.
Events categorized as \VH production are further divided into
events where the vector boson decays hadronically, \VhadH,
and events where the vector boson decays leptonically, \VlepH,
with the lepton ($\ell$) being either a muon or an electron. For the \VBF and \VhadH cases, two dedicated
DNN classifiers are trained, optimized for \VBF or \VhadH topologies,
respectively.
Each of these two classifiers produces two output discriminants used in the analysis.
One set, \DnnVBFbsm and \DVHhbsm, is optimized to enhance the separation between the
SM \PH signals and anomalous coupling signal hypotheses in the corresponding production process.
The other set, \DnnVBFbkg and \DVHhbkg, is optimized to
discriminate between the \PH signals and the nonresonant SM background.
Categories targeting \VH production with leptonic decays of the associated vector boson are divided
into three regions, depending on the number of reconstructed charged leptons.
Events with no reconstructed leptons are assigned to the \VH MET
categories. The region with two same-flavor leptons is assigned to
$\PZ(\Pell\Pell)\PH$ categories, while the region with exactly one reconstructed lepton to the
$\PW(\Pell\PGn)\PH$ ones.

To reduce the nonresonant background
contamination each region is further divided using the same BDT output developed for the STXS analysis
~\cite{HIG-19-015} (\DVHmbkg, \DVHlbkg, and \DVHllbkg) and to separate the SM
\VlepH events from several \VlepH scenarios with anomalous couplings (\DVHmbsm, \DVHlbsm, and \DVHllbsm).

In the case of the \Hgg analysis, in addition to the \MELA discriminants \DzggH and \DCPggH,
the standard diphoton multivariate analysis (MVA) is employed (\DggHbkg)
to suppress the nonresonant background, which increases the signal significance by about 30\%.
This MVA output is commonly used in several
analyses of the Higgs boson decay properties in the diphoton
channel~\cite{HIG-19-015}. Furthermore, a multiclass BDT
classifier is trained to separate the $CP$-even
and $CP$-odd \ggH signal processes from the inclusive background, consisting
of nonresonant and \PH background processes. Two output discriminants of the
classifier are used, one separating the inclusive background
from the other contributions (\Dggjetsbkg) and other targeting the $CP$-odd \ggH
signal process (\Dggjetsbsm).

The discriminant variables used in the \HVV and \Hgg analyses are reported
in Tables~\ref{tab:ac_discriminants_HVV} and~\ref{tab:ac_discriminants_Hgg}.

\begin{table}[ht!]

  \caption{
    List of discriminants for separating anomalous couplings from the SM
    contribution in the \HVV analysis. The third column indicates the targeted discrimination for
    that specific observable. Discriminants in this table are only used for event categorization.
  }
  \centering
  \cmsTable{
    \begin{tabular}{ccl}

      Process                  & Discriminant & Main goal                                                     \\
      \hline                                                                                                  \\
      \VBF                     & \DzVBF       & Separate between $CP$-even, $CP$-odd and mixed $CP$ scenarios \\
      \VBF                     & \DnnVBFbkg   & Separate \PH signal from nonresonant backgrounds              \\
      \VBF                     & \DnnVBFbsm   & Separate between SM \PH and several BSM \PH scenarios         \\
      [\cmsTabSkip]
      \hline                                                                                                  \\
      \VhadH                   & \DVHhbkg     & Separate \PH signal from nonresonant backgrounds              \\
      \VhadH                   & \DVHhbsm     & Separate between SM \PH and several BSM \PH scenarios         \\
      [\cmsTabSkip]
      \hline                                                                                                  \\
      $\PW(\Pell\PGn)\PH$-lep  & \DVHlbkg     & Separate \PH signal from nonresonant backgrounds              \\
      $\PW(\Pell\PGn)\PH$-lep  & \DVHlbsm     & Separate \PH signal from several BSM \PH scenarios            \\
      [\cmsTabSkip]
      \hline                                                                                                  \\
      $\PZ(\Pell\Pell)\PH$-lep & \DVHllbkg    & Separate \PH signal from nonresonant backgrounds              \\
      $\PZ(\Pell\Pell)\PH$-lep & \DVHllbsm    & Separate \PH signal from several BSM \PH scenarios            \\
      [\cmsTabSkip]
      \hline                                                                                                  \\
      $\PZ(\PGn\PGn)\PH$-MET   & \DVHmbkg     & Separate \PH signal from nonresonant backgrounds              \\
      $\PZ(\PGn\PGn)\PH$-MET   & \DVHmbsm     & Separate \PH signal from several BSM \PH scenarios            \\
      [\cmsTabSkip]                                                                                           \\
    \end{tabular}
    \label{tab:ac_discriminants_HVV}
  }
\end{table}

\begin{table}[ht!]
  \centering
  \caption{
    List of discriminants for separating anomalous couplings from the SM
    contribution in the \Hgg analysis. The third column indicates the targeted discrimination for
    that specific observable. For the \DzggH
    discriminant, the ``$\ensuremath{\Pg\Pg\PH}$" label indicates that this observable
    is constructed using matrix elements computed for the \ggH
    production process to differentiate it from the equivalent
    discriminant for the \VBF process (\DzVBF). Discriminants in this table are only used for event categorization.
  }
  \cmsTable{
    \begin{tabular}{ccl}

      Process & Discriminant & Main goal                                                                    \\
      \hline                                                                                                \\
      [\cmsTabSkip]
      \ggH    & \DzggH       & Separate between $CP$-even, $CP$-odd and mixed $CP$ scenarios                \\
      \ggH    & \DCPggH      & Differentiate the interference  between $CP$-even and $CP$-odd contributions \\
      \ggH    & \DggHbkg     & Separate \PH signal from nonresonant backgrounds                             \\
      \ggH    & \Dggjetsbkg  & Separate between (SM and $CP$-odd) \ggH + 2 jets signal from                 \\
              &              & resonant and nonresonant background                                          \\
      \ggH    & \Dggjetsbsm  & Separate between BSM $CP$-odd \ggH + 2 jets signal from SM and               \\
              &              & resonant and nonresonant backgrounds                                         \\
      [\cmsTabSkip]
    \end{tabular}
    \label{tab:ac_discriminants_Hgg}
  }
\end{table}

\section{Event selection and categorization}
\label{sec:event_selection}

To provide sensitivity to different production mechanisms and anomalous
couplings hypotheses, events are divided into various analysis categories.  Each
category is designed to select as many events as possible
from a given production mechanism, and maximize the
separation between SM and BSM hypotheses. Moreover, the selection
rejects as many continuum background events as possible,
to minimize the statistical uncertainty. This section
describes the categorization schemes used for
different event topologies and different spin-0 boson hypotheses,
applied to either the \HVV (Section~\ref{sec:VBF_VH_cat}) or
\Hgg (Section~\ref{sec:ggH_cat}) analyses.

\subsection{\texorpdfstring{\VBF}{VBF} and \texorpdfstring{\VH}{VH} event categories}
\label{sec:VBF_VH_cat}
The sensitivity to \HVV couplings comes from measuring \VBF or \VH production.
It is impossible to make reconstruction categories that are
pure in either \VBF or \VH production, especially from the dominant
signal production process, \ggH, but also from \ttH, which, despite its
small cross section, may have events with additional jets that populate the \VBF or hadronic \VH reconstructed
categories.

The analysis employs the orthogonal phase-space partition defined in the STXS framework of Ref.~\cite{HIG-19-015},
optimized to target the different Higgs boson production modes: \ggH, \ttH, \tH, \VBF, and \VH.
The categories corresponding to the resonant production modes (\ggH, \ttH, and \tH) are
identical to those used in the STXS analysis, while a dedicated categorization strategy
is adopted for the \VBF and \VH production modes to enhance the sensitivity to anomalous \HVV couplings.

\subsubsection{The \texorpdfstring{\VBF}{VBF} categories} \label{sec:vbfcategories}
The \VBF production process is characterized by the presence of two highly energetic jets,
with large dijet invariant mass \mjj and well separated in pseudorapidity, originating
from the fragmentation of the quarks in the collision.
Events where the dijet system is instead consistent with the
decay of a vector boson are categorized separately, as described later
in this section. No analysis categories are constructed to target the
zero- or one-jet \qqH STXS bins.
For the \VBF categories, events are required to have at least two jets with $\mjj > 350\GeV$.
The \pt-(sub)leading jet must have $\pt > 40\GeV$ ($\pt > 30\GeV$). Both jets are required to have $\abs{\eta} < 4.7$. This list of criteria is referred to as the ``\VBF preselection''.

\begin{figure}[ht!]
  \centering
  \includegraphics[width=0.45\textwidth]{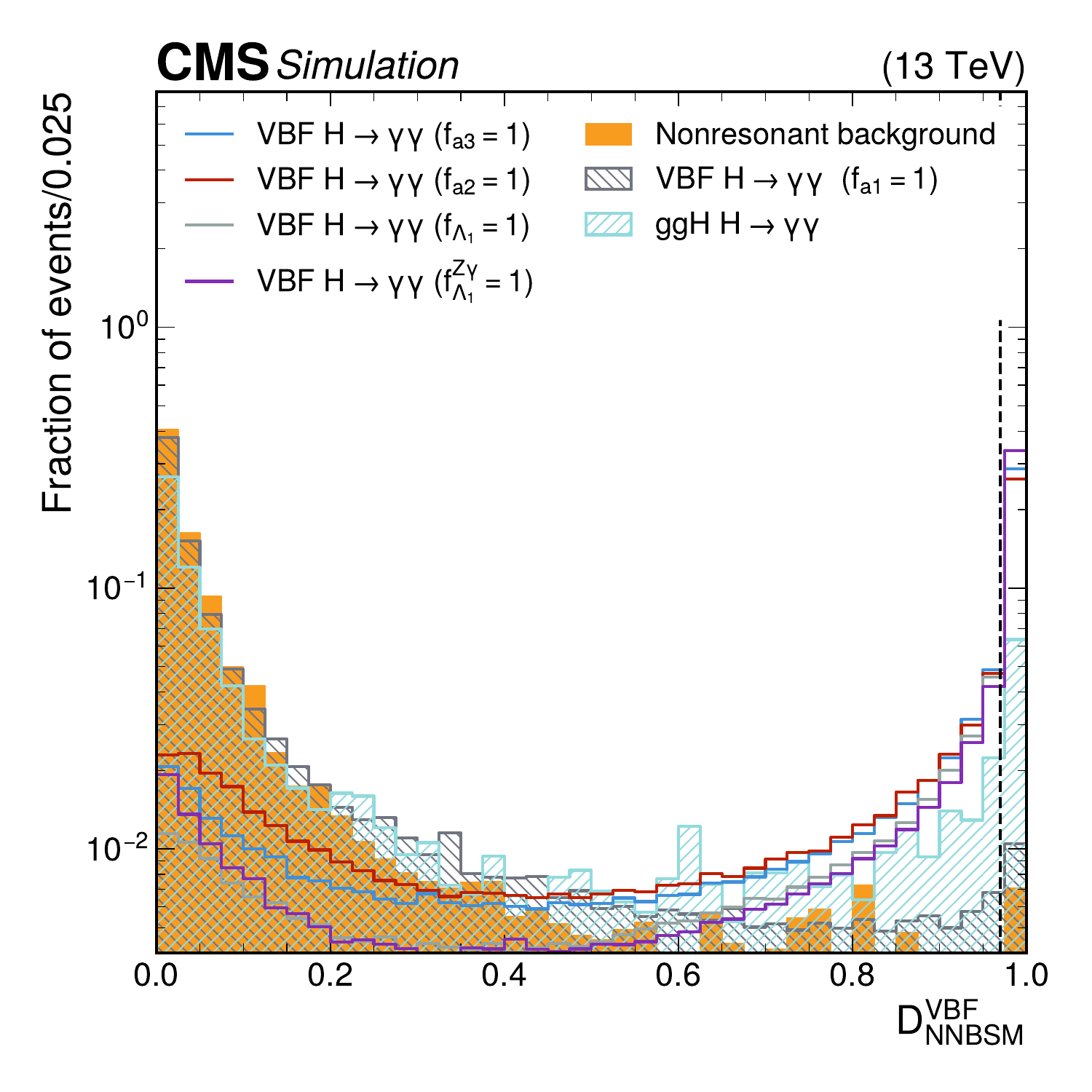}
  \includegraphics[width=0.45\textwidth]{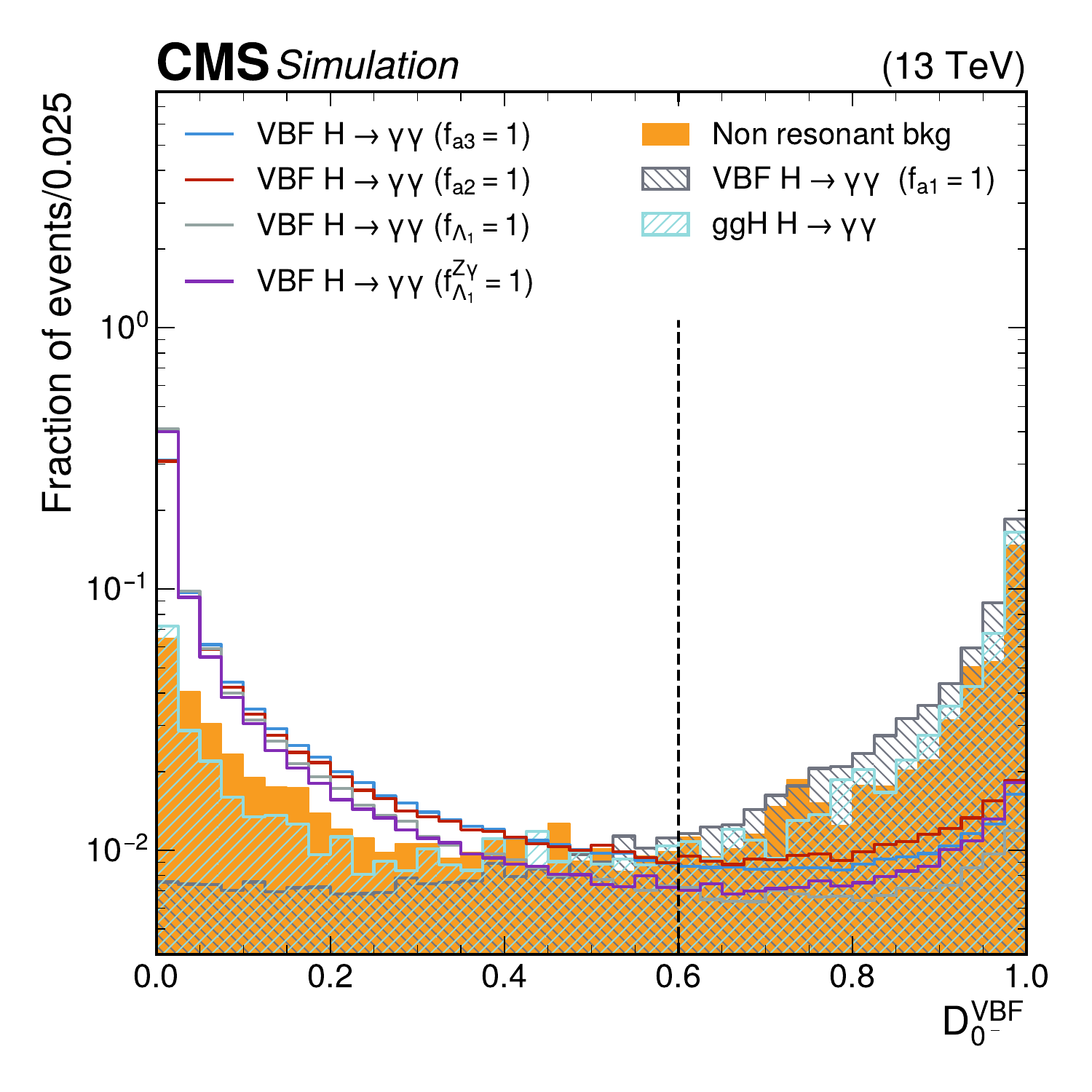}
  \caption{
    Distribution of the \DnnVBFbsm (left) and \DzVBF (right) discriminant for the SM \VBF signal and for four anomalous coupling hypotheses, shown together with the main resonant background (SM \ggH production), and the continuum diphoton background. The distributions are shown after the \VBF preselection described in the text and are normalized to the unit area. The vertical dashed lines indicate the category boundaries applied in the analysis. \label{fig:vbf_distr}}
\end{figure}

Events with the \VBF topology are divided into two bins in each of the three
discriminants: \DzVBF, \DnnVBFbkg, and \DnnVBFbsm, resulting in a
three-dimensional phase space composed of eight categories.
Several boundary conditions were tested in this three-dimensional space,
and the final categorization was chosen as the one yielding the highest sensitivity to a $CP$-odd signal.
The first variable,
\DzVBF, is a \MELA discriminant designed to separate $CP$-even and $CP$-odd scenarios,
while the other two variables, \DnnVBFbkg and \DnnVBFbsm, are two output classes
of a DNN trained to separate
the \VBF signal and background,
as well as the SM \VBF and BSM hypotheses.
Out of the eight selected categories,
only five are
populated by a significant number of events and will be used as \VBF analysis categories.
The events not selected
into those are used in other analysis categories.
The resulting analysis categories are named according
to their relative signal composition.
Categories enriched in the SM \VBF signal
are denoted as qqH SM-like, those enriched
in the BSM \VBF signal as qqH BSM-like, while
the remaining categories are referred to as \ggH-like.
Throughout this paper, these analysis categories are referred to as ``Tags.''
The tag names are given in
decreasing order of the expected ratio of signal-to-background events
(S/B, where S is the SM \PH and B the nonresonant background).
For example, the tag with the highest S/B targeting the
two-jet \VBF topology, \ggH-like bin, is denoted  \ggH-like Tag0.
Although the optimization process is performed to maximize the sensitivity
to the $CP$-odd anomalous coupling (\fC), it has been verified that this analysis
categorization also retains optimal sensitivity to other anomalous coupling scenarios
(\fL, \fLZg, \fB), because of the similar kinematic configuration of the \VBF events.
Therefore, this same categorization is adopted to enhance the analysis sensitivity across all anomalous coupling benchmarks. Figure~\ref{fig:vbf_distr} shows the distributions, normalized to unit area, of the \DnnVBFbsm and \DzVBF discriminants for the SM \VBF signal, four anomalous \HVV coupling hypotheses, the \ggH process, and the nonresonant background. The \DnnVBFbsm discriminant provides strong separation between signal and background, while \DzVBF is specifically designed to distinguish between the SM and anomalous coupling hypotheses. Defining categories with high purity for both the SM and BSM components is crucial for accurately constraining the effective cross section fractions.

The inputs to the DNN include various jet kinematic and angular
variables, as well as the $\pt/\mgg$ of each photon and angular
variables involving both jets and photons. These inputs for the SM
\VBF and \ggH  processes, as well as for BSM \VBF and non-\PH SM production
of two prompt photons, are taken from simulation. Since the modeling
of backgrounds where at least one of the two photons is a
misreconstructed jet is poor, predominantly due to the fact that very
few $\Pell$+jets and multijet simulated events pass the selection criteria,
a data-driven technique is employed to describe the nonresonant background.
For this reason, simulated
nonresonant background events are used exclusively
for the training of the multivariate discriminants and are not included in
the final fits.

\begin{figure}[ht!]
  \centering
  \includegraphics[width=0.45\textwidth]{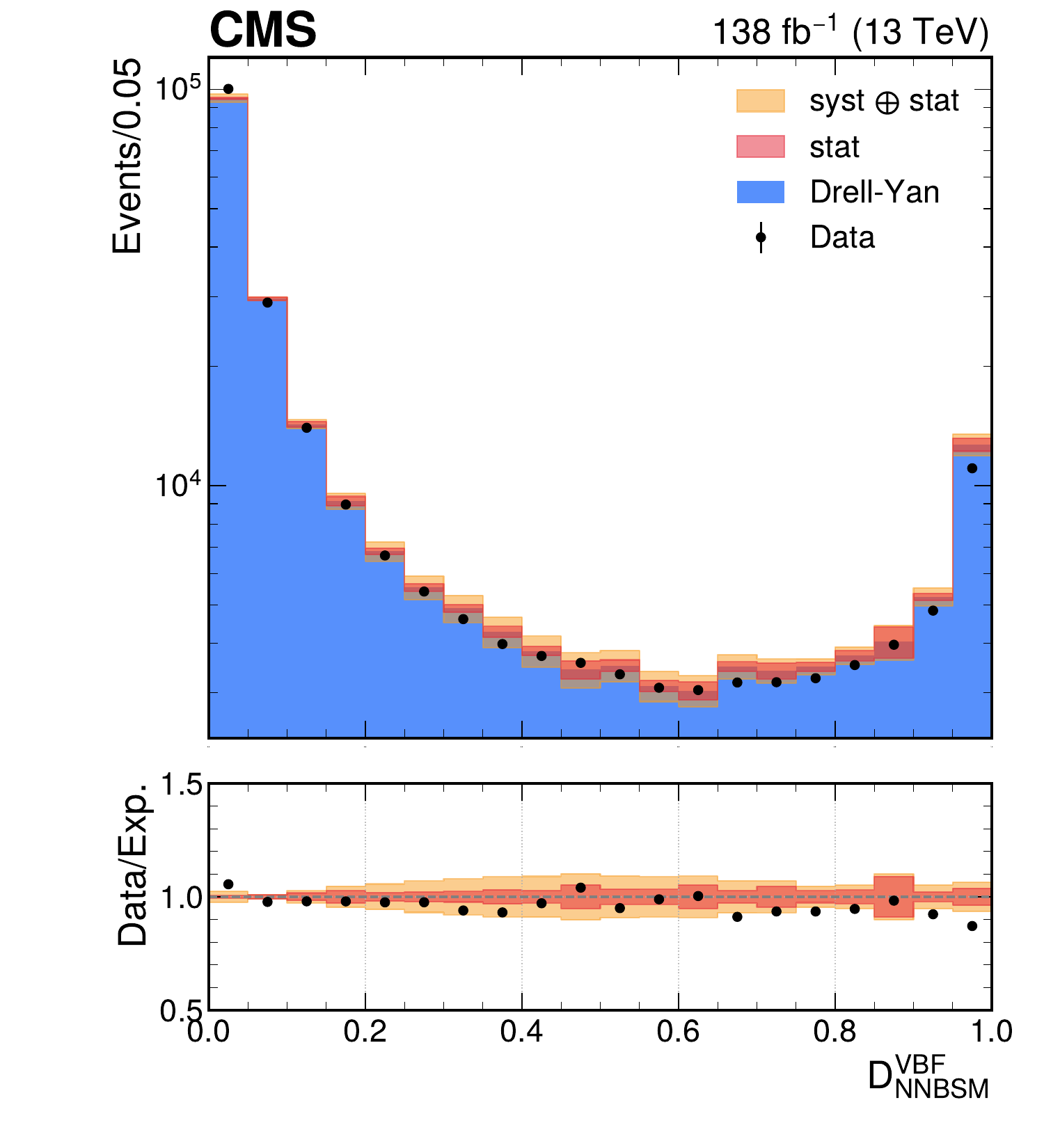}
  \includegraphics[width=0.45\textwidth]{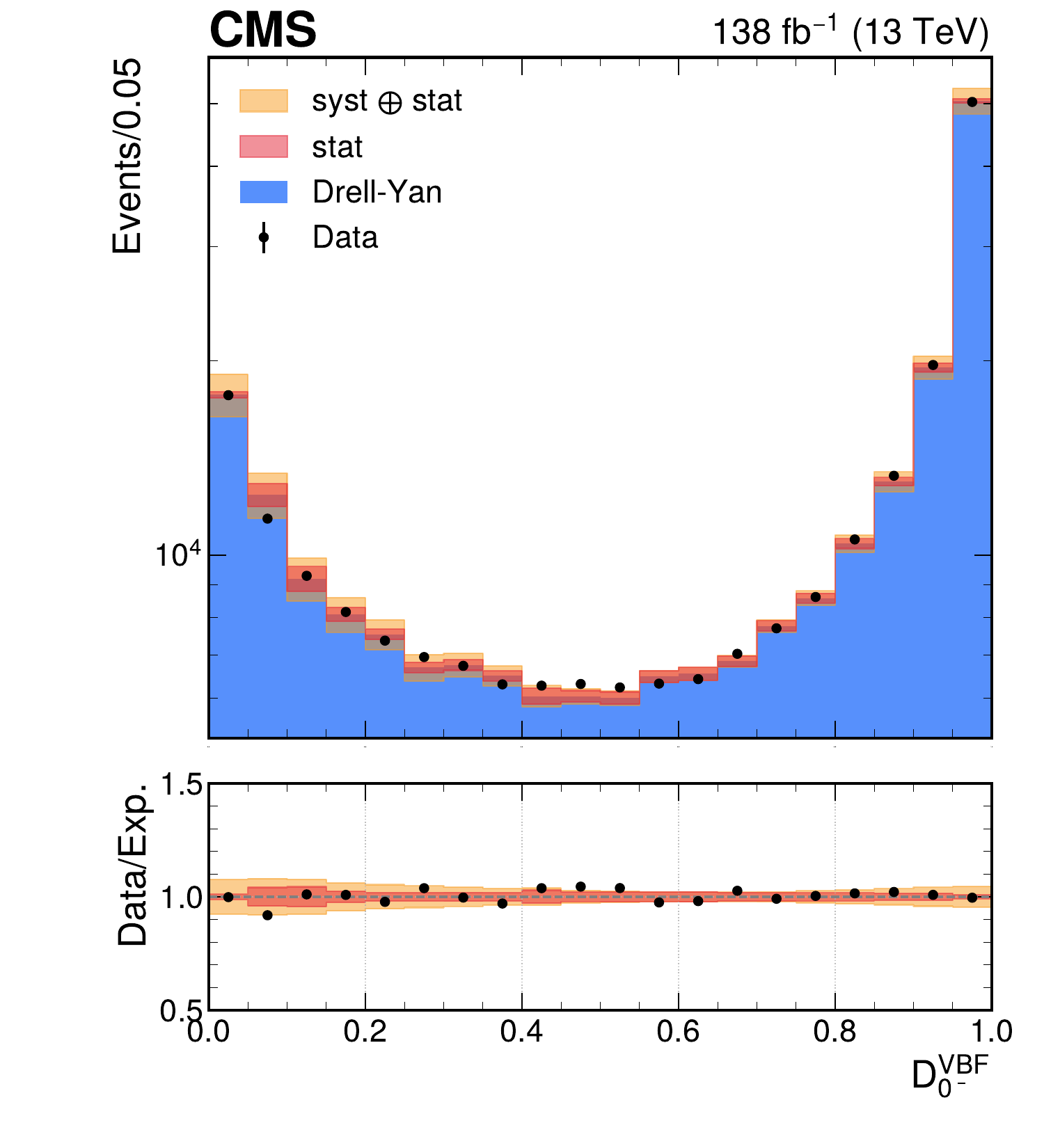}
  \caption{Distributions of the \DnnVBFbsm (left) and \DzVBF (right) outputs for simulation (blue filled histograms, normalized to the data integral) and DY data events (black markers).
    The corresponding ratio plots are shown in the bottom panels.
    The systematic uncertainty is estimated by comparing NLO and LO DY simulations, and is treated as a shape uncertainty. \label{fig:vbf_distr_Zee}}
\end{figure}

All variables used in the categorization are validated using DY events generated with an invariant mass requirement
of $m_{\Pell\Pell} > 50\ \text{GeV}$, where the electrons are reconstructed
as photons to mimic the \Hgamgam process. The corresponding distributions,
obtained by applying the same preselection criteria as in the signal region,
are shown in Fig.~\ref{fig:vbf_distr_Zee}. To ensure that the simulation accurately 
models the input variables used for the event categorization,
 all variables input to the classifier are validated using DY events. 
 This validation provides confidence that the classifier response 
 and the resulting category assignments are reliably described.
The definitions of the \VBF categories and expected signal and background yields
are shown in Tables~\ref{tab:qqh_categories} and~\ref{tab:vbf_yields}.
The ``qqH BSM-like Tag0'' and ``qqH BSM-like Tag1'' categories are those in which
the contribution from anomalous couplings is expected to be maximal.

\begin{table}[htbp]
  \centering
  \caption{Definition of the \VBF categories based on the values of the discriminants \DnnVBFbkg, \DzVBF  and \DnnVBFbsm. \label{tab:qqh_categories}}
  \begin{tabular}{lccc}

    Analysis category & \DnnVBFbkg & \DzVBF & \DnnVBFbsm \\
    \hline
    \ggH-like Tag0    & $>$0.05    & $>$0.6 & $<$0.97    \\
    \ggH-like Tag1    & $>$0.05    & $<$0.6 & $<$0.97    \\
    qqH BSM-like Tag0 & $<$0.05    & $<$0.6 & $>$0.97    \\
    qqH BSM-like Tag1 & $<$0.05    & $<$0.6 & $<$0.97    \\
    qqH SM-like Tag0  & $<$0.05    & $>$0.6 & $<$0.97    \\
  \end{tabular}

\end{table}

\begin{table}[htb!]
  \centering
  \caption{ The expected number of signal events in the case of
    SM \PH with $\mH=125\GeV$ in analysis categories targeting
    \VBF production (qqH). The fraction of the total number of events arising
    from the \VBF production process in each analysis category with respect to the total Higgs boson production is also
    provided.  Entries with values less than 0.1\% are not shown.
    The $\seff$, defined as half of the smallest interval containing 68.3\% of
    the \mgg distribution, is listed for each analysis category.  The
    final column shows the expected ratio of signal to
    signal-plus-background, S/(S+B), where S and B are the numbers of
    expected signal and background events in a $\pm1\seff$ window
    centered on $\mH$.
    \label{tab:vbf_yields}}
  \begin{tabular}{lcccc}

    \multirow{2}{*}{Analysis category} & \multicolumn{4}{c}{\PH(125) expected signal}                                                      \\
                                       & \multirow{1}{*}{yield}                       & \multicolumn{1}{c}{qqH} & \seff (\GeVns) & S/(S+B) \\
    \hline
    \ggH-like Tag0                     & 118.9                                        & 44\%                    & 1.86           & 0.07    \\
    \ggH-like Tag1                     & 64.2                                         & 23\%                    & 1.71           & 0.05    \\
    [\cmsTabSkip]
    qqH BSM-like Tag0                  & 11.3                                         & 12\%                    & 1.55           & 0.51    \\
    qqH BSM-like Tag1                  & 30.8                                         & 59\%                    & 1.67           & 0.45    \\
    [\cmsTabSkip]
    qqH SM-like Tag0                   & 79.1                                         & 75\%                    & 1.86           & 0.37    \\
  \end{tabular}
\end{table}

\subsubsection{\texorpdfstring{\VH}{VH}, with \texorpdfstring{\V}{V} hadronic decay categories} \label{sec:vhhadcategories}

Analysis categories with an enhanced contribution from hadronic \VH
production are constructed in a manner similar to those targeting
VBF-like dijet events. Events are required to contain at least two jets,
and the two jets with the highest transverse momentum are selected as the dijet candidate.
The selected jet pair is then required to satisfy a dedicated set of selection criteria.
The hadronic \VH preselection requires two jets within
$\abs{\eta} < 2.4$ and with $\pt > 30\GeV$, rejecting those
consistent with a pileup jet identification criterion.  
In addition, the reconstructed \mjj is
required to be consistent with a decay of a vector boson, $60 < \mjj <
  120\GeV$.

A DNN, referred to as the \VhadH DNN, is trained to distinguish
three separate classes of events: \PH events from other
production processes (\ggH, \VBF, \ttH, and \VH leptonic) or continuum nonresonant background, the
SM \VH signal and the BSM \VH signal, both with subsequent hadronic
decays of the \PV boson.
The DNN is trained using the photon transverse momentum normalized to the
diphoton invariant mass ($\pt/\mgg$), angular variables
involving jets and photons, and the dijet invariant mass.
Given the expected resonant dijet mass peak from vector boson, \mjj,
which is not present in any of the other backgrounds,
including all the other \PH productions, this is one of the most
powerful inputs in the \VhadH DNN.

Two of the output probabilities, called \DVHhbkg
and \DVHhbsm, represent the probabilities for an event to be
background or BSM signal, respectively, and are used to define analysis
categories.

Analysis categories based on portions of the
two-dimensional plane of the two discriminating variables \DVHhbkg
and \DVHhbsm are optimized, both in the number and in the shape,
maximizing the sensitivity to a $CP$-odd signal, resulting in five
categories, two dominated by BSM \VH signals, and two by SM \VH
signal.  The category boundaries are optimized independently, which leads to a configuration where one category is fully contained within another. Consequently, events satisfying the selection criteria of the inner category are not included in the outer one by construction.
The categories boundaries are summarized in Table~\ref{tab:vhhad_categories}.
The expected signal and background yields in each hadronic \VH
analysis category are shown in Table~\ref{tab:vhhad_yields}.  The ``\VhadH BSM Tag0'' and ``\VhadH BSM Tag1'' categories are those in which the contribution from anomalous couplings is expected to be maximal.

\begin{table}[htbp]
  \centering
  \caption{Definition of the \VhadH categories, \ie, \VH events where the vector boson decays hadronically, based on the values of the discriminants \DVHhbkg and \DVHhbsm.}
  \begin{tabular}{lcc}

    Analysis category                                & \DVHhbkg                 & \DVHhbsm \\
    \hline
    \VhadH SM Tag0                                   & $<$0.08                  & $<$0.56  \\
    \VhadH SM Tag1                                   & $0.08 < \DVHhbkg < 0.25$ & $<$0.45  \\
    \VhadH SM Tag2                                   & $0.25 < \DVHhbkg < 0.54$ & $<$0.29  \\
    \VhadH BSM Tag0                                  & $<$0.066                 & $>$0.89  \\
    \VhadH BSM Tag1 (excluding cat. \VhadH BSM Tag0) & $<$1.0                   & $>$0.75  \\
  \end{tabular}
  \label{tab:vhhad_categories}
\end{table}

\begin{table}[htb!]
  \centering
  \caption{ The expected number of signal events in the case of
    SM \PH with $\mH=125\GeV$ in analysis categories targeting
    \VH associated production in which the vector boson
    decays hadronically, shown for an integrated luminosity of
    138\fbinv.  The fraction of the total number of events arising
    from the \VH production process in each analysis category with respect to the total Higgs boson production is also
    provided.  Entries with values less than 0.1\% are not shown.
    The $\seff$, defined as half of the smallest interval containing 68.3\% of
    the \mgg distribution, is listed for each analysis category.  The
    last column shows the expected ratio of signal to
    signal-plus-background, S/(S+B), where S and B are the numbers of
    expected signal and background events in a $\pm1\seff$ window
    centered on $\mH$.
    \label{tab:vhhad_yields}}
  \centering
  \begin{tabular}{lcccc}

    \multirow{2}{*}{Analysis category} & \multicolumn{4}{c}{\PH(125) expected signal}                                                      \\
                                       & \multirow{1}{*}{yield}                       & \multicolumn{1}{c}{\VH} & \seff (\GeVns) & S/(S+B) \\
    \hline
    \VhadH BSM Tag0                    & 4.41                                         & 13\%                    & 1.72           & 0.30    \\
    \VhadH BSM Tag1                    & 11.8                                         & 20\%                    & 1.67           & 0.24    \\
    [\cmsTabSkip]
    \VhadH SM Tag0                     & 16.6                                         & 49\%                    & 1.69           & 0.13    \\
    \VhadH SM Tag1                     & 37.6                                         & 34\%                    & 1.70           & 0.07    \\
    \VhadH SM Tag2                     & 100.5                                        & 16\%                    & 1.63           & 0.05    \\
  \end{tabular}
\end{table}

\subsubsection{The \texorpdfstring{\VH}{VH}, with \texorpdfstring{\PV}{V} leptonic decay categories} \label{sec:vhlepcategories}

In addition to the EW production resulting in a hadronic final state
of the particles associated with the \PH production, additional
analysis categories target events in which the \PH is produced
in association with a \PW or \PZ vector boson that subsequently decays
leptonically.

The definition of the preselection for the \VH leptonic and \VH MET
categories is the same as in the earlier STXS
measurement~\cite{HIG-19-015}. The phase space with two same-flavor reconstructed
leptons in the final state targets the $\PZ(\Pell\Pell)\PH$ associated production, where $\Pell$ is either an electron or a muon.
Additional selection criteria for the events falling into the
$\PZ\PH$ leptonic tags require two leptons consistent
with the decay of a $\PZ$ boson, including that the dilepton mass
(\mll) is between 60 and 120\GeV. To select the $\PW(\Pell\PGn)\PH$
candidates, events with one reconstructed lepton are required. Additional selection
criteria are applied on the photon identification BDT to further reject background
events containing nonprompt photons, and on the invariant mass of the
reconstructed lepton with each photon to reduce the contamination of
$\PZ$+jets events with an electron misidentified as a photon. Events
passing this selection are collected in categories called
$\PW\PH$ leptonic tags.
To recover events from the leptonic decays of the \PW or \PZ boson, where one or more leptons are not reconstructed
because of inefficiencies or limited geometrical acceptance, and to gain sensitivity to the
$\PZ(\PGn\PGn)\PH$ production process, dedicated \VH MET tag categories are built. In addition to vetoing events with leptons,
$\ptmiss > 50\GeV$ is required and the azimuthal angle between the
diphoton system and \ptvecmiss must be greater than two radians.

For events satisfying selections targeting
the $\PZ(\Pell\Pell)\PH$, $\PW(\Pell\PGn)\PH$, and $\PZ(\PGn\PGn)\PH$ production
processes, dedicated multivariate discriminants are trained to separate
the possible BSM \VH signals from the SM \PH signals (both \VH
and other processes), and the continuum background.

\begin{figure}[!tbhp]
  \centering \includegraphics[width=0.45\textwidth]{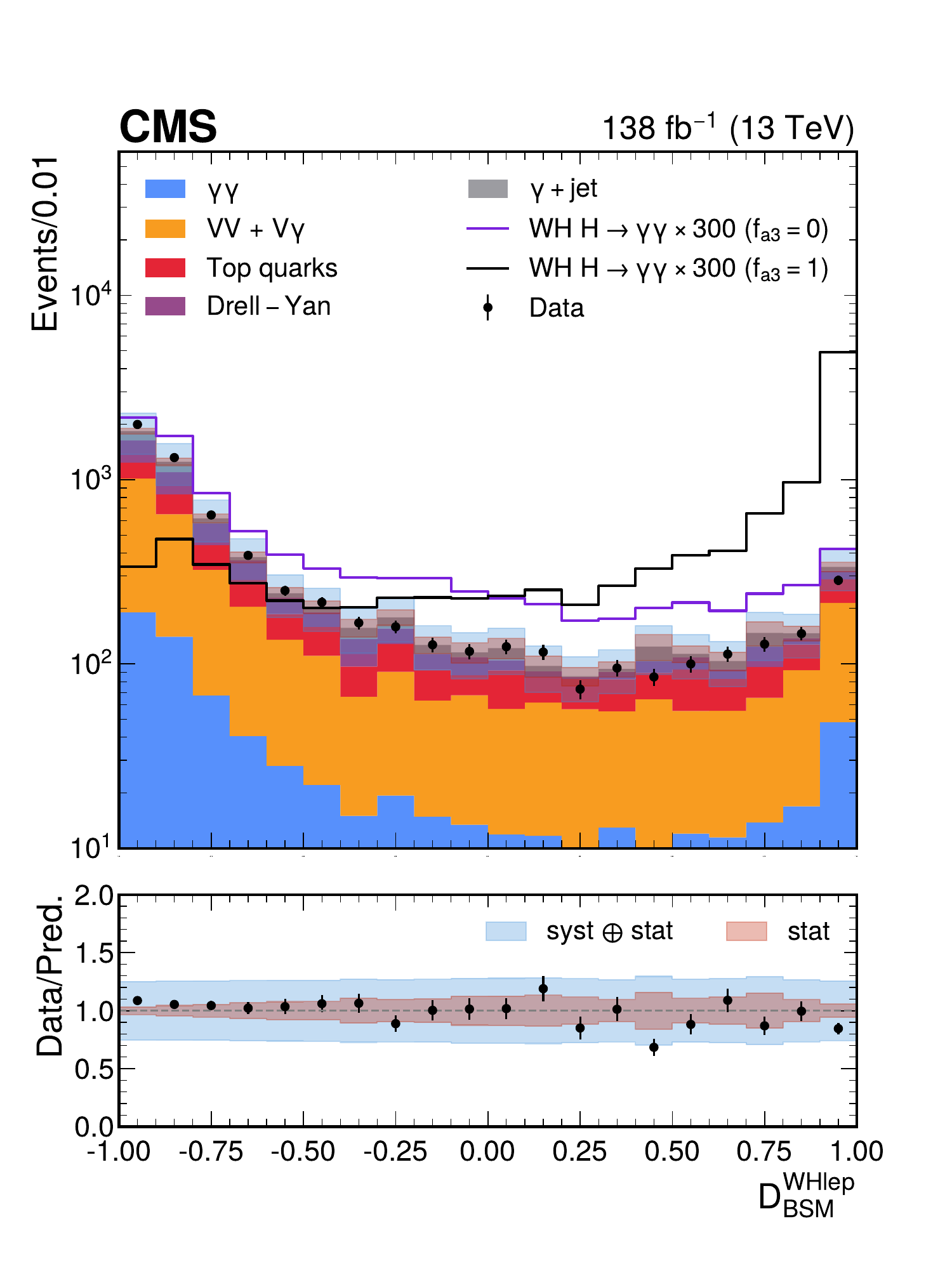}
  \centering \includegraphics[width=0.45\textwidth]{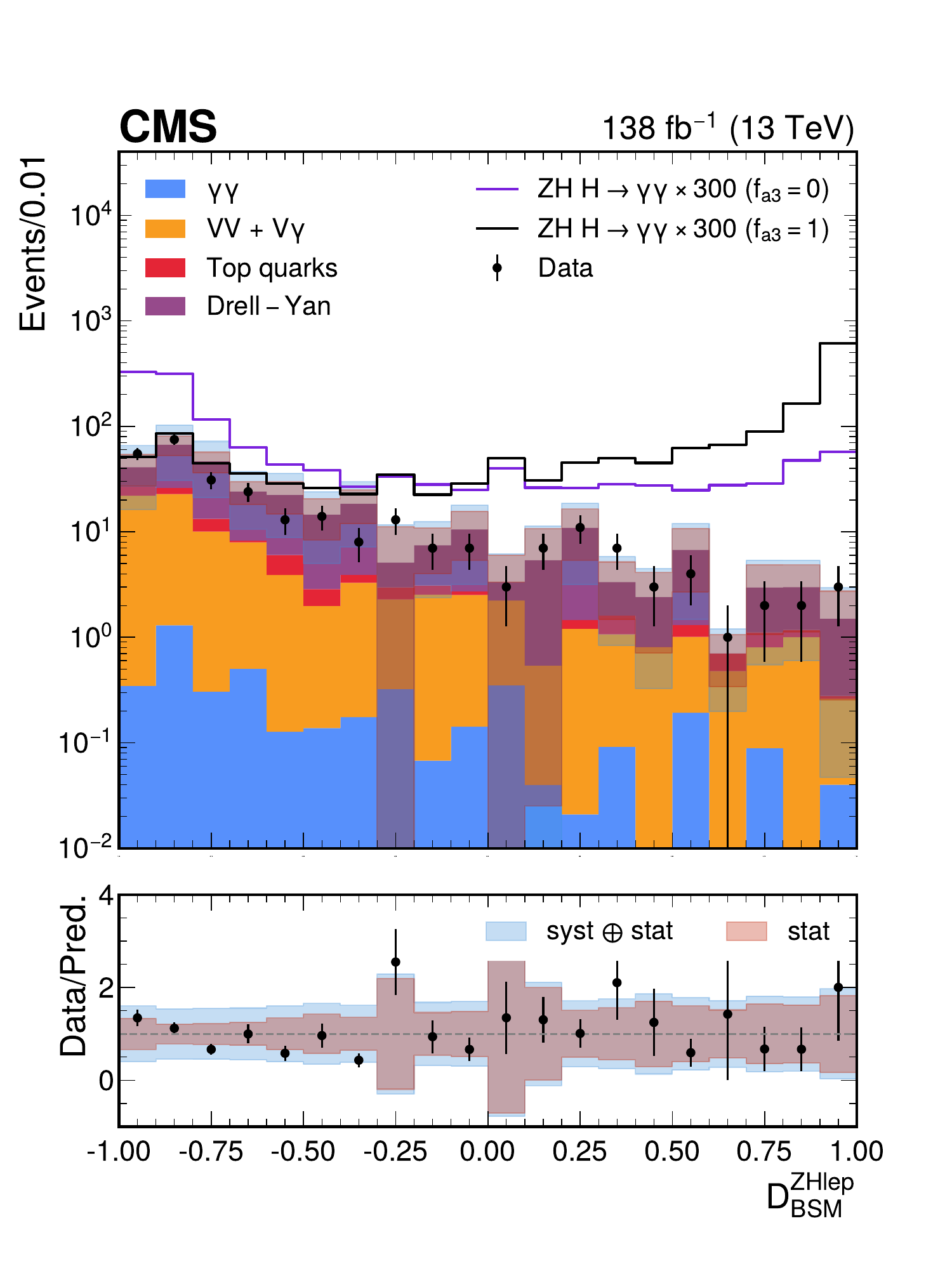}
  \centering \includegraphics[width=0.45\textwidth]{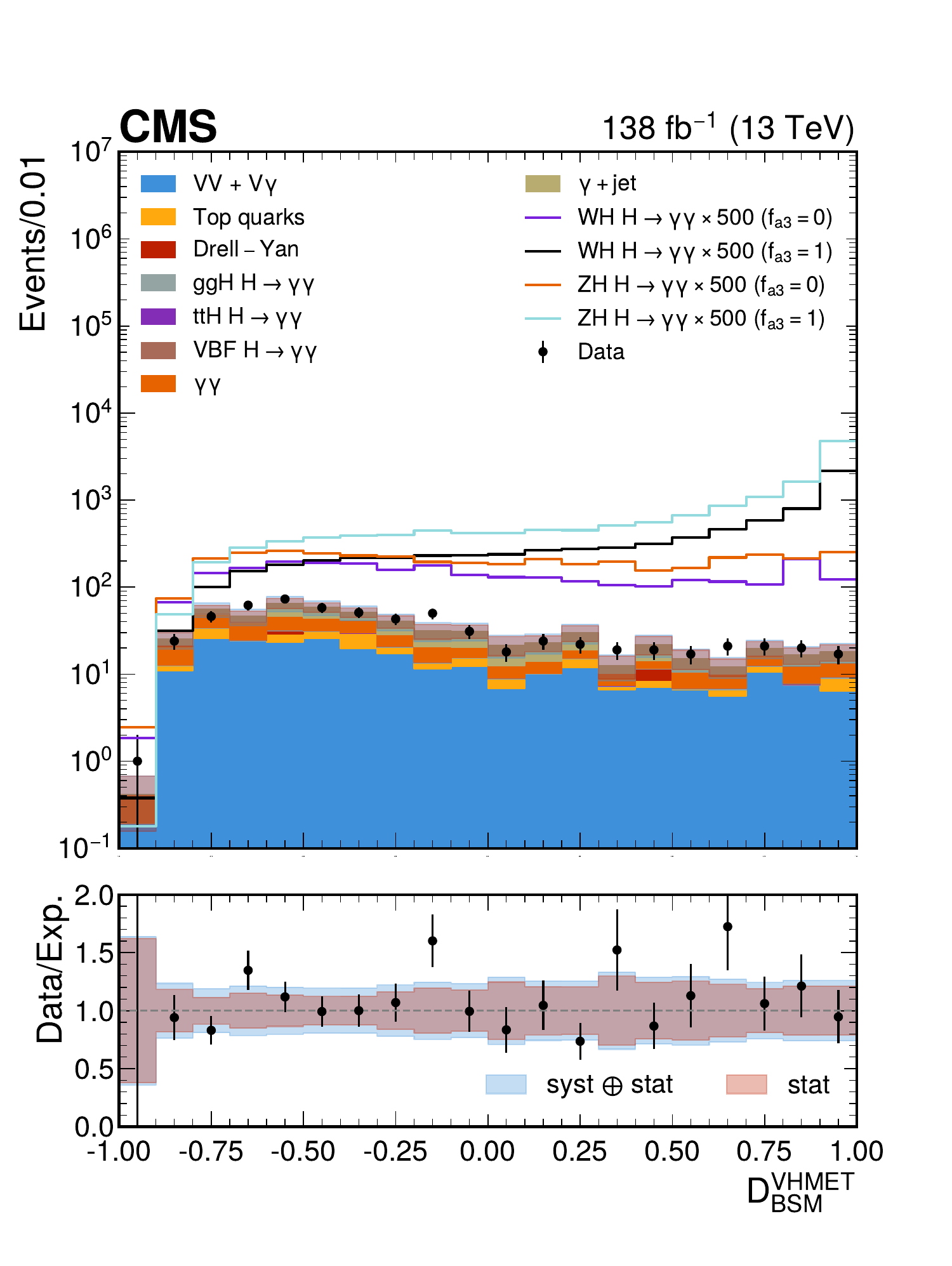}
  \caption{Output scores for the \VH leptonic
    BDTs,  \DVHlbsm (upper left), \DVHllbsm (upper right), \DVHmbsm (lower)
    trained to separate the SM \PH signal from $CP$-odd $(f_{a_3}=1)$ sample.
    The statistical uncertainty in the data points
    is denoted as vertical bars and that on the background simulation by
    the gray/blue bars. The simulated signal and background distributions are
    normalized to the luminosity of the data. To increase its
    visibility, after the normalization, the signal is scaled by a factor of
    either  300 or 500 for the different discriminants.  For the \DVHmbsm distribution,
    a requirement of $\DVHmbkg > 0.619$ is applied to exclude events not used in the analysis.
    A systematic uncertainty is assigned to the component of the nonresonant background determined from a fit to data,
    to account for discrepancies between data and simulation.
    \label{fig:vhl_distr}
  }

\end{figure}

Categories are constructed by using boundaries on two different
multivariate discriminants in each of the
three \ZH, \WH leptonic, and \VH MET tags. One discriminant,
referred to as
the STXS BDT, is the one defined in Ref.~\cite{HIG-19-015} to separate
the SM \PH in each of these production processes from the continuum
background.
It is referred as \DVHlbkg, \DVHllbkg, and \DVHmbkg for
the \WH, \ZH and \VH MET tags, respectively. The other is
the BDT trained to distinguish between SM and BSM signals.
It is referred as \DVHlbsm, \DVHllbsm, and \DVHmbsm for the three targeted
production processes.
For each production process one BDT is trained using input variables that are sensitive to
modifications of the \PH couplings to the vector bosons, mostly
kinematic properties of the photons, leptons, and jets present in the
event, including angular variables describing the separation between
the photons and leptons. In the case of the \VH MET tag, the \ptmiss
variable is added to the list of input variables of the BDT.
The distribution of the output scores for the \VH leptonic
BDTs, \DVHlbsm, \DVHllbsm, \DVHmbsm, trained to separate the SM \PH signal from the $CP$-odd $(f_{a_3}=1)$ sample, are shown in Fig.~\ref{fig:vhl_distr}.

The two-dimensional phase space is partitioned in rectangular bins, as described in Table~\ref{tab:vhlep_categories},
maximizing the statistical significance for an anomalous $CP$-odd
coupling. Since the number of events in these categories can be small because of the lower production rates in leptonic final states, the background estimation from data sidebands may become unreliable. To mitigate this, a requirement is imposed that each category must contain at least 10 events.  From this procedure,  the optimal boundary and the number of categories are determined resulting in a total of four categories for the \WH leptonic production, two categories for the \ZH leptonic production, and an additional four categories for the \VH MET tag.

The expected signal and background yields in each leptonic \VH analysis
category are shown in Table~\ref{tab:vhlep_yields}.

\begin{table}[htbp]
  \renewcommand{\arraystretch}{1.3}
  \centering
  \caption{Definition of the \VlepH categories based on the values of the discriminants \DVHbkg  and \DVHbsm.}
  \begin{tabular}{lcc}

    Analysis category           & \DVHbkg range                & \DVHbsm range              \\
    \hline
    $\PW(\Pell\PGn)\PH$ Tag0    & $0.385 < \DVHlbkg < 1.00$    & $0.79 < \DVHlbsm < 1.00$   \\
    $\PW(\Pell\PGn)\PH$ Tag1    & $0.385 < \DVHlbkg < 1.00$    & $-0.68 < \DVHlbsm < 0.79$  \\
    $\PW(\Pell\PGn)\PH$ Tag2    & $0.125 < \DVHlbkg < 0.385$   & $0.89 < \DVHlbsm < 1.00$   \\
    $\PW(\Pell\PGn)\PH$ Tag3    & $0.125 < \DVHlbkg < 0.385$   & $-0.68 < \DVHlbsm < 0.89$  \\
    $\PZ(\Pell\Pell)\PH$ Tag0   & $0.229 < \DVHllbkg < 1.00$   & $-0.68 < \DVHllbsm < 1.00$ \\
    $\PZ(\Pell\Pell)\PH$ Tag1   & $-0.135 < \DVHllbkg < 0.229$ & $-0.16 < \DVHllbsm < 1.00$ \\
    $\PV(\mathrm{MET})\PH$ Tag0 & $0.798 < \DVHmbkg < 1.00$    & $0.86 < \DVHmbsm < 1.00$   \\
    $\PV(\mathrm{MET})\PH$ Tag1 & $0.798 < \DVHmbkg < 1.00$    & $-1.00 < \DVHmbsm < 0.86$  \\
    $\PV(\mathrm{MET})\PH$ Tag2 & $0.619 < \DVHmbkg < 0.798$   & $0.92 < \DVHmbsm < 1.00$   \\
    $\PV(\mathrm{MET})\PH$ Tag3 & $0.619 < \DVHmbkg < 0.798$   & $-1.00 < \DVHmbsm < 0.92$  \\
  \end{tabular}

  \label{tab:vhlep_categories}
\end{table}

\begin{table}[htb!]
  \renewcommand{\arraystretch}{1}
  \centering
  \caption{\label{tab:vhlep_yields} The expected number of signal events in the case of
    SM \PH with $\mH=125\GeV$ in analysis categories targeting
    \VH associated production in which the vector boson
    decays leptonically, shown for an integrated luminosity of
    138\fbinv.  The fraction of the total number of events arising
    from the \VH production process in each analysis category with respect to the total Higgs boson production is also
    provided.  Entries with values less than 0.1\% are not shown.
    The $\seff$, defined as half of the smallest interval containing 68.3\% of
    the \mgg distribution, is listed for each analysis category.  The
    last column shows the expected ratio of signal to
    signal-plus-background, S/(S+B), where S and B are the numbers of
    expected signal and background events in a $\pm1\seff$ window
    centered on $\mH$.
  }
  \centering

  \begin{tabular}{lcccc}

    \multirow{2}{*}{Analysis category} & \multicolumn{4}{c}{\PH(125) expected signal}                                                      \\
                                       & \multirow{1}{*}{yield}                       & \multicolumn{1}{c}{\VH} & \seff (\GeVns) & S/(S+B) \\
    \hline
    $\PW(\Pell\PGn)\PH$ Tag0           & 1.4                                          & 93\%                    & 1.82           & 0.60    \\
    $\PW(\Pell\PGn)\PH$ Tag1           & 5.8                                          & 98\%                    & 1.96           & 0.56    \\
    $\PW(\Pell\PGn)\PH$ Tag2           & 0.4                                          & 64\%                    & 1.83           & 0.15    \\
    $\PW(\Pell\PGn)\PH$ Tag3           & 3.6                                          & 87\%                    & 1.90           & 0.18    \\
    [\cmsTabSkip]
    $\PZ(\Pell\Pell)\PH$ Tag0          & 1.2                                          & 99\%                    & 1.91           & 0.45    \\
    $\PZ(\Pell\Pell)\PH$ Tag1          & 0.2                                          & 82\%                    & 2.15           & 0.06    \\
    [\cmsTabSkip]
    $\PV(\mathrm{MET})\PH$ Tag0        & 1.1                                          & 96\%                    & 2.06           & 0.45    \\
    $\PV(\mathrm{MET})\PH$ Tag1        & 2.2                                          & 96\%                    & 2.06           & 0.40    \\
    $\PV(\mathrm{MET})\PH$ Tag2        & 1.2                                          & 45\%                    & 1.46           & 0.31    \\
    $\PV(\mathrm{MET})\PH$ Tag3        & 6.7                                          & 80\%                    & 2.05           & 0.18    \\
  \end{tabular}
\end{table}

\subsection{The \texorpdfstring{\ggH}{ggH} event categories}
\label{sec:ggH_cat}
The analysis targeting the \ggH production process aims to optimize the sensitivity to anomalous
couplings of the Higgs boson to gluons.
This is done by employing the kinematic correlations
of the final state particles in events where the Higgs boson is produced
via gluon fusion in association with two jets (\ggH + 2 jets) radiated from
the initial partons, and have kinematics similar to those of the \VBF production.

\begin{figure}[!htbp]
  \centering
  \includegraphics[width=0.45\textwidth]{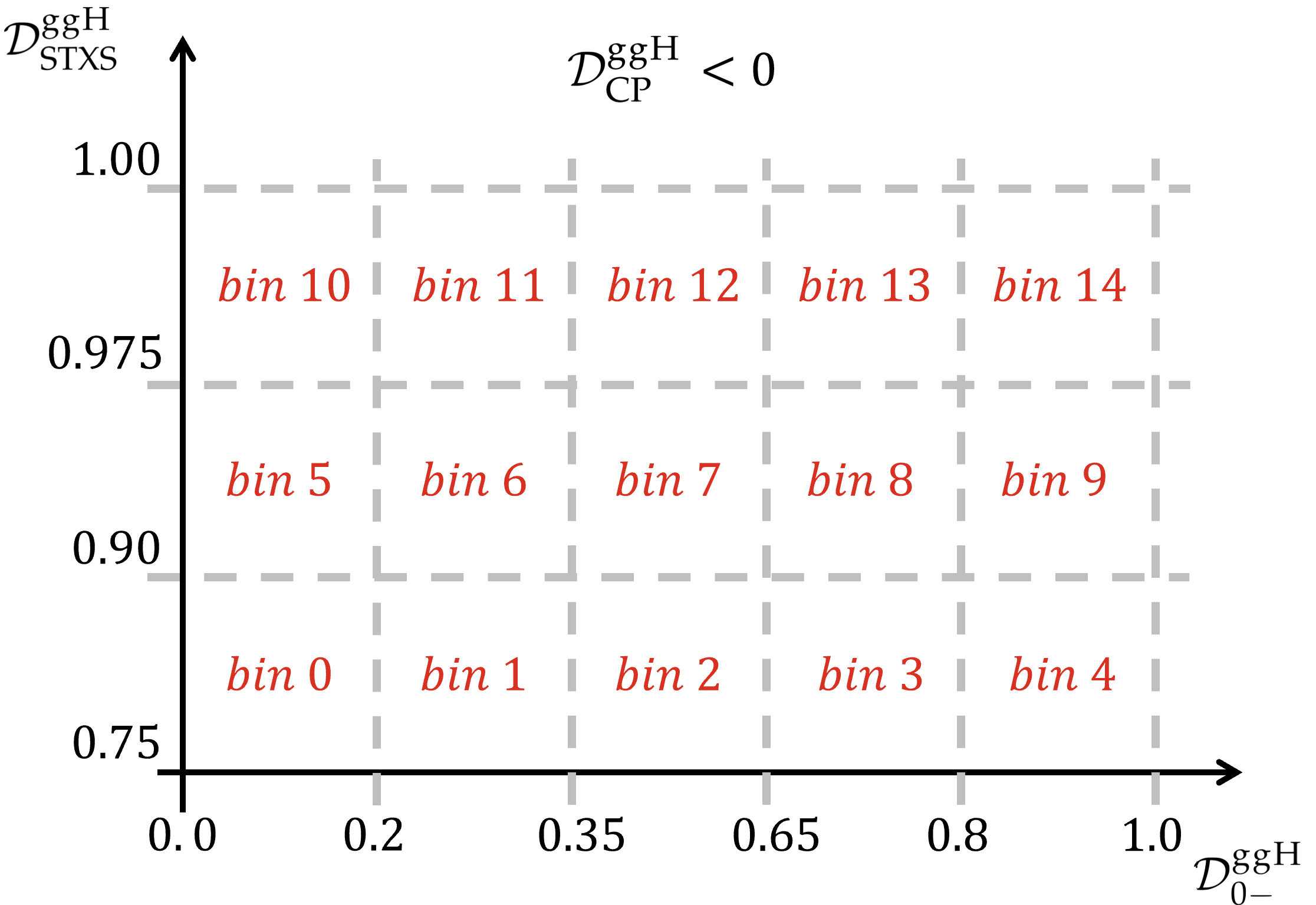}
  \includegraphics[width=0.45\textwidth]{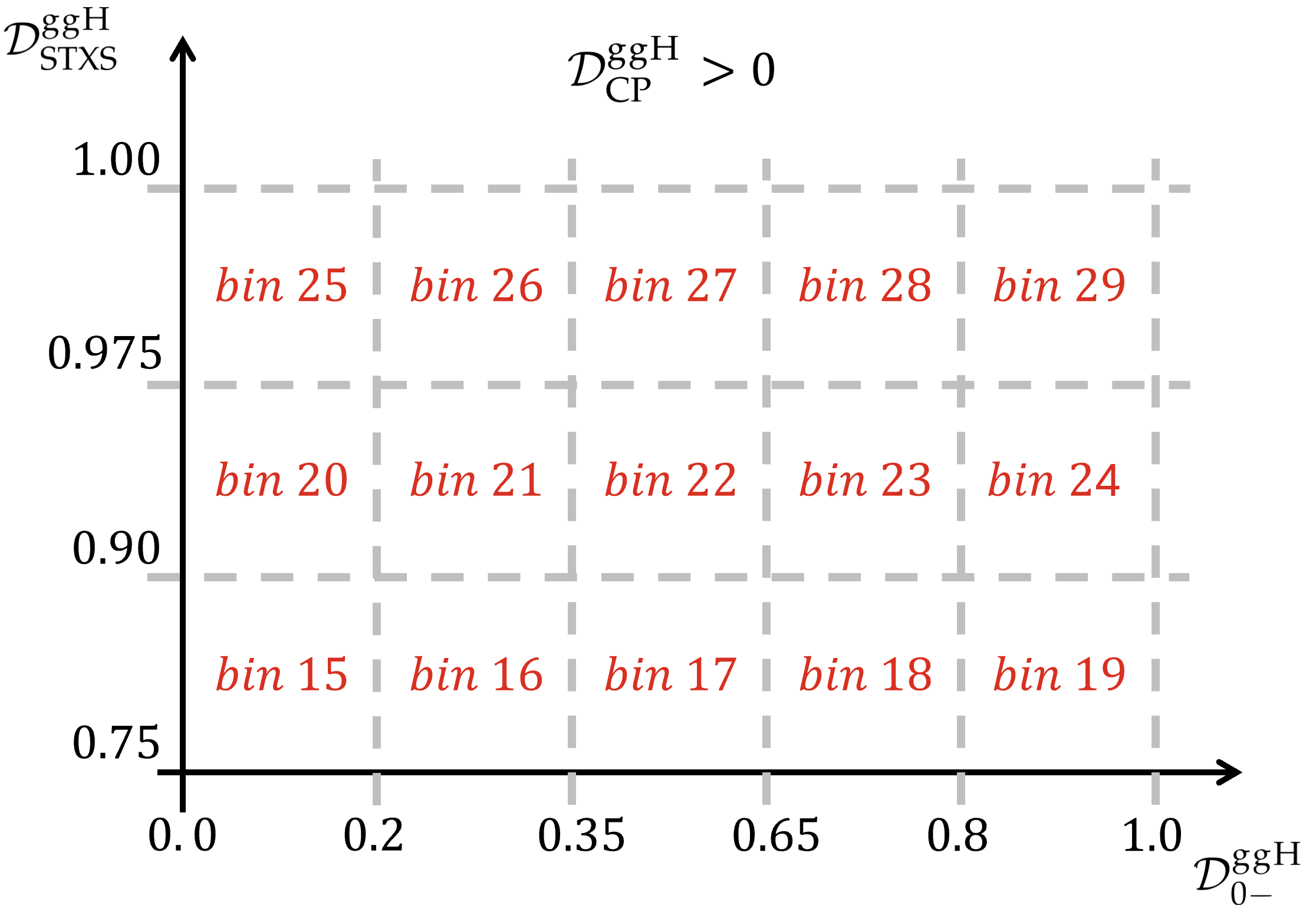} \\
  \caption{Definition of the \Hgg analysis categories defined in bins of \DzggH and \DggHbkg
    for negative (left) and positive (right) values of \DCPggH.
    \label{fig:binning_ggh}}
\end{figure}

To enrich the \ggH categories in \VBF-like events, two jets with
$\pt > 30\GeV$ are required. Contrary to the \VBF selection used in the \HVV
analysis, a selection on \mjj is not applied, to enhance the presence of the signal,
which tends to have \mjj values lower than the ones expected in \VBF processes.
To maximize the analysis sensitivity to anomalous \Hgg couplings, events are
divided into distinct categories. As in the \HVV analysis, this subdivision is
achieved by utilizing discriminating variables aimed at distinguishing between
various processes present in the \VBF-like phase space: \ggH SM
events, \ggH $CP$-odd events, \VBF events, and nonresonant
$\PGg\PGg$ events. As mentioned in Section~\ref{sec:kinematics}, this is
achieved through a three-dimensional optimization approach utilizing \MELA
discriminants \DzggH and \DCPggH, and the standard diphoton MVA
\DggHbkg~\cite{HIG-19-015}.
\begin{figure*}[!tbhp]
  \centering
  \includegraphics[width=0.45\textwidth]{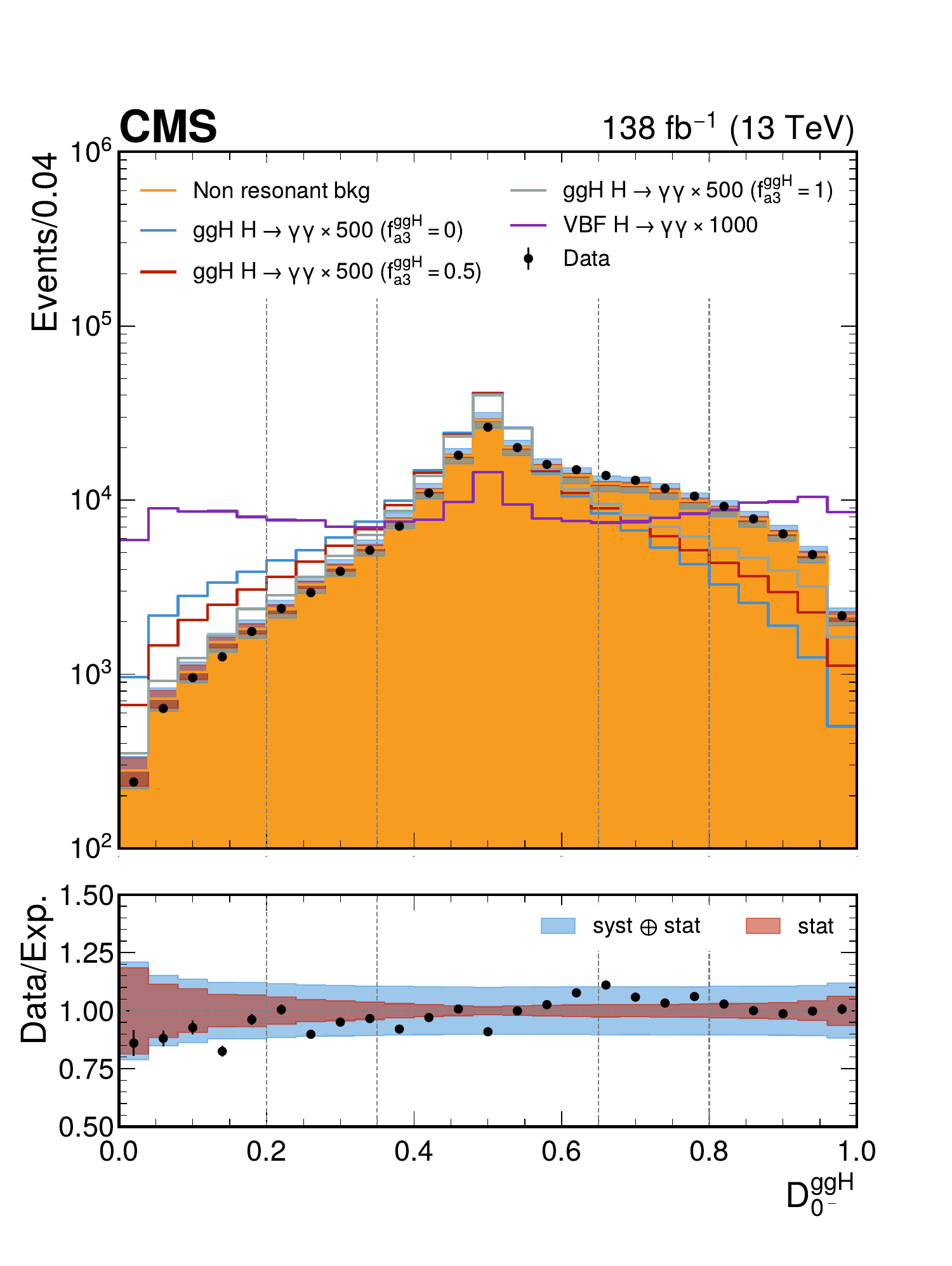}
  \includegraphics[width=0.45\textwidth]{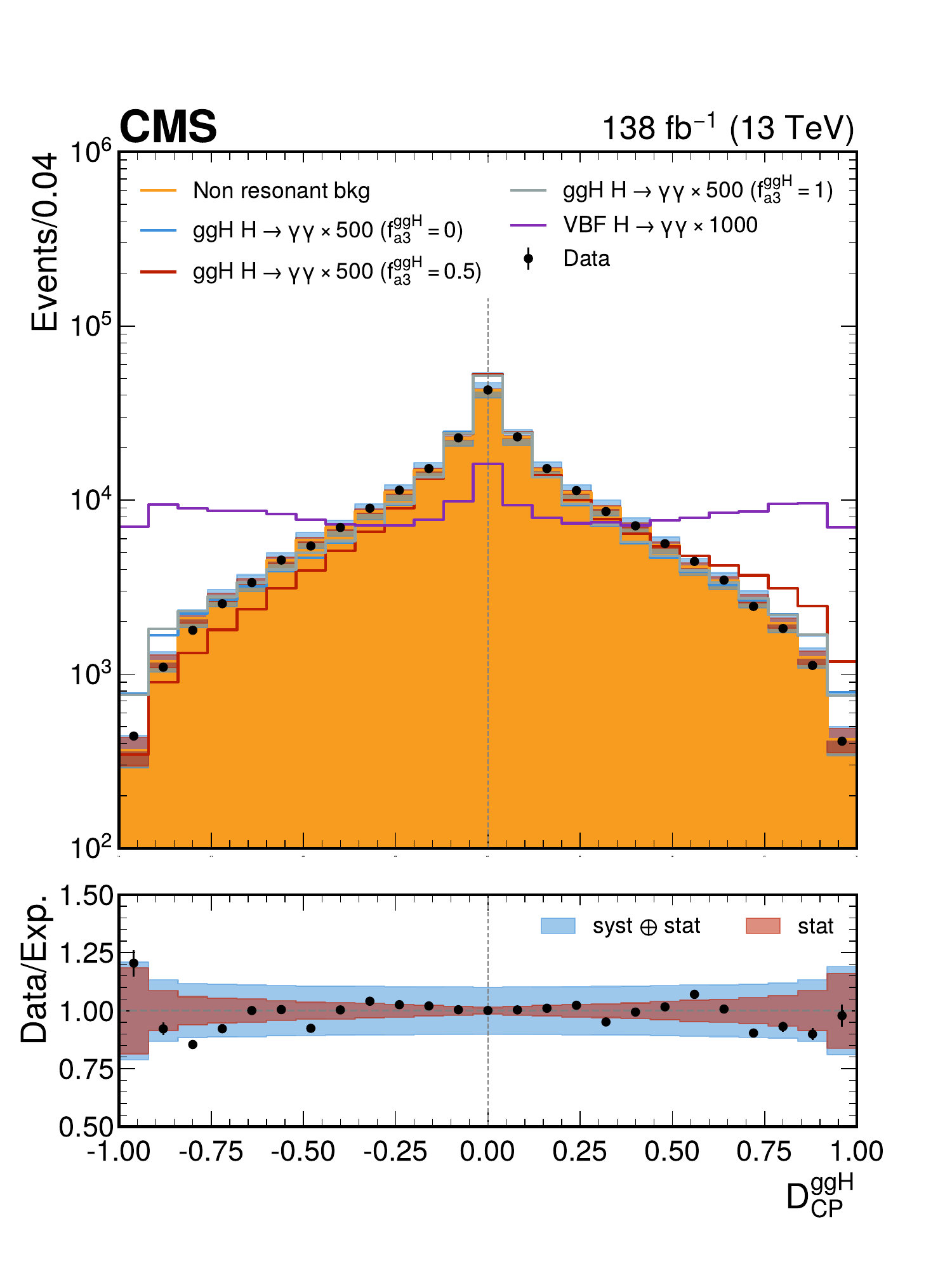}

  \caption{Signal and background distributions for the \MELA discriminants
    \DzggH (left) and \DCPggH (right) used in the \Hgg analysis.
    Events are required to have two jets with $\pt > 30\GeV$.
    The nonresonant background is normalized to the data.
    The dashed vertical lines indicate the bin boundaries applied in the analysis.
    A 10\% systematic uncertainty is assigned to the component of the
    nonresonant background, obtained from a fit to data, to account for
    discrepancies between data and simulation.
    \label{fig:discr_ggH}}
\end{figure*}
The categories are chosen to maximize
the sensitivity to a $CP$-odd signal. As a result, 30 bins are
defined: five bins in the \MELA variable \DzggH to separate \ggH SM
and BSM processes; three bins in the standard diphoton MVA \DggHbkg to reduce
the continuum background; and two bins
in the \MELA discriminant \DCPggH to be sensitive to the interference term and thus the sign of \fG.
Moreover, a dedicated \ggH + 2 jets multiclass BDT is trained to further separate the $CP$-even and $CP$-odd \ggH signal
processes from the inclusive background
(nonresonant background and other \PH production processes).
Selections on the dedicated
\ggH + 2 jets background classifier (\Dggjetsbkg), separating the inclusive
background from the other contributions, and on the classifier targeting the
$CP$-odd  \ggH signal process (\Dggjetsbsm) are applied in each of the 30 bins to further
suppress the background and isolate anomalous contributions.
Designed
to isolate the signal events in the \ggH + 2 jets topology, this multiclass BDT ensures better performance than the standard
diphoton MVA alone, which was trained inclusively on all Higgs boson production processes.
Figure~\ref{fig:discr_ggH} shows the signal and background distributions of the discriminants used in the \Hgg analysis, where events are selected according to the criteria described in Section~\ref{sec:reconstruction} and have two jets with $\pt > 30\GeV$. The definition of the 30 bins used in the \Hgg analysis is summarized in Fig.~\ref{fig:binning_ggh}. Their expected signal and background yields are reported in Table~\ref{tab:ggh_yields}.

\begin{table}[htb!]
  \centering
  \caption{ The expected number of signal events in the case of
    SM \PH with $\mH=125\GeV$ in analysis categories targeting
    \ggH production associated with two jets, shown for an integrated luminosity of
    138\fbinv.
    The fraction of the total number of events arising
    from the \ggH production process in each analysis category with respect to the total Higgs boson production is also
    provided.  Entries with values less than 0.1\% are not shown.
    The $\seff$, defined as half of the smallest interval containing 68.3\% of
    the \mgg distribution, is listed for each analysis category.  The
    last column shows the expected ratio of signal to
    signal-plus-background, S/(S+B), where S and B are the numbers of
    expected signal and background events in a $\pm1\seff$ window
    centered on $\mH$.
    \label{tab:ggh_yields}}
  \begin{tabular}{lcccc}

    \multirow{2}{*}{Analysis category} & \multicolumn{4}{c}{\PH(125) expected signal}                                                       \\
                                       & \multirow{1}{*}{yield}                       & \multicolumn{1}{c}{\ggH} & \seff (\GeVns) & S/(S+B) \\
    \hline
    ggH 0                              & 5.4                                          & 39\%                     & 2.03           & 0.07    \\
    ggH 1                              & 6.4                                          & 62\%                     & 2.04           & 0.04    \\
    ggH 2                              & 37.5                                         & 81\%                     & 2.09           & 0.04    \\
    ggH 3                              & 5.0                                          & 75\%                     & 2.18           & 0.04    \\
    ggH 4                              & 3.7                                          & 66\%                     & 2.16           & 0.07    \\
    ggH 5                              & 13.2                                         & 34\%                     & 1.77           & 0.17    \\
    ggH 6                              & 17.4                                         & 60\%                     & 1.78           & 0.09    \\
    ggH 7                              & 114.0                                        & 77\%                     & 1.75           & 0.08    \\
    ggH 8                              & 16.3                                         & 70\%                     & 1.80           & 0.09    \\
    ggH 9                              & 10.8                                         & 60\%                     & 1.82           & 0.16    \\
    ggH 10                             & 9.9                                          & 29\%                     & 1.58           & 0.37    \\
    ggH 11                             & 13.5                                         & 59\%                     & 1.55           & 0.27    \\
    ggH 12                             & 99.4                                         & 72\%                     & 1.58           & 0.26    \\
    ggH 13                             & 12.4                                         & 63\%                     & 1.59           & 0.28    \\
    ggH 14                             & 9.4                                          & 46\%                     & 1.65           & 0.39    \\
    ggH 15                             & 5.5                                          & 37\%                     & 2.03           & 0.07    \\
    ggH 16                             & 6.5                                          & 61\%                     & 2.02           & 0.04    \\
    ggH 17                             & 37.2                                         & 80\%                     & 2.10           & 0.03    \\
    ggH 18                             & 5.0                                          & 74\%                     & 2.08           & 0.04    \\
    ggH 19                             & 3.7                                          & 64\%                     & 2.04           & 0.07    \\
    ggH 20                             & 13.5                                         & 36\%                     & 1.74           & 0.18    \\
    ggH 21                             & 17.5                                         & 60\%                     & 1.76           & 0.09    \\
    ggH 22                             & 113.1                                        & 77\%                     & 1.76           & 0.08    \\
    ggH 23                             & 16.3                                         & 70\%                     & 1.73           & 0.09    \\
    ggH 24                             & 11.2                                         & 59\%                     & 1.84           & 0.15    \\
    ggH 25                             & 9.8                                          & 29\%                     & 1.56           & 0.38    \\
    ggH 26                             & 13.5                                         & 58\%                     & 1.58           & 0.26    \\
    ggH 27                             & 97.8                                         & 73\%                     & 1.58           & 0.25    \\
    ggH 28                             & 12.4                                         & 63\%                     & 1.54           & 0.28    \\
    ggH 29                             & 9.1                                          & 46\%                     & 1.60           & 0.40    \\
  \end{tabular}
\end{table}

\section{Statistical procedure}
\label{sec:sig_bkg}

The statistical procedure used in this analysis is identical to that
described in Ref.~\cite{Khachatryan:2014jba}, and follows the CMS \textsc{Combine} tool~\cite{combine}. A likelihood
function is defined for each analysis category using analytic models
to describe the diphoton invariant mass (\mgg)
distribution  of signal and background events,
with nuisance parameters to account for the experimental and
theoretical systematic uncertainties. The signal and background models derived in each category
are described respectively in Sections~\ref{sec:signalModel} and~\ref{sec:background}.

To extract the signal production
cross sections and the anomalous couplings effective fractions, $\vec{f}=(\fB, \fC, \fL, \fLZg)$
for the \HVV analysis and $f^{\ggH}_{a3}$ for the \Hgg one,
a simultaneous extended maximum
likelihood fit~\cite{Barlow:1990vc} to the \mgg spectrum
is performed in the range $100 < \mgg < 180\GeV$. The fit is performed simultaneously
across all analysis categories with a dedicated signal template for each process
(\ggH, \ttH, \VBF, \WH, and \ZH) and the four data-taking periods (2016 pre-VFP, 2016 post-VFP,
2017, and 2018).

The primary vertex associated with the diphoton candidate
is assigned using a multivariate algorithm (the vertex identification BDT), based on
observables related to tracks recoiling against the diphoton system~\cite{Aaboud:2018xdt}. A second vertex-related
multivariate discriminant, referred to as the vertex probability BDT, estimates the probability that
the vertex selected by the vertex identification BDT lies within 1\cm of the vertex from which the
diphoton system originated.
This criterion is motivated by the fact that, if the interaction point is reconstructed to within
about 1\cm of the true production vertex, the diphoton mass resolution is dominated by the ECAL
energy resolution, while the contribution from the uncertainty in the photon opening angle is negligible.
Conversely, an incorrect vertex assignment degrades the diphoton mass resolution because of the
misreconstruction of the opening angle between the two photons, leading to a poorer invariant mass resolution.
If the interaction point is known to better than about 1 cm, the resolution is dominated by the ECAL
energy contribution only. If the interaction point is incorrectly assigned, the resolution worsens
substantially because of the effect on the calculation of the opening angle between the photons, and thus the invariant mass.

When the available sample size is sufficient, event categories are further
subdivided into two vertex scenarios, depending on whether the diphoton vertex is correctly
identified, defined as being within 1\cm of the true interaction point.

The likelihood function is defined as a product of conditional probabilities over all categories $i$~\cite{HIG-20-007}:
\begin{equation}
  \mathcal{L}(\text{data} | \muf, \muV, \vec{f}, \theta) = \prod_{i}\prod_{j}\mathrm{Poisson}(n_j^{i}|s_{j}^{i}(\muf, \muV, \vec{f}, \theta)+b_{j}^{i}(\theta))\cdot p(\tilde{\theta} | \theta),
  \label{eq:likelihoodDensity}
\end{equation}

where $n_{j}$ is the observed number of data events in the $j^{\text{th}}$ \mgg bin. The $s_{j}$ and $b_{j}$ values represent the expected SM Higgs boson contributions from all the production processes and the nonresonant background processes, respectively. They are functions of the set of nuisance parameters $\theta$ that correspond to systematic uncertainties, and the parameters of interest (POIs) that modify the \PH signal processes: \muf, \muV, and $\vec{f}$.
The parameters \muf and \muV are the \PH signal strength modifiers
that scale the \ggH{+}\ttH and \VBF{+}\WH{+}\ZH cross sections, respectively.
Note that \muV does not impose any
SM-like ratio between \VBF and \VH inside the experimental acceptance.
The relative response of each production mode to anomalous HVV interactions is
fully determined by the Monte Carlo simulation through the process-dependent SM and BSM cross sections  used in the signal model.
The $\vec{f}$ term represents the set of anomalous coupling parameters that modify the distributions of the \ggH or \VBF{+}\VH signals.
Finally, the $p(\tilde{\theta}|\theta)$ term represents the set of external constraints from auxiliary measurements of the nuisance parameters $\tilde{\theta}$.
The systematic uncertainties that affect only the normalization of the signal and background processes are assigned log-normal external constraints, while the shape-altering systematic uncertainties are assigned Gaussian external constraints.
The negative log-likelihood is defined as
\begin{equation}
  \dNLL=-2\ln{\frac{\mathcal{L}(\text{data} | \muf, \muV, \vec{f}, \theta)}{\mathcal{L}(\text{data} | \hat{\muf}, \hat{\muV}, \vec{\hat{f}}, \hat{\theta})}},
\end{equation}
where $\hat{\muf}$ and $\hat{\muV}$ are the best fit values of the signal strength modifiers,
which are profiled in the fit, and $\hat{\theta}$ is the maximum likelihood estimate of the nuisance parameters.
In the \HVV analysis, the POIs of the likelihood fit are the effective cross section fractions $\vec{f}$,
which are tested one at a time, setting the values of all other anomalous coupling parameters to zero.
In the \Hgg analysis the POI is \fG, while \fC is fixed to the SM value ($\fC=0$). The parameter $\vec{\hat{f}}$
represents the best fit values of the POIs that maximize the likelihood. In all fits, \mH is fixed
to its most precisely measured value of 125.38\GeV~\cite{HggMass}.

In this analysis, as in many other effective field theory studies, the conditions of Wilks' theorem~\cite{Wilks} are
violated because of the quadratic dependence of the signal yield on the coupling parameters~\cite{Bernlochner:2022oiw,CMS:2025ugn},
which prevents a correct estimation of the confidence level (CL) using the likelihood-ratio test statistic.
The likelihood ratio intervals for the quadratic model may thus
undercover or overcover.
Nevertheless, for consistency with previous analyses, we quote confidence
intervals by defining the 68\% and 95\% CL as the regions where $\dNLL < 1$ and 3.84, respectively.

\subsection{Signal model}\label{sec:signalModel}
\begin{figure}[!tbhp]
  \centering
  \includegraphics[width=0.49\textwidth]{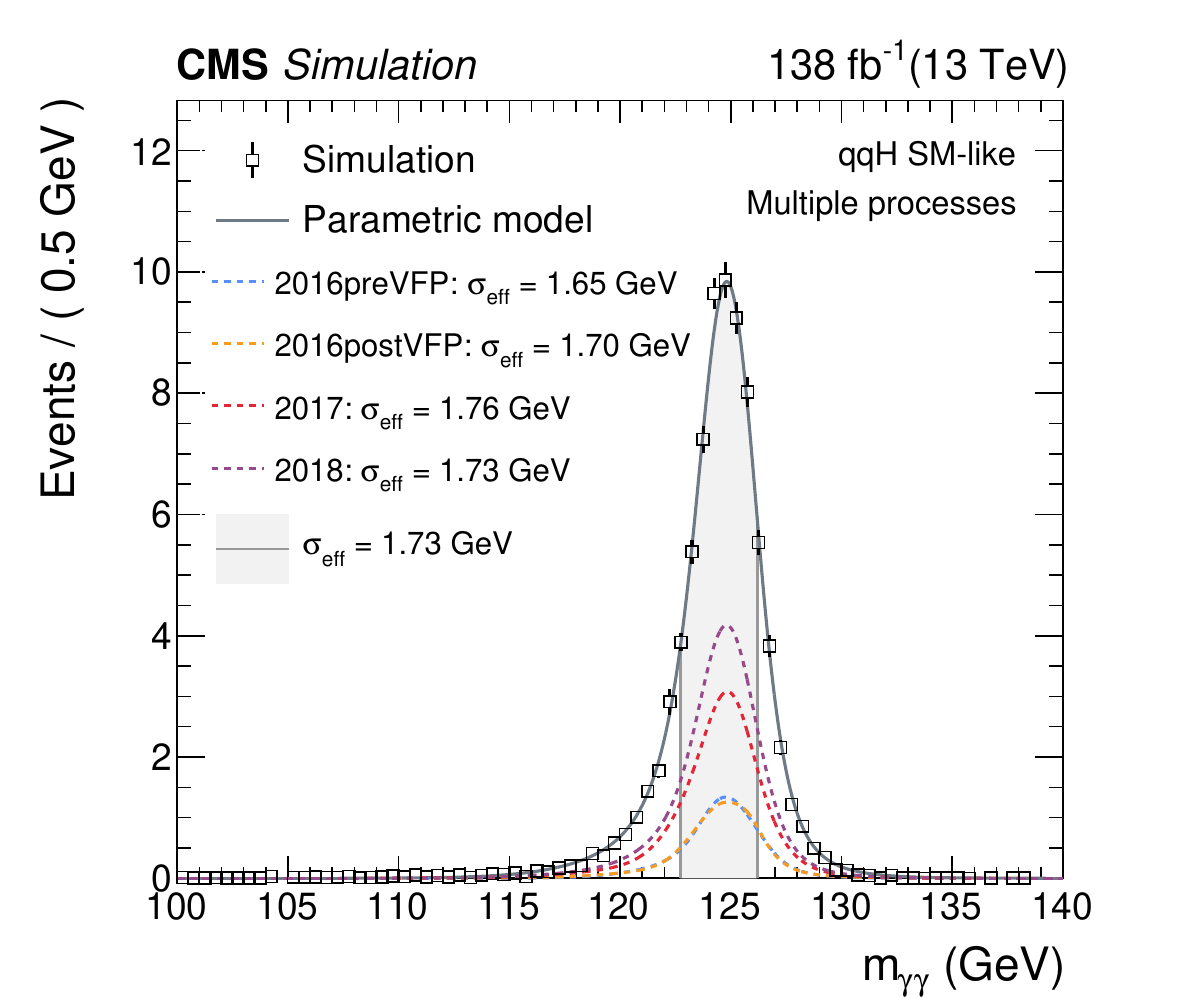}
  \includegraphics[width=0.49\textwidth]{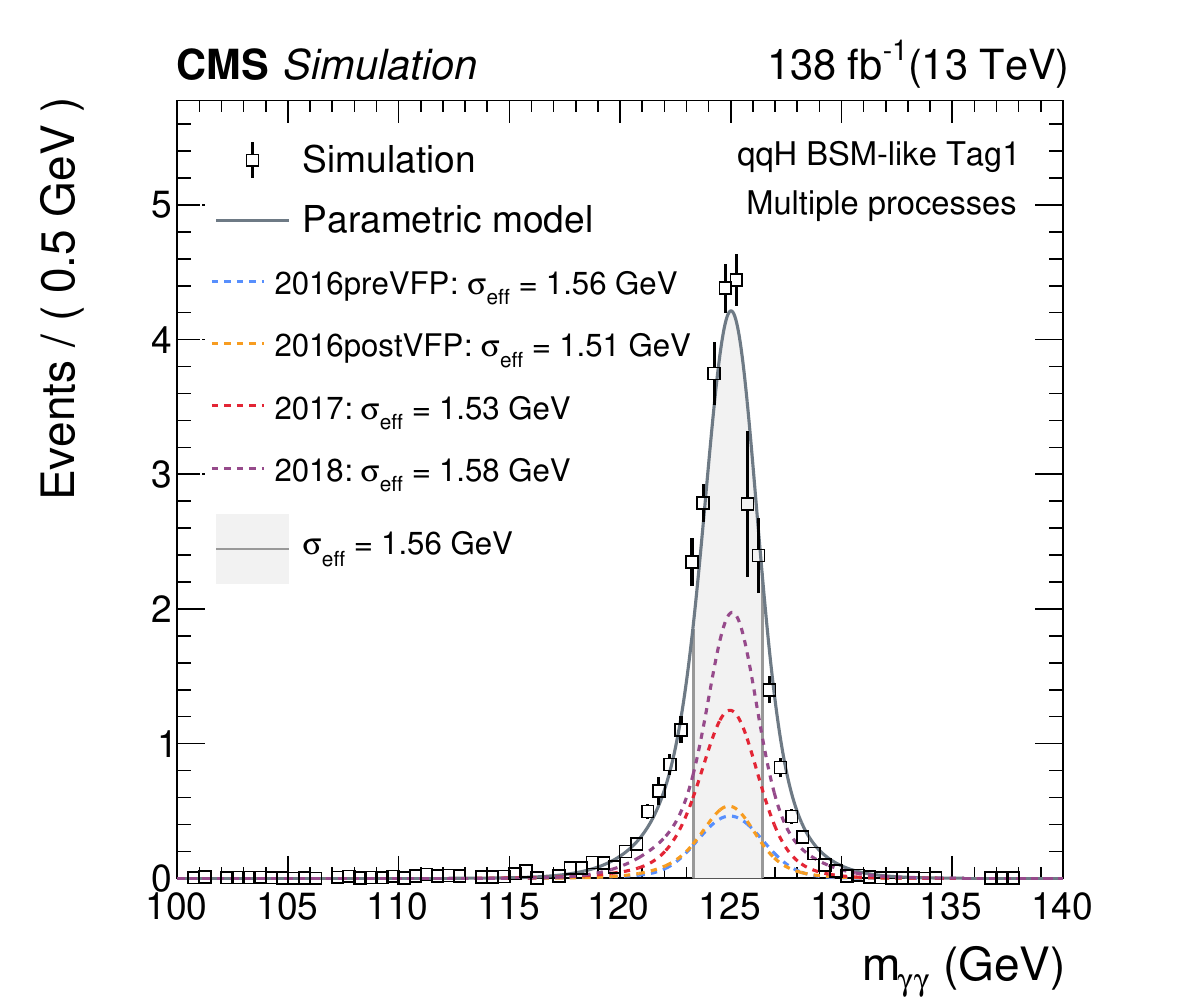}

  \caption{Examples of fits to the \mgg distribution for SM signal samples with $\mH = 125\GeV$ are shown for the luminosity-weighted average of the four
    data-taking periods, in two categories targeting \VBF process, one dominated by SM-like (left) events and the other by BSM-like (right) events.
    Different Higgs boson production processes are summed according to their expected SM cross sections.
    The points represent simulation events weighted by their respective event weights.
    Colored lines represent the individual signal models for each data-taking year.
    The effective mass resolution (\seff) of the \mgg distribution is also indicated in the figure. \label{fig:SigBkg_SigPlots}}
\end{figure}

The parametrized model for the signal of a Higgs boson decaying into two photons
is derived for each of the categories described in the previous section.
The distribution of events in
\mgg from the signal simulation that fall into a
respective category is fitted with a sum of Gaussian probability density
functions. The optimal number of Gaussian distributions is determined with the help
of a dedicated selection algorithm, based on an $\mathcal{F}$-test~\cite{fTest}.
The signal model is derived
separately for events where the vertex has been correctly located
(within 1\cm of the true vertex) and events where the vertex
location is not close enough to the true vertex to be considered
right, if the sample statistics allows it.
The final fit function for each analysis category is obtained by summing the individual functions corresponding to the correct and incorrect vertex scenarios.
Figure~\ref{fig:SigBkg_SigPlots} displays the signal models for each year separately, focusing on two categories dominated by \VBF BSM and SM events.
The \seff is defined as half the width of the smallest interval containing 68.3\% of the \mgg distribution.
It is worth noting that BSM events are characterized by higher \pt of the two photons, which leads to improved mass resolution and, consequently, a smaller \seff.

\subsection{Background model}
\label{sec:background}

The model used to describe the background is extracted from data using
the discrete profiling method~\cite{Envelope,HggRun1}.  This technique
quantifies the systematic uncertainty associated with the choice of a
particular analytic function to fit the background \mgg distribution
among a family of functions.
The minimum order within each family of functions is determined
through an ${F}$-test~\cite{fTest} and a requirement
on the goodness-of-fit to the data.  The choice of the background
function is treated as a discrete nuisance parameter in the likelihood
fit to the data.

When fitting these functions to the \mgg distribution, the value of
twice the negative logarithm of the likelihood (\dNLL) is minimized.
A penalty term is added to the \dNLL to account for the number of
unconstrained parameters in each candidate function.  When measuring
a given parameter of interest, the discrete profiling
method minimizes the overall \dNLL considering all allowed functions
for each analysis category.  Checks are performed to ensure that
describing the background \mgg distribution in this way introduces
negligible bias in the final results, with the estimated bias amounting
to approximately 20\% of the total uncertainty.

\section{Systematic uncertainties}
\label{sec:Systematics}

Several sources of systematic uncertainties are taken into account in the analysis
and are described in more detail in Ref.~\cite{HIG-19-015}.
Their impact on the measurement of the anomalous coupling parameters is small compared
to the statistical uncertainty, mainly because many systematic contributions are strongly
reduced in the cross section ratios.

The uncertainties related to the background estimation derived from data are addressed
using the discrete profiling method, as previously described, which propagates the uncertainty
on the choice of function through the fits.

The signal model is subject to several systematic uncertainties, which are addressed
through two distinct approaches. When an uncertainty affects the shape of the \mgg
distribution, it is treated in the signal model as a nuisance parameter that can alter
the mean and width of the Gaussian functions. These shape-related uncertainties are
primarily experimental and arise from differences between the photon and electron initiated showers,
which are used to derive the energy scale corrections.
The dominant contributions in this category come from uncertainties in the photon
energy scale and resolution, which impact the corrections applied to photon energies
in data and the resolution in simulation, uncertainties from shower shape corrections,
introduced to account for imperfect modeling of electromagnetic showers in simulation,
and uncertainties from inaccuracies in the modeling of material upstream of the ECAL,
which particles traverse before reaching the calorimeter. These sources account
for approximately 13--17\% of the total systematic uncertainty, depending on
the effective cross section ratio.

In contrast, uncertainties that do not influence the shape of the \mgg distribution
are treated as log-normal variations in the event yields.
The dominant systematic uncertainty related to \mgg scale arises from the non-linearity of the photon energy scale.
A relative uncertainty of 0.25\% is assigned to account for potential
differences in the energy scale between the Higgs boson mass of interest and the reference point at 125.28\GeV, used in the fit.

These are associated
with either experimental or theoretical sources. Among the dominant experimental
contributions are uncertainties in the jet energy scale and resolution.
These affect the event categorization significantly, as many of the kinematic
discriminants used in the analysis rely on jet-related variables, leading to
event migration across categories. The impact of jet energy uncertainties amounts
to about 15\% of the total systematic uncertainty. Other relevant sources include
uncertainties in photon identification (1--5\%) and in the per-photon energy
resolution estimate (0.2--6\%), the latter being derived from the photon energy regression.
Additional experimental uncertainties, such as those from trigger and preselection
efficiencies, lepton identification, \ptmiss, and pileup jet identification,
contribute an overall uncertainty of approximately 1--5\%. Theoretical
uncertainties have a small effect on the results, as they largely cancel
in the cross section ratios. The only exception is the uncertainty
related to the imperfect knowledge of the proton's PDFs, which contributes
significantly to the measurement of the \HVV effective cross section ratios
(up to 20\%). This systematic uncertainty impacts the \pt distribution of jets,
a key observable in the \HVV analysis. This is particularly relevant
because \HVV BSM events tend to produce particles with higher \pt compared to SM
events.

The impact of each uncertainty source is evaluated separately for each analysis category.
Most experimental uncertainties are treated as uncorrelated across different data-taking years,
except for partial correlations in the integrated luminosity and jet energy correction
uncertainties.

\section{Results}
\label{sec:results}

The signal strength modifiers $\vec{\mu} = (\muf$, $\muV)$, controlling the rate of the Higgs boson production processes,
and the set of anomalous coupling parameters $\vec{f}$,
that potentially modify the \ggH and \VBF{+}\VH signal distributions, are simultaneously constrained in
each production process considered.  In the following, we describe the separate
measurements of $\vec{f} = \left(\fB, \fC, \fL, \fLZg\right)$ and $\fG$,  in the \HVV and
\Hgg analyses, respectively, resulting from the binned
maximum likelihood fit to the data obtained by combining all categories
for the different channels and periods.

\subsection{Results of the \texorpdfstring{\HVV}{HVV} analysis}

\begin{figure}[!htbp]
  \centering
  \includegraphics[width=0.49\textwidth]{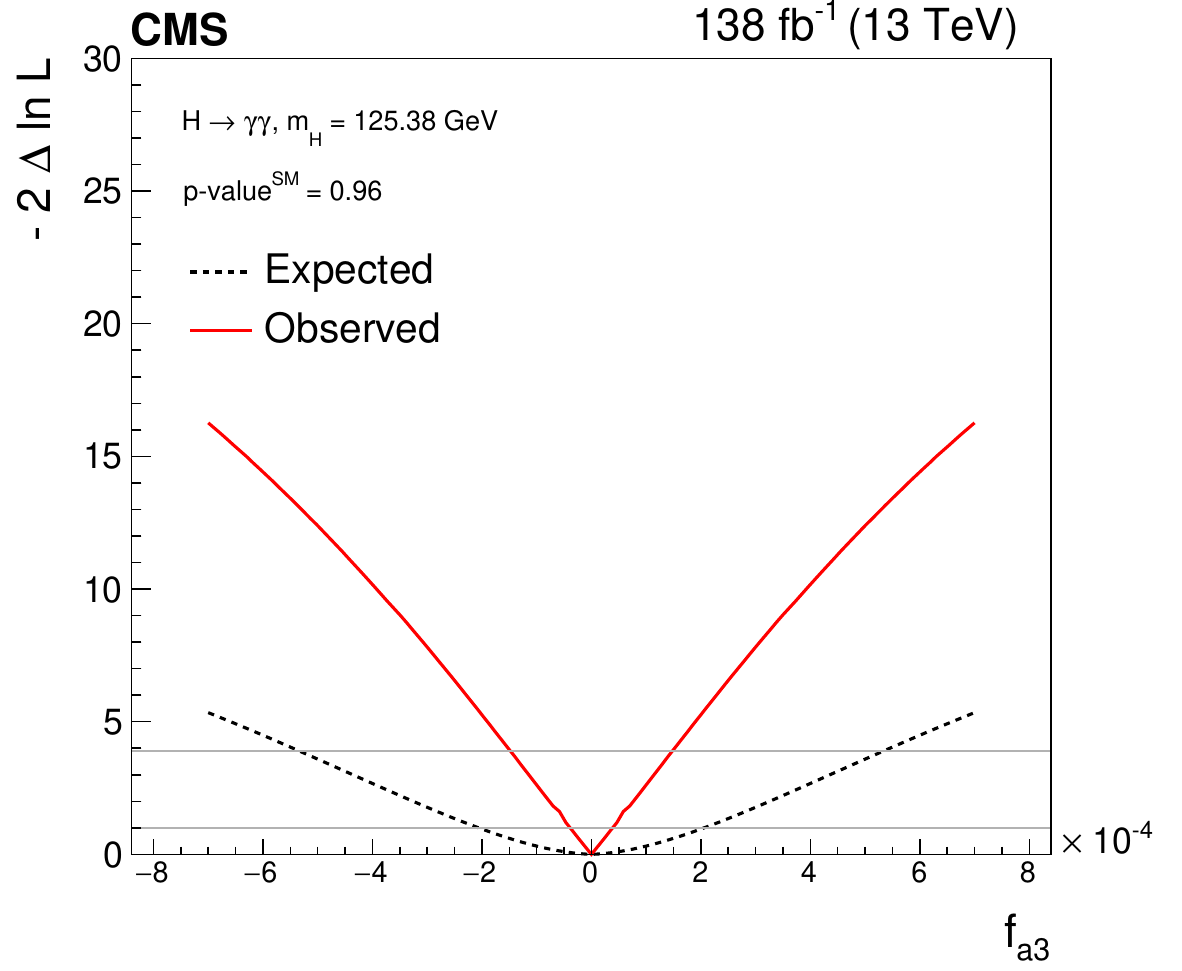}
  \includegraphics[width=0.49\textwidth]{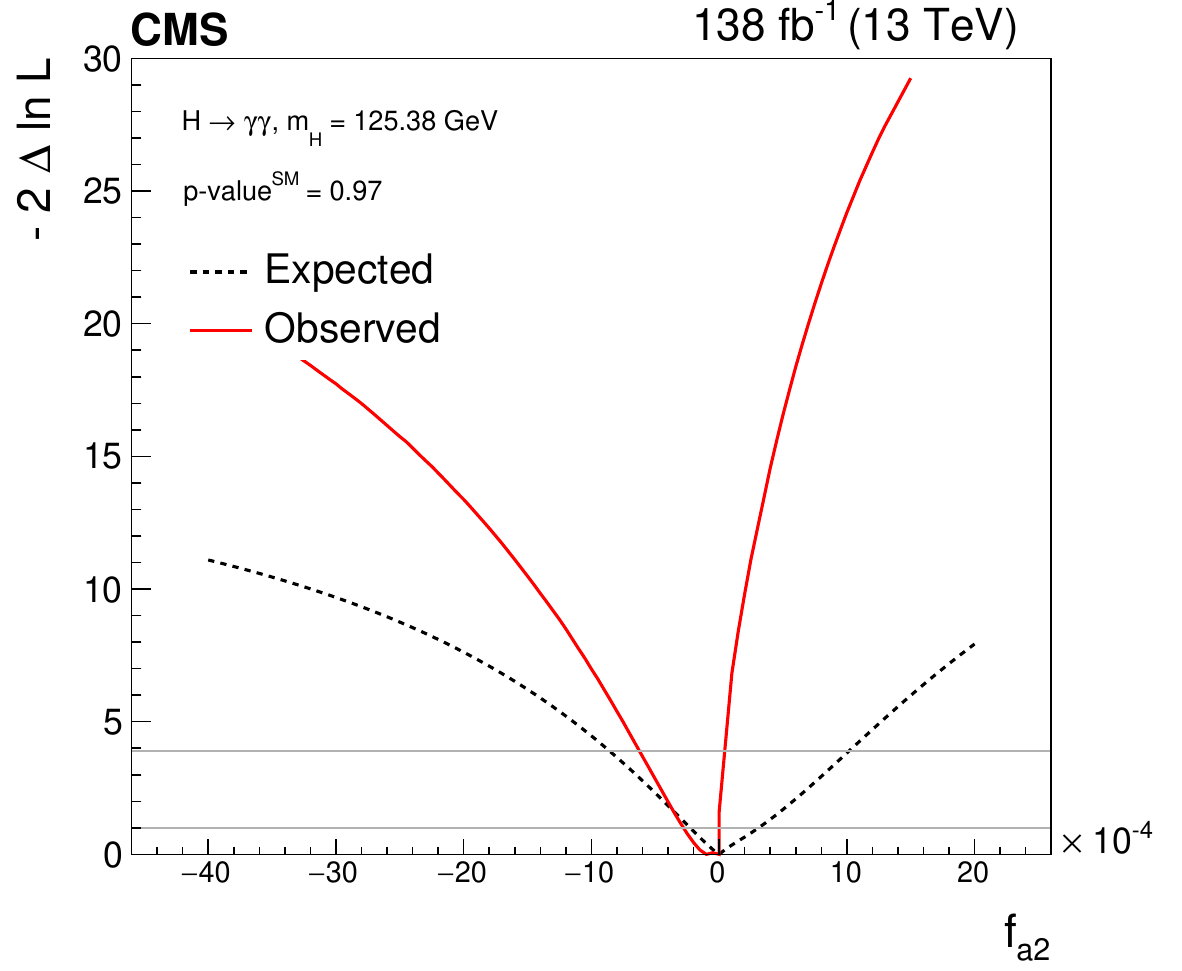} \\
  \includegraphics[width=0.49\textwidth]{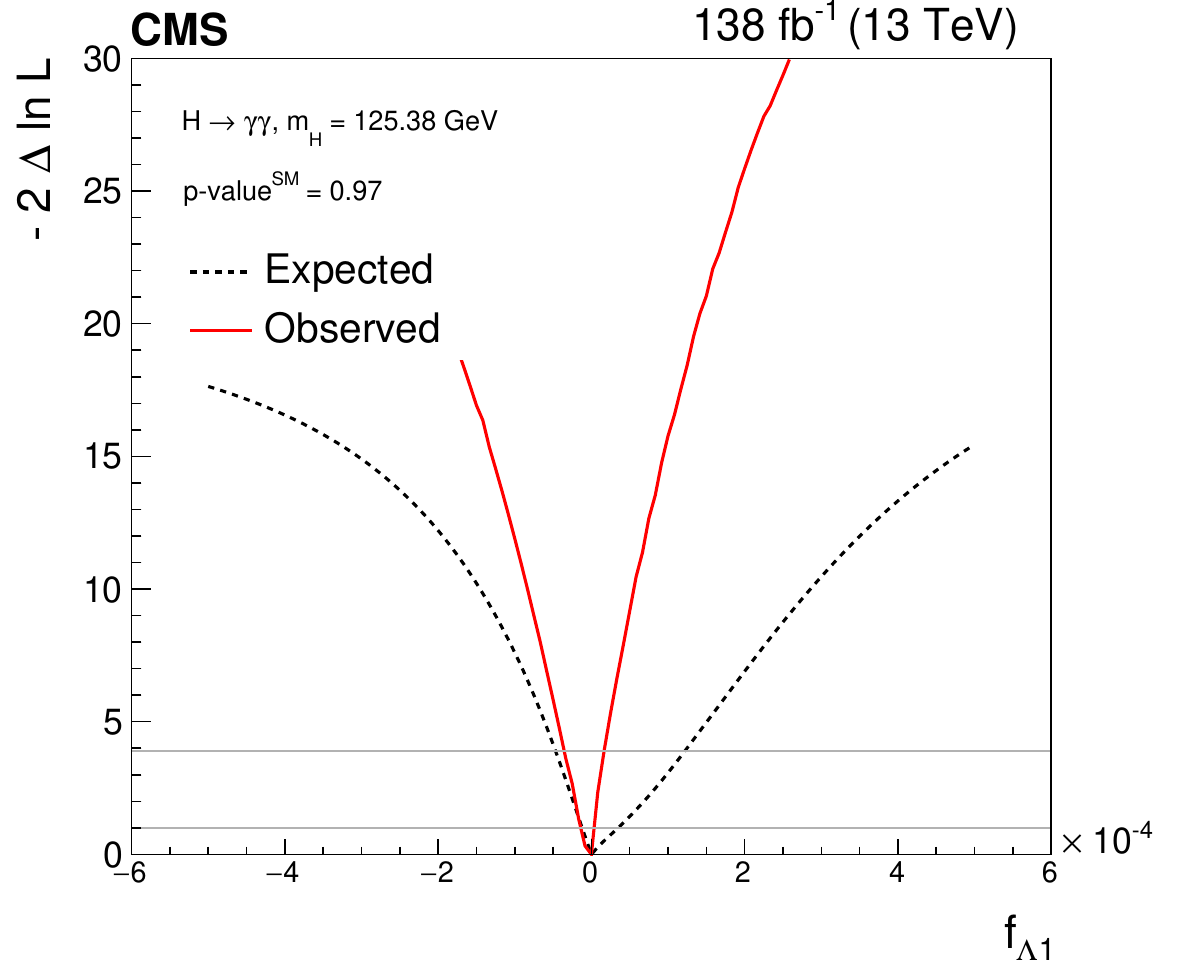}
  \includegraphics[width=0.49\textwidth]{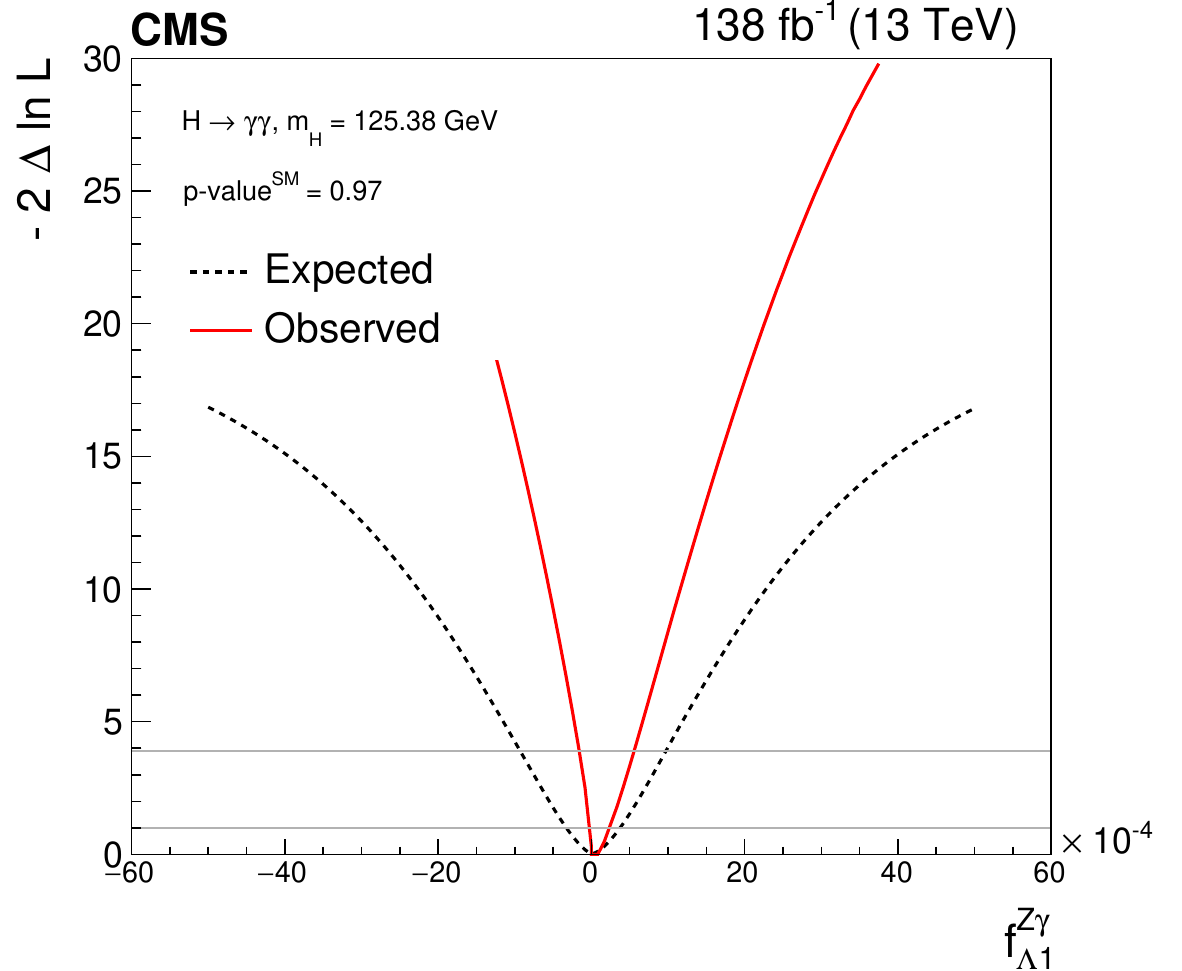}
  \caption{Likelihood scan for the expected and observed constraints of the \HVV anomalous couplings parameters.}
  \label{fig:fai_exp}
\end{figure}

The four $f_i$ parameters describing the anomalous \HVV couplings, as
defined in Eq.~(\ref{eq:formfact-fullampl-spin0}) and Eq.~(\ref{eq:fa_definitions}), are estimated using the likelihood function defined in
Eq.~(\ref{eq:likelihoodDensity}).
Figure~\ref{fig:fai_exp} shows the likelihood profiles for the expected and observed constraints
on the
$\vec{f} = \left(\fB, \fC, \fL, \fLZg\right)$ parameters
using the \VBF and \VH categories. The reported $p$-values are obtained from a
goodness-of-fit test performed using the saturated model test statistic.

\begin{figure}[!htbp]
  \centering
  \includegraphics[width=0.49\textwidth]{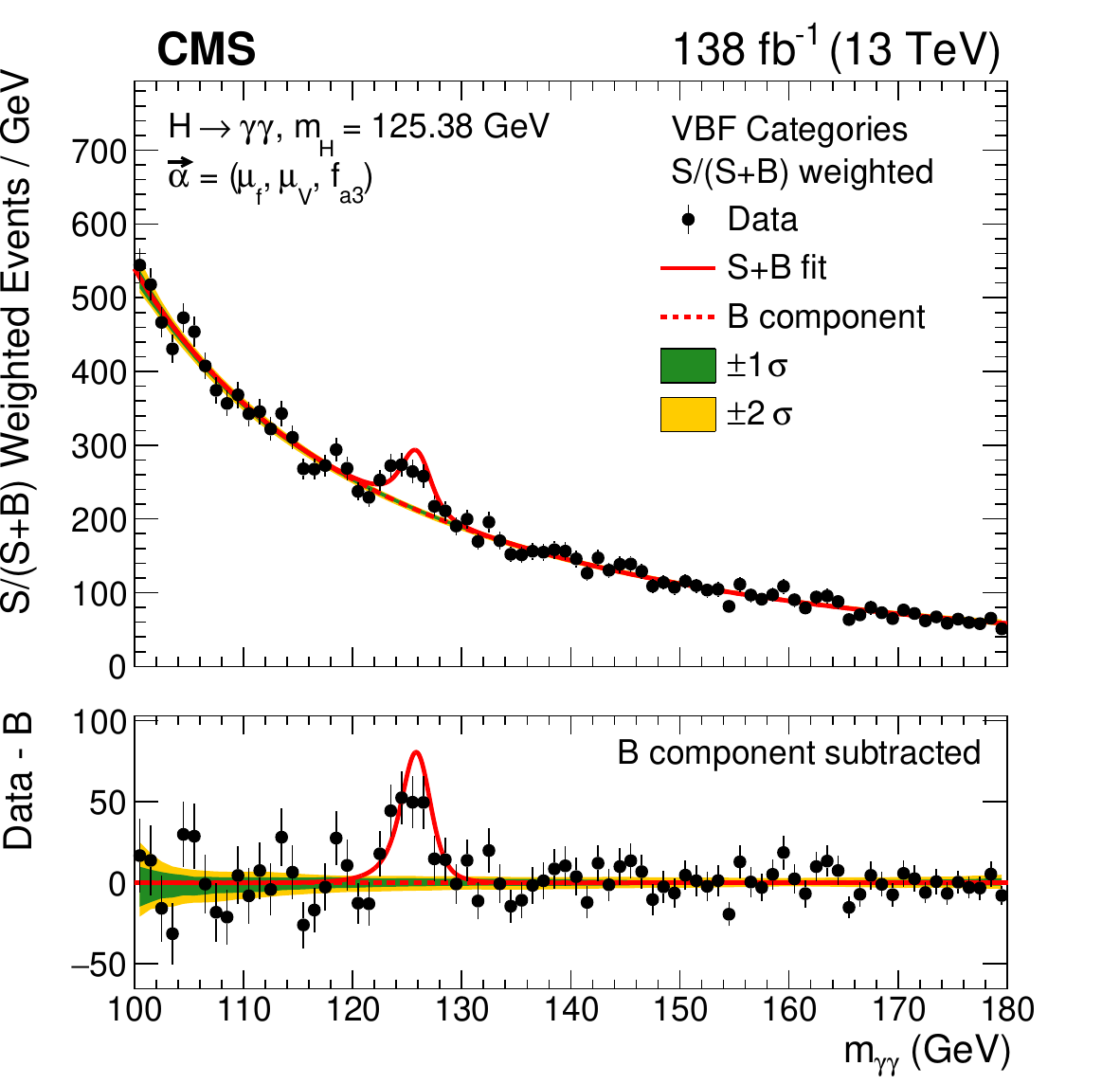}
  \includegraphics[width=0.49\textwidth]{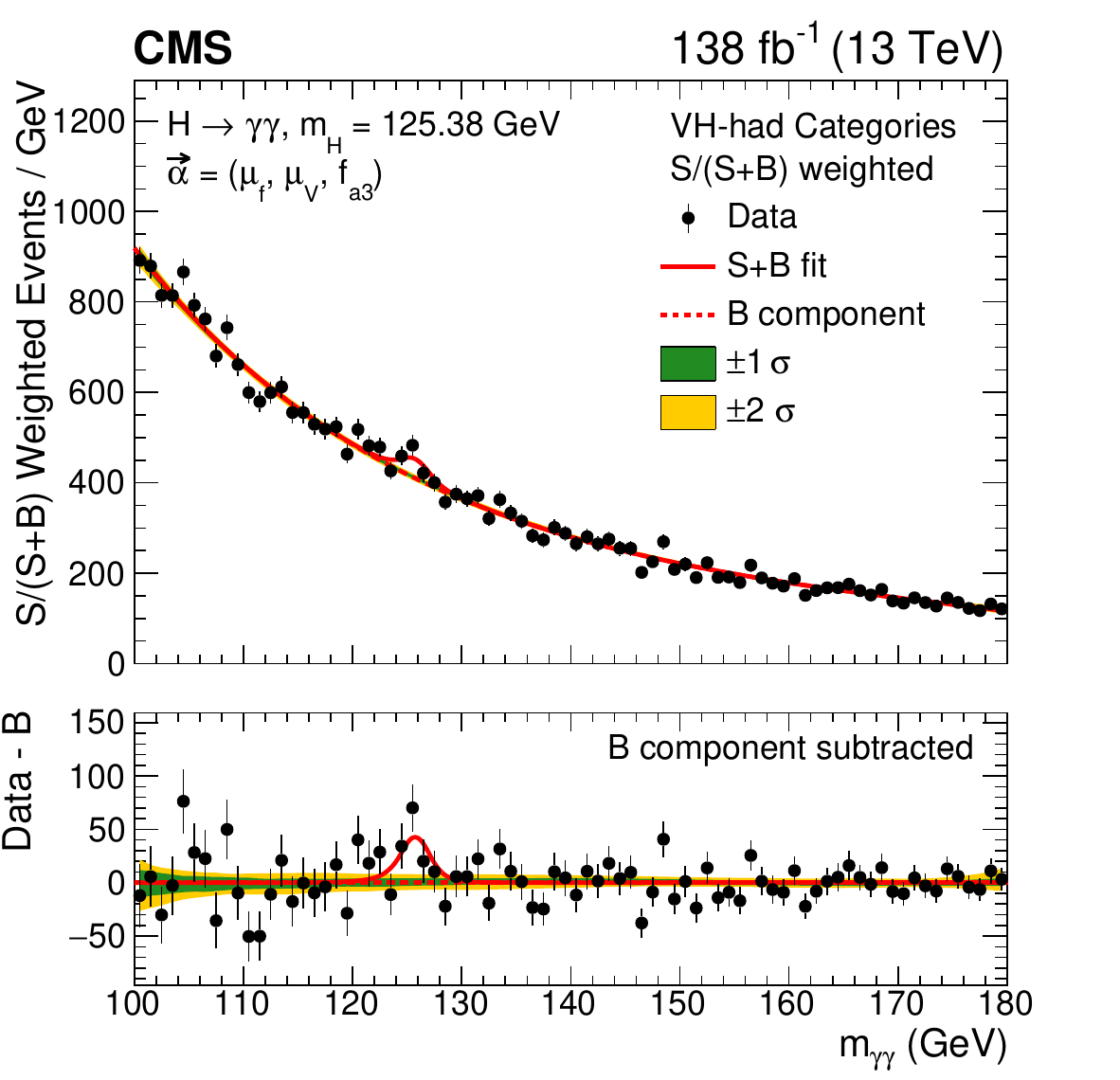}\\
  \includegraphics[width=0.49\textwidth]{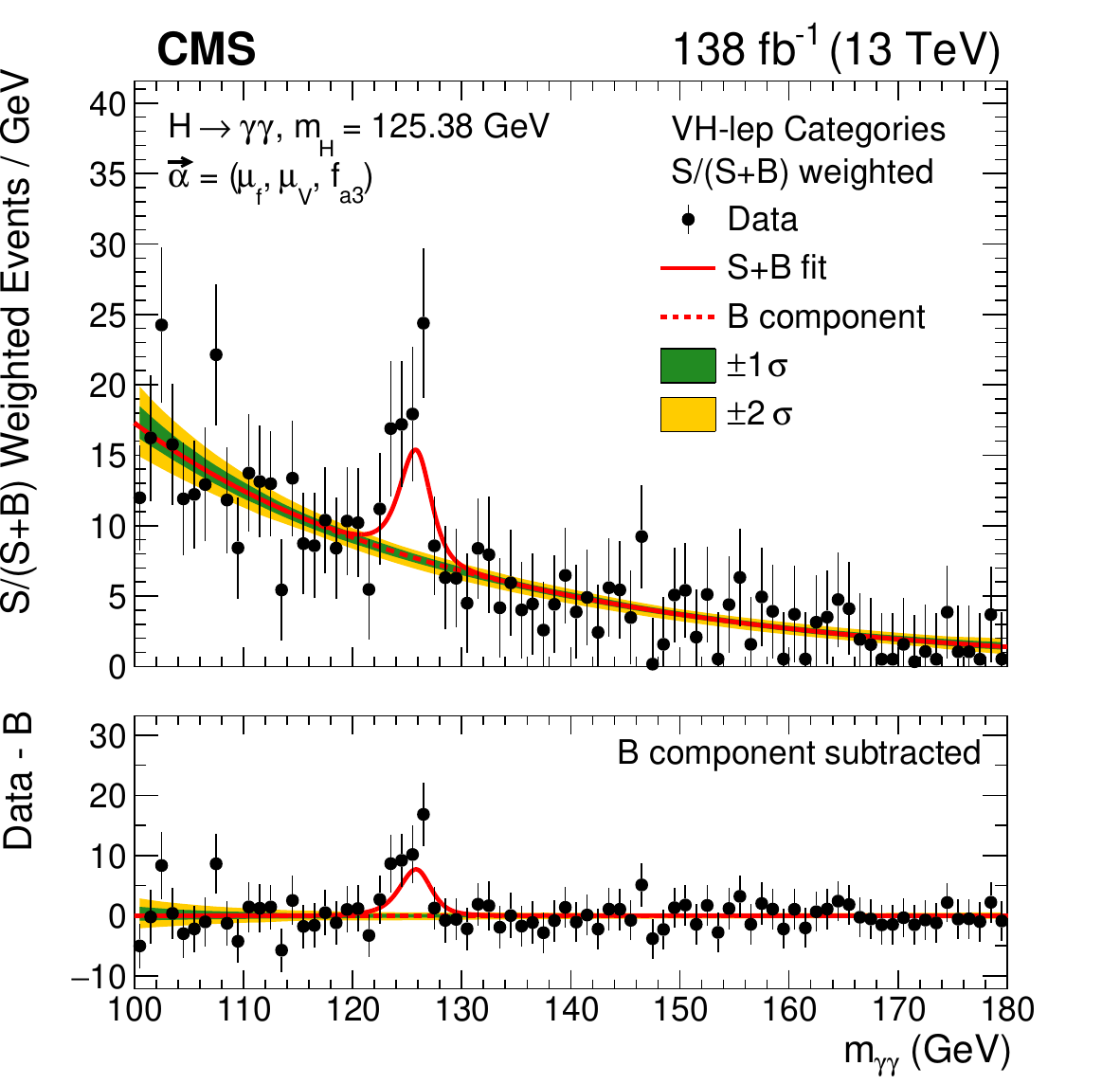}
  \includegraphics[width=0.49\textwidth]{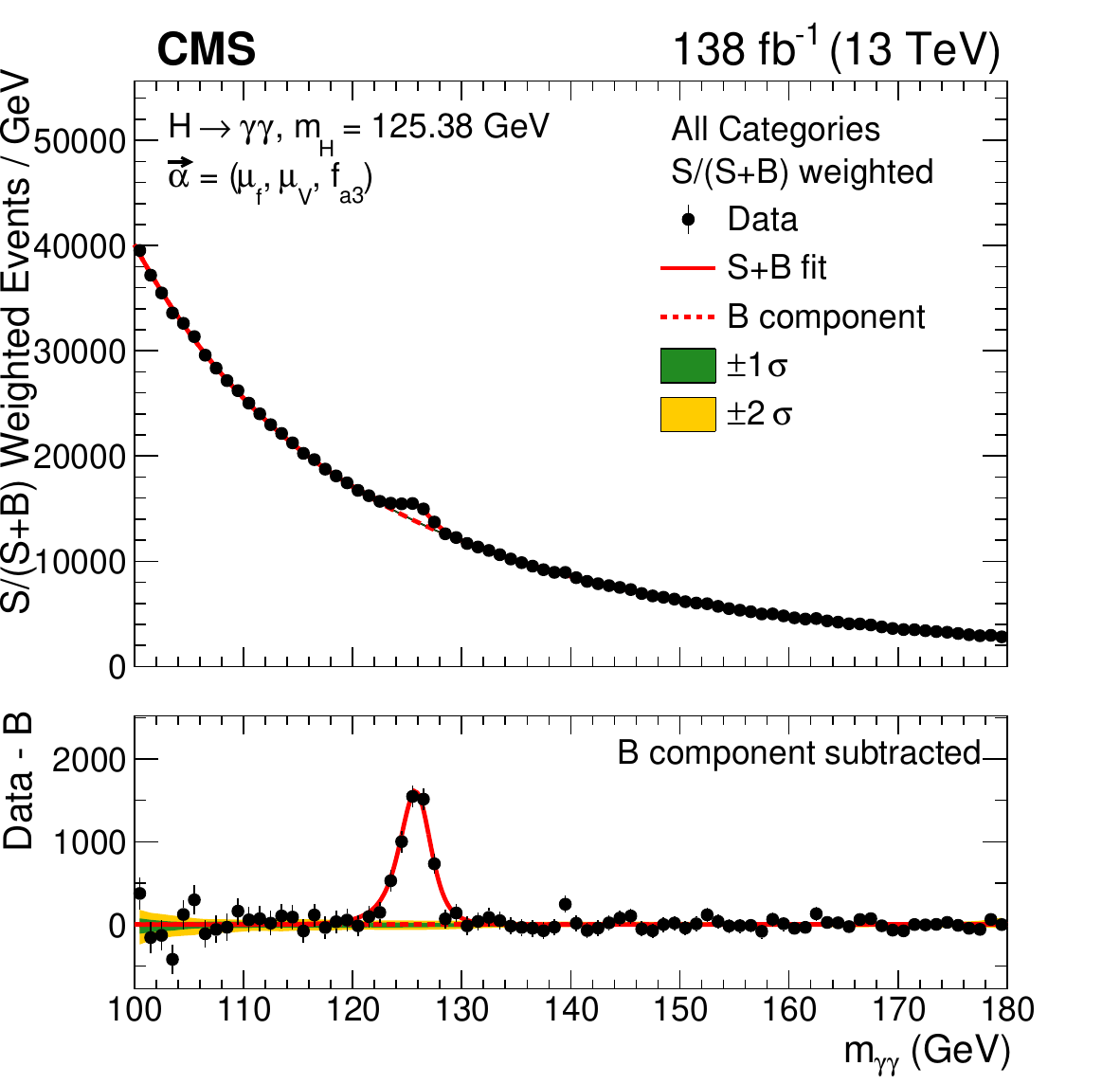}
  \caption{
    Invariant mass distributions are presented separately for categories optimized for \VBF process (upper left),
    hadronic \VH (upper right), and leptonic \VH (lower left).
    The best fit signal-plus-background model is shown overlaid on the S/(S+B)-weighted
    distribution of the data points (black) from the fit to the $f_{a3}$ anomalous coupling parameter.
    The vector $\vec{\alpha}$  denotes the set of parameters allowed to float in the fit.
    The lower right plot shows the combined
    distribution across all categories. Here, S and B represent the expected number of signal and background
    events in the mass peak region. The green and yellow bands correspond to the one and two standard
    deviation uncertainties on the background component of the fit. The solid red line indicates the
    total signal-plus-background prediction, while the dashed red line represents the background-only
    contribution. The lower panel in each plot displays the residuals obtained by subtracting the
    background component from the data.}  \label{fig:results_sPlusBplot_Cats}
\end{figure}

The results are also reported in Tables~\ref{tab:fai_exp} and ~\ref{tab:fai_exp_95}.  The results are obtained considering possible
modifications of the \HVV couplings, with respect to the SM, of one type
at a time, while fixing all the other anomalous coupling parameters to zero.

The signal strength modifiers $\vec{\mu} = (\muf, \mu_V)$ are treated as free parameters
in the fit. In the $f_{a3}$ case, the extracted values are $\mu_V = 1.37^{+0.26}_{-0.22}$ and
$\muf = 1.05^{+0.12}_{-0.10}$. For the $f_{a2}$, $f_{\Lambda 1}$, and $f_{\Lambda 1}^{Z\gamma}$ fits,
the results for $\mu_V$ and \muf are similar with those obtained in the $f_{a3}$ case,
given the common categorization and preselection across the fits, and are compatible with the standard model expectation
with uncertainties similar to the ones reported in the dedicated STXS measurements~\cite{HIG-19-015}.

\begin{figure}[!htbp]
  \centering
  \includegraphics[width=0.5\textwidth]{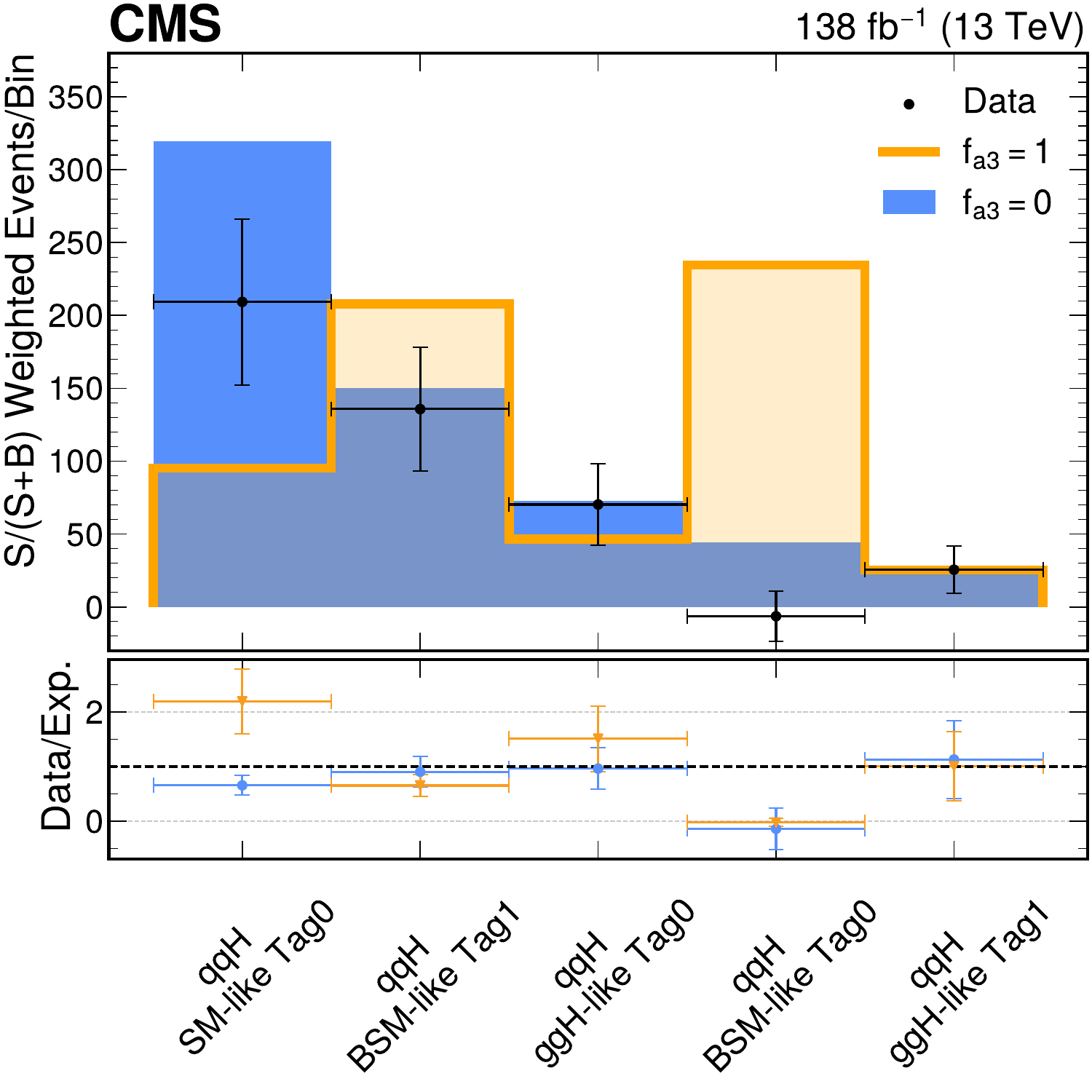}\\
  \caption{ Distribution of events weighted by S/(S+B), using bins optimized for the \VBF process.
    Here, S denotes the sum of all resonant signal events and B represents the nonresonant background.
    The plot shows the event yields in each bin within the mass window
    $m_\PH - \sigma_{eff} < \mgg < m_\PH +  \sigma_{eff}$, for both the full BSM
    hypothesis (orange) and the SM hypothesis (blue).
    The data points (black dots with error bars) indicate the observed events in the same mass window, after background subtraction, and include statistical uncertainties.
    \label{fig:cat_yield}}
\end{figure}

In Table~\ref{tab:fai_exp_95} we also report the comparison with previous analyses in other decay channels, showing that this work provides
stronger 95\% CL constraints
on the parameters $\fB$,$\fL$ and $\fLZg$ than the best CMS results published so far.
Figure~\ref{fig:results_sPlusBplot_Cats} shows the  best fit to data of the signal-plus-background model for each production process, for the
\HVV analysis categories that are most pure in BSM signals and for all the \HVV categories merged together.
As shown in Fig.~\ref{fig:fai_exp}, the observed curve reaches a higher sensitivity than expected. This can be attributed to two main factors:
first, the fitted value of $\muV$ is higher than expected, mainly due to an excess in the \VH channels, still compatible with the SM expectation;
second, the categories with the highest sensitivity to BSM couplings exhibit a deficit in the observed data.
This is illustrated in Fig.~\ref{fig:cat_yield}, which represents the distribution of the events in data and for two different \fC hypotheses (one for a pure $CP$-even state, and one for a pure $CP$-odd state) in categories optimized for the \VBF production process.

\begin{table}[!htbp]
  \centering
  \caption{ Summary of expected and observed
    results in terms of the best fit value and the 68\% CL intervals
    \label{tab:fai_exp}.}
  \begin{tabular}{lcc}
    \multirow{2}{*}{Parameter} & Observed [$\times 10^{-4}$] & Expected [$\times 10^{-4}$] \\
                               & \Hgamgam ($68\%$ CL)        & \Hgamgam ($68\%$ CL)        \\ \hline
    \fC                        & $0.00^{+0.39}_{-0.39}$      & $[+2.1,-2.1]$               \\
    \fB                        & $-0.81_{-2.6}^{+0.65}$      & $[-2.3,+3.1]$               \\
    \fL                        & $-0.014_{-0.14}^{+0.032}$   & $[-0.12,+0.35]$             \\
    \fLZg                      & $0.83_{-0.92}^{+1.5}$       & $[-3.3,+3.7]$               \\
  \end{tabular}
\end{table}

\begin{table}[!htbp]
  \caption{ Summary of expected and observed \HVV anomalous coupling parameter results in terms
    of the 95 \% CL intervals
    for the \HVV analysis described in this paper and, for comparison, from the combination of
    \Hllll + \HTT channels in Ref.~\cite{HIG-20-007}.}\label{tab:fai_exp_95}
  \centering
  \begin{tabular}{lcc}
                 {Parameter} & Observed (Expected) [$\times 10^{-4}$] & Observed (Expected) [$\times 10^{-4}$]               \\
                             & \Hgamgam ($95\%$ CL)                   & $ \PH\to4\ell$ + $\PH\to\tau^+\tau^-$ ($95\%$ CL)~\cite{HIG-20-007} \\ \hline
                 \fC         & $[-1.5,1.5]$\, ($[-5.3,5.3]$)          & $[-0.1,8.8]$\, $([-2.1,2.1])$                                       \\
                 \fB         & $[-5.5,0.4]$\, ($[-8.6,10]$)           & $[-10,25]$\, $([-11,12])$                                           \\
                 \fL         & $[-0.35,0.16]$\, ($[-0.47,1.2]$)       & $[-2.2,1.6]$\, $([-1.1,3.8])$                                       \\
                 \fLZg       & $[-2.4,4.7]$\, ($[-9.3,9.7]$)          & $[-27,41]$\, $([-26,25])$                                           \\
  \end{tabular}

\end{table}

\subsection{Results of the \texorpdfstring{\Hgg}{Hgg} analysis}
\begin{figure}[!htbp]
  \centering

  \includegraphics[width=0.49\textwidth]{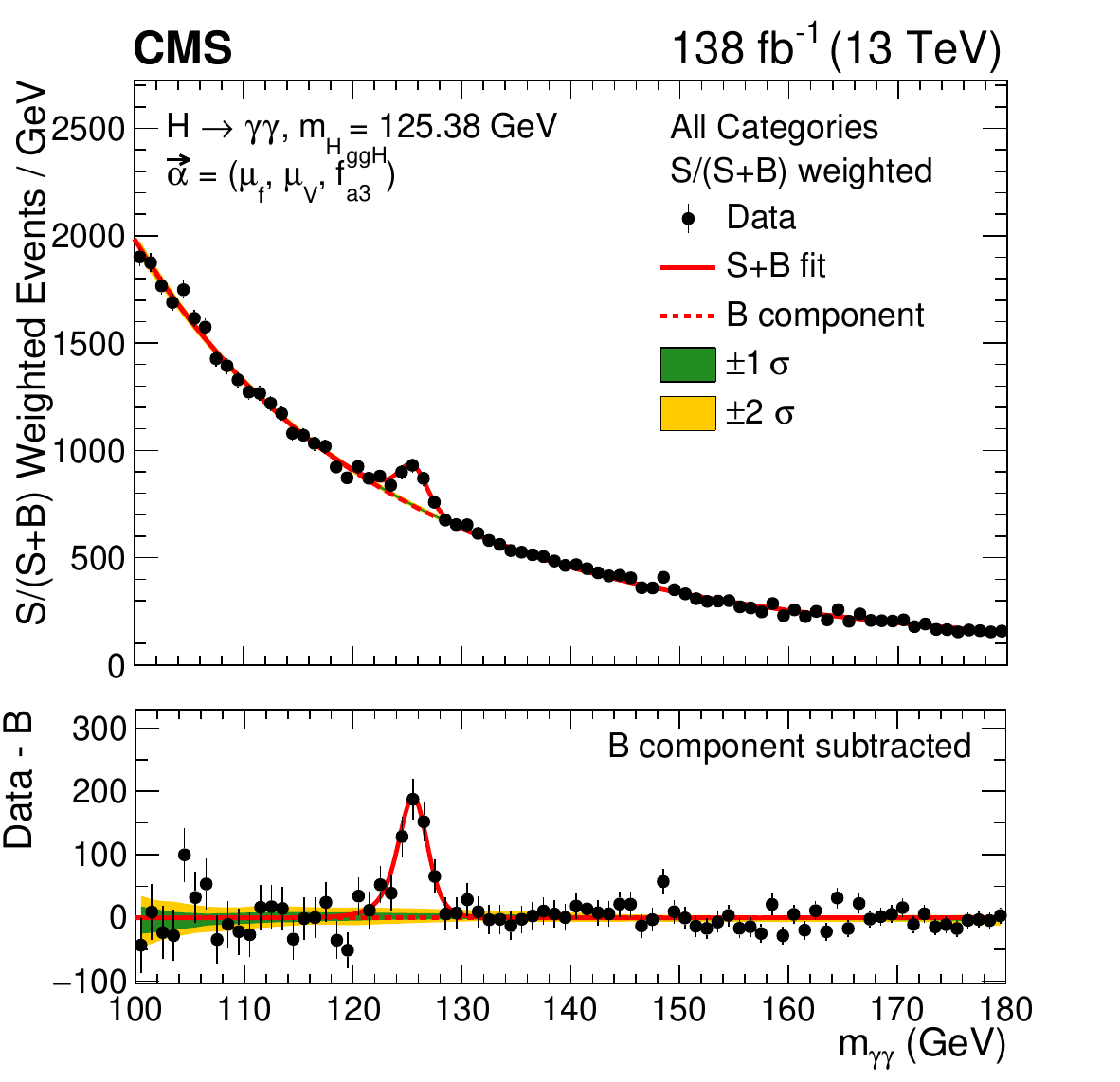}
  \caption{
    Data points (black) and signal-plus-background model fit for the sum of
    all the \Hgg analysis categories weighted by S/(S+B),
    where S is the number of signal events in the mass peak and B is the nonresonant background.
    The vector $\vec{\alpha}$  denotes the set of parameters allowed to float in the fit.
    The one standard deviation (green) and two standard
    deviation (yellow) bands show the uncertainties in the background
    component of the fit.  The solid red line shows the total
    signal-plus-background contribution, whereas the dashed red line
    represents the background component only.  The lower panel shows the residuals
    after subtraction of the background
    component. }
  \label{fig:sigbkgfit_obs}
\end{figure}

In the \Hgg case, the same fitting approach of the \HVV analysis is used, with \fG as
main POI, which parametrizes the fraction of anomalous $CP$-odd Higgs boson
coupling to gluons. The signal strength  parameters $\vec{\mu} = (\muf$, $\muV)$
are fully floated, while \fC is fixed to zero. Figure~\ref{fig:sigbkgfit_obs} shows
the best fit to data of the signal-plus-background model for the sum of
all the \Hgg analysis categories projected onto the \mgg  observed spectrum.

\begin{figure}[!htbp]
  \centering
  \includegraphics[width=0.49\textwidth]{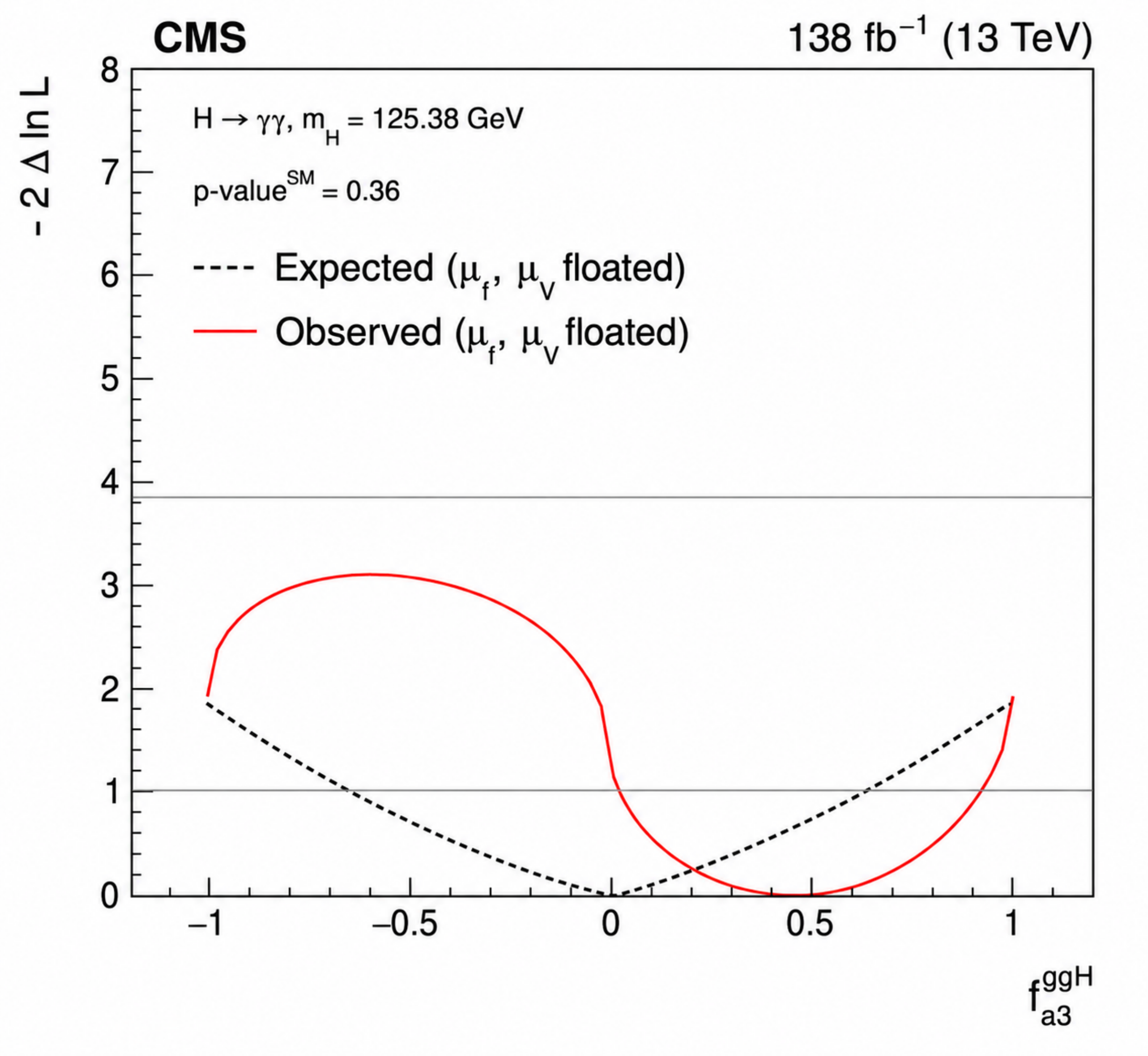}
  \includegraphics[width=0.49\textwidth]{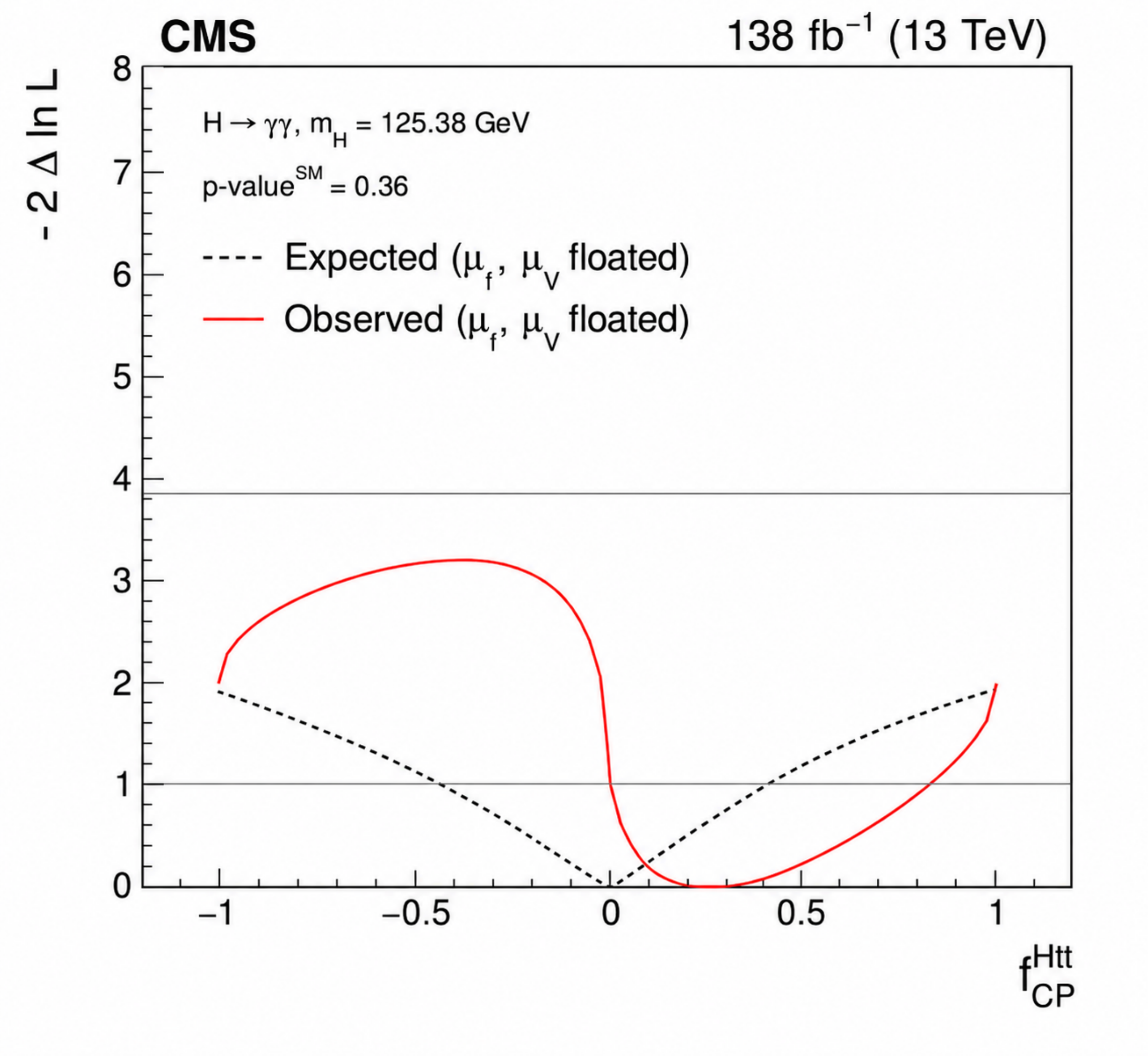}
  \caption{Likelihood profile for the expected and observed constraints on $CP$-odd
    anomalous coupling parameters: \fG (left) and $f_{CP}^{\Htt}$ (right). \label{fig:fai_obs_ggh} }
\end{figure}

The expected and observed likelihood profiles for the POI \fG with $\vec{\mu} = (\muf$, $\muV)$
profiled in the fit are shown in Fig.~\ref{fig:fai_obs_ggh} (left) for the full Run~2 data set.
The fitted $\vec{\mu}$ values are $\muf = 0.82^{+0.25}_{-0.27}$ and $\muV = 1.22^{+0.55}_{-0.55}$,
both consistent with the SM expectation.
As expected the uncertainty on \muf in this analysis is larger with respect to the \HVV one,
since the phase space analysed by the former is restricted to events with two additional jets.
The reported $p$-values are obtained from a
goodness-of-fit test performed using the saturated model test statistic.
The observed one-dimensional constraint on the anomalous coupling parameter \fG is found to be
$0.45^{+0.46}_{-0.42}\stat^{+0.10}_{-0.08}\syst$ at 68\% CL. As expected, the relative contribution
of the systematic uncertainties on the measurement is secondary with respect to the statistical uncertainty.

The difference in shape between the observed and expected distributions arises from the fact that, by construction,
the value of the negative log-likelihood at $\fG = \pm 1$ is constrained to be the same.
These points correspond to scenarios where $a_2 = 0$ and $a_3 = \pm 1$, in which the BSM contribution is maximal. Being proportional to the square of the amplitude,
the cross sections for $a_3 = +1$ and $a_3 = -1$ are expected to be identical.
The observed data appear to favor a value around 0.5, which indicates maximal destructive interference between
the SM and BSM contributions, while a value of $-0.5$ would correspond to maximal constructive interference.
These two opposite scenarios are well illustrated by the likelihood trend, which exhibits a parabolic shape with
a minimum around 0.5 and a maximum around $-0.5$.
In addition, a constraint
on $f_{CP}^{\Htt}$ of $0.26^{+0.57}_{-0.25}$  at 68\% CL is also derived and shown in Fig.~\ref{fig:fai_obs_ggh} (right).

Table~\ref{tab:sensitivity_comparison} illustrates the constraining power of
this analysis in relation to the \Hgg measurements in the other decay channels,
\Hllll~\cite{CMS-HIG-19-009} and \HTT~\cite{Sirunyan:2019htt},
as well as in other processes for the \Hgamgam channels,
\ie, \tH and \ttH, as described in Ref.~\cite{Chatrchyan:2020htt}.
The expected constraint on $\fG$ at 68\% CL is $[-0.65, 0.63]$, significantly more
stringent than the \ggH \Hllll measurement $[-1,1]$ and
a factor of 2 smaller than the corresponding results obtained with the \HTT
decay mode combined with \Hllll~\cite{HIG-20-007} $[-0.26,0.26]$
search (due mainly to larger branching fraction).
In addition, the constraining power on $f_{CP}^{\Htt}$ is equivalent to the combination of the
\ggH, \tH, and \ttH measurements in the \Hllll channels.

\begin{table*}[!tb]
  \caption{
    Summary of results in terms of the best fit values for the $\fG$ and $f_{CP}^{\Htt}$ parameters with the best fit values and allowed
    68\% CL and 95\% CL intervals.
    The \fG constraints obtained in this work are
    compared to those obtained in the \tH and \ttH \Hgamgam~\cite{Chatrchyan:2020htt},
    \ggH \Hllll~\cite{CMS-HIG-19-009},
    $\PH\to\PW\PW$~\cite{CMS:2024bua} and \HTT~\cite{HIG-20-007} channels, respectively.
    The most stringent constraint on \fG has been obtained in ~\cite{HIG-20-007}
    from the combination of the \HTT and \Hllll~\cite{CMS-HIG-19-009} decay channels.
    The interpretation of the $\fG$ result under the assumption of the top quark dominance
    in the gluon fusion loop is presented
    in terms of $f_{CP}^{\Htt}$.
    The most stringent constraint on $f_{CP}^{\Htt}$ comes from
    ~\cite{HIG-20-007}, where the \ggH, \tH and \ttH measurements are
    combined in the  \HTT, \Hllll \cite{CMS-HIG-19-009}, and \Hgamgam ~\cite{Chatrchyan:2020htt}
    decay channels.}
  \label{tab:sensitivity_comparison}
  \centering
  \renewcommand{\arraystretch}{1.15}
  \cmsTable{
    \begin{tabular}{clrrrr}
      \multirow{2}{*}{{Parameter}}   &
      \multirow{2}{*}{{Scenario}}    &
      \multicolumn{2}{c}{{Observed}} &
      \multicolumn{2}{c}{{Expected}}                                                                                                                                                   \\
                                     &                                                                & {68\% CL}                & {95\% CL}       & {68\% CL}       & {95\% CL}       \\
      \hline

      ~~~~\multirow{4}{*}{$ \fG \begin{cases} \\\\\\ \\ \end{cases}$ }
                                     & \ggH ($4\Pell$)~\cite{CMS-HIG-19-009}                          & $-0.04^{+1.04}_{-0.96}$  & $[-1, 1]$       & $[-1, 1]$       & $[-1, 1]$       \\
                                     & \ggH (\PW\PW)~\cite{CMS:2024bua}                               & $0.03^{+0.72}_{-0.38}$   & $[-1, 1]$       & $[-1, 1]$       & $[-1, 1]$       \\
                                     & \ggH ($\PGt\PGt$,$4\Pell$)~\cite{HIG-20-007}                   & $0.07^{+0.32}_{-0.07}$   & $[-0.15, 0.89]$ & $[-0.26, 0.26]$ & $[-1, 1]$       \\
                                     & {\ggH ($\PGg\PGg$)}                                            & {$0.45^{+0.47}_{-0.43}$}
                                     & {$[-1, 1]$}                                                    & {$[-0.65,0.63]$}         & {$[-1, 1]$}                                         \\
      [\cmsTabSkip]

      ~~~~\multirow{8}{*}{$ f_{CP}^{Htt} \begin{cases} \\ \\ \\ \\ \\ \\ \\ \end{cases}$ }
                                     & \tH,~\ttH ($4\Pell$)~\cite{CMS-HIG-19-009}                     & $0.88^{+0.12}_{-1.88}$   & $[-1, 1]$       & $[-1, 1]$       & $[-1, 1]$       \\
                                     & \tH,~\ttH ($\PGg\PGg$)~\cite{Chatrchyan:2020htt}               & $0.00\pm0.33$            & $[-0.67, 0.67]$ & $[-0.49, 0.49]$ & $[-0.82, 0.82]$ \\
                                     & \tH,~\ttH ($4\Pell,~\PGg\PGg$)~\cite{Chatrchyan:2020htt}       & $0.00\pm0.33$            & $[-0.67, 0.67]$ & $[-0.48, 0.48]$ & $[-0.81, 0.81]$ \\
                                     & \ggH ($4\Pell$)~\cite{CMS-HIG-19-009}                          & $-0.01^{+1.01}_{-0.99}$  & $[-1, 1]$       & $[-1, 1]$       & $[-1, 1]$       \\
                                     & \ggH,~\tH,~\ttH ($4\Pell$)~\cite{CMS-HIG-19-009}               & $-0.56^{+1.56}_{-0.44}$  & $[-1, 1]$       & $[-0.47, 0.47]$ & $[-1, 1]$       \\
                                     & \ggH,~\tH,~\ttH ($4\Pell,~\PGg\PGg$)~\cite{Chatrchyan:2020htt} & $-0.04^{+0.38}_{-0.36}$  & $[-0.69, 0.68]$ & $[-0.30, 0.30]$ & $[-0.70, 0.70]$ \\
                                     & \ggH ($\PGt\PGt$,$4\Pell$+ $\PGg\PGg$)~\cite{HIG-20-007}       & $0.03^{+0.17}_{-0.03}$   & $[-0.07, 0.51]$ & $[-0.12, 0.12]$ & $[-0.49, 0.49]$ \\
                                     & {\ggH ($\PGg\PGg$)}                                            & {$0.26^{+0.57}_{-0.25}$}
                                     & {$[-1, 1]$}
                                     & {[$-0.43, 0.41$]}
                                     & {$[-1, 1]$}                                                                                                                                     \\
      [\cmsTabSkip]
    \end{tabular}}
\end{table*}

\section{Summary}
\label{sec:Summary}

A search for possible anomalous interactions between the Higgs boson (H)
and vector bosons and gluons, including potential $CP$-violating effects, has been presented.
The search is based on proton--proton collision data at $\sqrt{s} = 13\TeV$ collected by the CMS experiment
corresponding to an integrated luminosity of 138\fbinv.

The analysis targets the Higgs boson candidates
reconstructed in the diphoton decay channel in events produced
via vector boson fusion and associated production with a vector boson.
For the first time, anomalous couplings between the Higgs
boson and vector boson are studied in this decay channel.
The observed limits provide the most stringent constraints
on some of the targeted effective cross-section fractions
among the CMS results published to date.

This analysis also investigates anomalous Higgs boson coupling
to gluons in events produced via gluon fusion in association with two jets.
The observed constraints on the effective cross section fraction for a
$CP$-odd anomalous H-gluon coupling are comparable to the best CMS
constraints obtained so far in other decay channels.

Since systematic uncertainties largely cancel in the considered
cross-section fractions, all measurements are currently limited by
statistical precision.

\begin{acknowledgments}
  We congratulate our colleagues in the CERN accelerator departments for the excellent performance of the LHC and thank the technical and administrative staffs at CERN and at other CMS institutes for their contributions to the success of the CMS effort. In addition, we gratefully acknowledge the computing centers and personnel of the Worldwide LHC Computing Grid and other centers for delivering so effectively the computing infrastructure essential to our analyses. Finally, we acknowledge the enduring support for the construction and operation of the LHC, the CMS detector, and the supporting computing infrastructure provided by the following funding agencies: SC (Armenia), BMBWF and FWF (Austria); FNRS and FWO (Belgium); CNPq, CAPES, FAPERJ, FAPERGS, and FAPESP (Brazil); MES and BNSF (Bulgaria); CERN; CAS, MoST, and NSFC (China); MINCIENCIAS (Colombia); MSES and CSF (Croatia); RIF (Cyprus); SENESCYT (Ecuador); ERC PRG and PSG, TARISTU24-TK10 and MoER TK202 (Estonia); Academy of Finland, MEC, and HIP (Finland); CEA and CNRS/IN2P3 (France); SRNSF (Georgia); BMFTR, DFG, and HGF (Germany); GSRI (Greece); MATE and NKFIH (Hungary); DAE and DST (India); IPM (Iran); SFI (Ireland); INFN (Italy); MSIT and NRF (Republic of Korea); MES (Latvia); LMTLT (Lithuania); MOE and UM (Malaysia); BUAP, CINVESTAV, CONACYT, LNS, SEP, and UASLP-FAI (Mexico); MOS (Montenegro); MBIE (New Zealand); PAEC (Pakistan); MSHE, NSC, and NAWA (Poland); FCT (Portugal); MESTD (Serbia); MICIU/AEI and PCTI (Spain); MOSTR (Sri Lanka); Swiss Funding Agencies (Switzerland); MST (Taipei); MHESI (Thailand); TUBITAK and TENMAK (T\"{u}rkiye); NASU (Ukraine); STFC (United Kingdom); DOE and NSF (USA).

  \hyphenation{Rachada-pisek} Individuals have received support from the Marie-Curie program and the European Research Council and Horizon 2020 Grant, contract Nos.\ 675440, 724704, 752730, 758316, 765710, 824093, 101115353, 101002207, 101001205, and COST Action CA16108 (European Union); the Leventis Foundation; the Alfred P.\ Sloan Foundation; the Alexander von Humboldt Foundation; the Science Committee, project no. 22rl-037 (Armenia); the Fonds pour la Formation \`a la Recherche dans l'Industrie et dans l'Agriculture (FRIA) and Fonds voor Wetenschappelijk Onderzoek contract No. 1228724N (Belgium); the Beijing Municipal Science \& Technology Commission, No. Z191100007219010, the Fundamental Research Funds for the Central Universities, the Ministry of Science and Technology of China under Grant No. 2023YFA1605804, the Natural Science Foundation of China under Grant No. 12535004, and USTC Research Funds of the Double First-Class Initiative No.\ YD2030002017 (China); the Ministry of Education, Youth and Sports (MEYS) of the Czech Republic; the Shota Rustaveli National Science Foundation (Georgia); the Deutsche Forschungsgemeinschaft (DFG), among others, under Germany's Excellence Strategy -- EXC 2121 ``Quantum Universe" -- 390833306, and under project number 400140256 - GRK2497; the Hellenic Foundation for Research and Innovation (HFRI), Project Number 2288 (Greece); the Hungarian Academy of Sciences, the New National Excellence Program - \'UNKP, the NKFIH research grants K 131991, K 138136, K 143460, K 143477, K 147557, K 146913, K 146914, K 147048, TKP2021-NKTA-64, and 2025-1.1.5-NEMZ\_KI-2025-00004, and MATE KKP and KKPCs Research Excellence and Flagship Research Groups grants (Hungary); the Council of Science and Industrial Research, India; ICSC -- National Research Center for High Performance Computing, Big Data and Quantum Computing, FAIR -- Future Artificial Intelligence Research, and CUP I53D23001070006 (Mission 4 Component 1), funded by the NextGenerationEU program, the Italian Ministry of University and Research (MUR) under Bando PRIN 2022 -- CUP I53C24002390006, PRIN PRIMULA 2022RBYK7T (Italy); the Latvian Council of Science; the Ministry of Science and Higher Education, project no. 2022/WK/14, and the National Science Center, contracts Opus 2021/41/B/ST2/01369, 2021/43/B/ST2/01552, 2023/49/B/ST2/03273, and the NAWA contract BPN/PPO/2021/1/00011 (Poland); the Funda\c{c}\~ao para a Ci\^encia e a Tecnologia (Portugal); the National Priorities Research Program by Qatar National Research Fund; MICIU/AEI/10.13039/501100011033, ERDF/EU, ``European Union NextGenerationEU/PRTR", projects PID2022-142604OB-C21, PID2022-139519OB-C21, PID2023-147706NB-I00, PID2023-148896NB-I00, PID2023-146983NB-I00, PID2023-147115NB-I00, PID2023-148418NB-C41, PID2023-148418NB-C42, PID2023-148418NB-C43, PID2023-148418NB-C44, PID2024-158190NB-C22, RYC2021-033305-I, RYC2024-048719-I, CNS2023-144781, CNS2024-154769 and Plan de Ciencia, Tecnolog{\'i}a e Innovaci{\'o}n de Asturias, Spain; the Chulalongkorn Academic into Its 2nd Century Project Advancement Project, the National Science, Research and Innovation Fund program IND\_FF\_68\_369\_2300\_097, and the Program Management Unit for Human Resources \& Institutional Development, Research and Innovation, grant B39G680009 (Thailand); the Eric \& Wendy Schmidt Fund for Strategic Innovation through the CERN Next Generation Triggers project under grant agreement number SIF-2023-004; the Kavli Foundation; the Nvidia Corporation; the SuperMicro Corporation; the Welch Foundation, contract C-1845; and the Weston Havens Foundation (USA).
\end{acknowledgments}

\bibliography{auto_generated}

\cleardoublepage \appendix\section{The CMS Collaboration \label{app:collab}}\begin{sloppypar}\hyphenpenalty=5000\widowpenalty=500\clubpenalty=5000
\cmsinstitute{Yerevan Physics Institute, Yerevan, Armenia}
{\tolerance=6000
A.~Hayrapetyan, V.~Makarenko\cmsorcid{0000-0002-8406-8605}, A.~Tumasyan\cmsAuthorMark{1}\cmsorcid{0009-0000-0684-6742}
\par}
\cmsinstitute{Institut f\"{u}r Hochenergiephysik, Vienna, Austria}
{\tolerance=6000
W.~Adam\cmsorcid{0000-0001-9099-4341}, L.~Benato\cmsorcid{0000-0001-5135-7489}, T.~Bergauer\cmsorcid{0000-0002-5786-0293}, M.~Dragicevic\cmsorcid{0000-0003-1967-6783}, P.S.~Hussain\cmsorcid{0000-0002-4825-5278}, M.~Jeitler\cmsAuthorMark{2}\cmsorcid{0000-0002-5141-9560}, N.~Krammer\cmsorcid{0000-0002-0548-0985}, A.~Li\cmsorcid{0000-0002-4547-116X}, D.~Liko\cmsorcid{0000-0002-3380-473X}, M.~Matthewman, J.~Schieck\cmsAuthorMark{2}\cmsorcid{0000-0002-1058-8093}, R.~Sch\"{o}fbeck\cmsAuthorMark{2}\cmsorcid{0000-0002-2332-8784}, M.~Shooshtari\cmsorcid{0009-0004-8882-4887}, M.~Sonawane\cmsorcid{0000-0003-0510-7010}, N.~Van~Den~Bossche\cmsorcid{0000-0003-2973-4991}, W.~Waltenberger\cmsorcid{0000-0002-6215-7228}, C.-E.~Wulz\cmsAuthorMark{2}\cmsorcid{0000-0001-9226-5812}
% ...
\par}
\vskip\cmsinstskip
\dag:~Deceased\\
$^{1}$Also at Yerevan State University, Yerevan, Armenia\\
$^{2}$Also at TU Wien, Vienna, Austria\\
% ...
\end{sloppypar}
%%% END EDITABLE REGION %%%
% skeleton_end
\end{document}